\documentclass[12pt,a4paper]{article}
\usepackage{geometry}
\usepackage{jheppub} 
\usepackage{comment}
\usepackage[english]{babel}
\usepackage[labelfont=bf]{caption}
\usepackage{empheq}
\usepackage{bbm}
\usepackage{float}
\usepackage{comment}
\usepackage[utf8]{inputenc}
\usepackage{natbib} 
\usepackage{mathrsfs}
\usepackage[mathscr]{eucal}
\usepackage{amsmath}
\usepackage{amsfonts}
\usepackage[most]{tcolorbox}
\usepackage{amssymb}
\usepackage{graphicx}
\usepackage{scalerel}
\usepackage{multirow}
\usepackage{cancel}
\usepackage{slashed}
\usepackage{booktabs}
\usepackage{braket}
\usepackage{hyperref}
\usepackage{bm}
\usepackage{multicol}
\usepackage{multirow}
\usepackage{wasysym}

\DeclareCaptionLabelFormat{andtable}{#1~#2  \&  \tablename~\thetable} 

\allowdisplaybreaks 
\usepackage{fancyhdr}
\newcommand{\cL}{\mathcal{L}}
\usepackage{graphicx,scalerel}

\usepackage[shortlabels]{enumitem}
\setlist{itemsep=.1em,topsep=.5em}
\SetEnumerateShortLabel{i}{\textit{\roman*}}

\numberwithin{equation}{section}
\numberwithin{figure}{section}
\numberwithin{table}{section}

\usepackage[noindentafter]{titlesec} 
\DeclareMathAlphabet{\mathsfit}{OT1}{lmss}{m}{sl}
\DeclareMathAlphabet{\mathsfbf}{OT1}{lmss}{bx}{n}
\DeclareMathAlphabet{\mathsfbfit}{OT1}{lmss}{bx}{sl}
\DeclareMathVersion{chaptermath}
\SetSymbolFont{operators}{chaptermath}{OT1}{cmbr}{m}{n}
\SetSymbolFont{letters}{chaptermath}{OML}{cmbrm}{b}{it}
\SetSymbolFont{symbols}{chaptermath}{OMS}{cmbrs}{m}{n}
\DeclareMathVersion{subsectionmath}
\SetSymbolFont{operators}{subsectionmath}{OT1}{cmbr}{m}{n}
\SetSymbolFont{letters}{subsectionmath}{OML}{cmbrm}{m}{it}
\SetSymbolFont{symbols}{subsectionmath}{OMS}{cmbrs}{m}{n}

\titleformat{\chapter}{\bfseries \huge  }{\thechapter. }{-1pt}{}{}
\titlespacing{\chapter}{0pt}{10pt plus 1pt minus 1pt}{10pt plus 1pt minus 1pt}

\titleformat{\section}{\large \bfseries   }{\thesection}{10pt}{}{}
\titlespacing{\section}{0pt}{15pt plus 1pt minus 1pt}{5pt plus 1pt minus 1pt}
\titleformat{\subsection}{\normalsize \bfseries  }{\thesubsection}{10pt}{}{}
\titlespacing{\subsection}{0pt}{15pt plus 1pt minus 1pt}{5pt plus 1pt minus 1pt}
\titleformat{\subsubsection}{\normalsize \bfseries    }{\thesubsubsection}{10pt}{}{}
\titlespacing{\subsubsection}{0pt}{15pt}{5pt}

\begin{document}

\renewcommand{\headrulewidth}{0pt}
\renewcommand{\footrulewidth}{0pt}

\begin{flushright}
MITP-23-046\\
August 31, 2023
\end{flushright}

\thispagestyle{empty}
\begin{center}
\vspace{0.8cm}
{\Large\bf \bm{$K^\pm\to\pi^\pm a$} at Next-to-Leading Order in\\ 
Chiral Perturbation Theory and\\[3mm]
Updated Bounds on ALP Couplings}

\vspace{1cm}
{Claudia Cornella$^a$, Anne Mareike Galda$^a$, Matthias Neubert$^{a,b}$, and Daniel Wyler$^c$}\\[0.2cm]
{\em ${}^a$\,PRISMA$^+$ Cluster of Excellence {\em \&} MITP, 
Johannes Gutenberg University\\ 
55099 Mainz, Germany\\[2mm]
${}^b$\,Department of Physics \& LEPP, Cornell University, Ithaca, NY 14853, U.S.A.\\[2mm]
${}^c$\,Physik-Institut, Universit\"at Z\"urich, CH-8057 Z\"urich, Switzerland}
\end{center}
\vspace{3mm}

\begin{center}
\textbf{Abstract}
\end{center}
\vspace{-0.2cm}

\begin{abstract}

The weak decays $K^\pm\to\pi^\pm a$ offer a powerful probe of axion-like particles (ALPs). In this work, we provide a comprehensive analysis of these processes within chiral perturbation theory, extending existing calculations by including complete next-to-leading order (NLO) contributions and isospin-breaking corrections at first order in $(m_d-m_u)$. We show that the consistent incorporation of ALPs in the QCD and weak chiral Lagrangians requires a non-trivial extension of the corresponding operator bases, which we describe in detail. Furthermore, we show that in the presence of an ALP the so-called ``weak mass term'', which is unobservable in the Standard Model, is non-redundant already at leading order. We find that NLO corrections associated with flavor-violating ALP couplings modify the leading-order result by a few percent, with negligible uncertainties. NLO corrections proportional to flavor-conserving ALP couplings lead to potentially larger corrections, which, however, are accompanied by sizable uncertainties mainly due to the currently limited knowledge of various low-energy constants. We study how these corrections impact bounds on the ALP couplings, first model independently, and then specializing to the case of an ALP with flavor-universal couplings in the UV. Our findings confirm that the decays $K^\pm\to\pi^\pm a$ provide the strongest particle-physics constraints for $m_a\lesssim 300$\,MeV. In addition, we point out that these bounds have interesting implications for the ALP couplings to nucleons, which were so far only constrained by astrophysical measurements and non-accelerator experiments.

\end{abstract}

\newpage\maketitle 
\vspace{-4mm}

\section{Introduction}

Axions and axion-like particles (collectively referred to as ALPs in this work) are light pseudoscalar bosons arising in a large class of well-motivated extensions of the Standard Model (SM). They offer an elegant solution to the strong CP problem through the Peccei--Quinn mechanism \cite{Peccei:1977hh,Weinberg:1977ma,Wilczek:1977pj,Bardeen:1977bd,Kim:1979if,Shifman:1979if,Dine:1981rt,Zhitnitsky:1980tq} and may also provide insights into the  flavor structure underlying the SM \cite{Davidson:1981zd, Calibbi:2016hwq,Ema:2016ops}. More generally, ALPs can emerge as pseudo Nambu--Goldstone bosons in models featuring explicit global symmetry breaking. Their mass is protected by an approximate shift symmetry, allowing them to be much lighter than the symmetry-breaking scale. Due to this symmetry, ALP couplings to SM particles arise at dimension-five order and higher, and are therefore naturally suppressed by powers of the scale of global symmetry breaking. This makes the ALP a prominent example of a weakly coupled new-physics particle. In light of the absence of any discoveries of new heavy degrees of freedom at the LHC and at flavor factories, there has been renewed theoretical interest in ALP models. Search strategies for ALPs include a large variety of cosmological observations \cite{Cadamuro:2011fd,Millea:2015qra}, astrophysical measurements \cite{Payez:2014xsa,Jaeckel:2017tud}, and collider probes \cite{Mimasu:2014nea,Jaeckel:2015jla,Knapen:2016moh,Brivio:2017ije,Bauer:2017ris,Bauer:2018uxu}, as well as precision studies of flavor-violating transitions in the quark and lepton sectors \cite{Batell:2009jf,Freytsis:2009ct,Dolan:2014ska,MartinCamalich:2020dfe,Bauer:2019gfk,Bauer:2021mvw}.

Low-energy weak-interaction processes impose some of the most stringent bounds on ALP couplings to SM particles. This was realized shortly after the axion was originally proposed \cite{Bardeen:1978nq,Antoniadis:1981zw,Krauss:1986bq,Bardeen:1986yb} and has been further explored in recent years \cite{Freytsis:2009ct,Batell:2009jf,Dobrich:2018jyi,Cornella:2019uxs,Guerrera:2021yss}. In \cite{Georgi:1986df}, Georgi, Kaplan and Randall derived the effective chiral Lagrangian accounting for interactions between a light ALP (with mass below the scale of chiral symmetry breaking, $\mu_\chi=4\pi F_\pi\approx 1.6$\,GeV) and light pseudoscalar mesons, enabling a model-independent description that does not rely on the details of Peccei--Quinn symmetry breaking. Despite the long history of the subject, even several recent studies on weak decays, such as $K^\pm\to\pi^\pm a$ and $\pi^\pm\to e^\pm\nu_e a$, have omitted the contributions of relevant Feynman diagrams, leading to incomplete expressions for the decay amplitudes (see e.g.\ \cite{Bjorkeroth:2018dzu,Ertas:2020xcc,Gori:2020xvq}). Often, the amplitudes were derived starting from the amplitude for an analogous decay process involving a $\pi^0$ or $\eta$ meson, and then accounting for the (kinetic) mixing of the ALP with these neutral mesons by means of mixing angles $\theta_{\pi^0 a}$ and $\theta_{\eta_8 a}$, neglecting the fact that these mixing angles are unphysical and depend on the parameters of the chiral rotation used to eliminate the ALP--gluon coupling in the effective chiral Lagrangian. 

Recently, this problem has been reanalyzed and an omission in the representation of the weak-interaction quark currents in the effective theory has been corrected \cite{Bauer:2021wjo}, with important consequences. The chiral Lagrangian for weak-interaction processes established in \cite{Bauer:2021wjo} predicts a result for the $K^\pm\to\pi^\pm a$ decay rates approximately 37 times larger than the result obtained in \cite{Georgi:1986df} and used in several later works. Based on this finding, a detailed phenomenological study has shown that the decays $K^\pm\to\pi^\pm a$ and $K_L\to\pi^0 a$ impose the strongest constraints on the ALP couplings to gluons and light quarks in the region where $m_a$ is lighter than about 300\,MeV \cite{Bauer:2021mvw}. These calculations were performed at leading order (LO) in chiral perturbation theory, making it challenging to assign them a theoretical uncertainty. Given the significance of the subject, it is important to push the calculations to next-to-leading order (NLO), which is the primary objective of the present work. 

This paper is organized as follows: In Section~\ref{sec:effective_Lagrangians} we outline how to consistently incorporate the ALP in the strong- and weak-interaction chiral Lagrangians at $\mathcal{O}(p^2)$ and $\mathcal{O}(p^4)$. We extend the existing operator bases for the $\mathcal{O}(p^4)$ Lagrangians by including new operators which exist only in the presence of an ALP. We show that the ``weak mass term'', which is unobservable in the SM, yields a non-vanishing contribution to the $K^\pm\to\pi^\pm a$ decay amplitudes starting at LO in the chiral expansion. We also comment on how the $\eta'$ meson could be consistently included in the chiral Lagrangian in the presence of an ALP. In Section~\ref{sec:decayamplitude} we use these results to compute the $K^\pm\to\pi^\pm a$ decay amplitudes at NLO in the chiral expansion, including first-order isospin-breaking corrections in the mass difference $(m_u-m_d)$ at LO. In Section~\ref{sec:Phenomenology} we give detailed numerical results for the magnitude of the NLO corrections and their theoretical uncertainties. We apply our findings to derive updated bounds on the ALP couplings to gluons and quarks at low energies. In the context of a model with a flavor-universal ALP, we obtain updated bounds on the ALP couplings to gauge bosons and quarks at the high (UV) scale of Peccei--Quinn symmetry breaking. Finally, we comment on the implications of such bounds for the ALP couplings to nucleons. The main results and conclusions are summarized in Section~\ref{sec:conclusions}. Details of the construction of the operator basis for the weak-interaction chiral Lagrangian at $\mathcal{O}(p^4)$ are presented in Appendix~\ref{app:A}, and in Appendix~\ref{app:B} we collect explicit expressions detailing the calculations in Section~\ref{sec:decayamplitude}. These can be also found in a \texttt{Mathematica} notebook attached to the arXiv version of this paper.

\section{Effective Lagrangians}
\label{sec:effective_Lagrangians}

\subsection{Effective ALP interactions at low energies}
\label{sec:effective_alp_interactions}

Our starting point is the effective Lagrangian describing the interactions of the ALP with the particles of the SM just above the scale of spontaneous breaking of the QCD chiral symmetry $G_\chi=SU(3)_L\times SU(3)_R\to SU(3)_V$, i.e.\ $\mu_\chi=4\pi F_\pi\approx 1.6$\,GeV. At this scale, the  active quark flavors are the light quarks, $u,d,s$, which we collectively denote by $q=(u,d,s)$. The relevant effective Lagrangian up to dimension-5 order is
\begin{align}
   \cL_{\mathrm{eff}} 
   &= \cL_{\mathrm{QCD}} + \frac12(\partial_\mu a)(\partial^\mu a) 
    - \frac{m_{a,0}^2}{2}\,a^2 
    + c_{GG}\,\frac{\alpha_s}{4\pi}\, \frac{a}{f}\,G_{\mu \nu}^a\tilde{G}^{\mu \nu,a} + c_{\gamma \gamma}\,\frac{\alpha}{4\pi}\,\frac{a}{f}\, F_{\mu \nu}\tilde{F}^{\mu \nu} \notag\\
   &\quad + \frac{\partial_\mu a}{f} \left( \bar q_L k_Q\gamma^\mu q_L 
    + \bar q_R k_q\gamma^\mu q_R \right) ,
\label{eq:lag_alp_UV}
\end{align}
where $F_{\mu\nu}$ and $G_{\mu\nu}^a$ are the field-strength tensors of the photon and the gluon fields, respectively. The corresponding dual field-strength tensors are defined by $\tilde X^{\mu\nu}=\frac12\epsilon^{\mu\nu\alpha\beta} X_{\alpha \beta}$ for $X=F,G$ (with $\epsilon^{0123}=+1$). The parameter $f$ is the ALP decay constant. The quantities $k_Q$ and $k_q$ are hermitian $3\times 3$ matrices parametrizing the ALP couplings to left-handed and right-handed quarks, respectively. We omit ALP interactions with leptons as they are irrelevant for this study. The QCD Lagrangian is given by
\begin{align}
\begin{aligned}
   \cL_{\mathrm{QCD}} 
   = - \frac{1}{4} G_{\mu \nu}^a G^{\mu \nu,a} + \bar{q} i\slashed{D} q 
    - \left( \bar q_L\, m_q\, q_R  + \mathrm{h.c.} \right) , 
\label{eq:lag_QCD_light_flavors}
\end{aligned}
\end{align}
and we choose to work in the mass basis, where $m_q=\mathrm{diag}(m_u,m_d,m_s)$. 

In terms of the ALP--quark couplings as defined in \cite{Bauer:2020jbp}, our fermion couplings are given by
\begin{align}
\label{eq:UVcouplings}
   k_Q = \begin{pmatrix}
    [k_{U}]_{11} & 0 & 0 \\ 
    0 & [k_{D}]_{11} & [k_{D}]_{12} \\   
    0 & [k_{D}]_{21} & [k_{D}]_{22}
   \end{pmatrix} \,, \qquad 
   k_q = \begin{pmatrix}
    [k_{u}]_{11} & 0 & 0 \\ 
    0 & [k_{d}]_{11} & [k_{d}]_{12} \\   
    0 & [k_{d}]_{21} & [k_{d}]_{22}
   \end{pmatrix} \,.
\end{align}
The matrices $k_U$ and $k_D$ are connected via the CKM matrix $V$ via $k_D=V^\dagger k_U V$. Disregarding two-loop QED effects, the scale dependence of the fermion couplings is dictated by the evolution equations \cite{Bauer:2020jbp}
\begin{align}
   \frac{d}{d\ln\mu}\,k_q(\mu) = - \frac{d}{d\ln\mu}\,k_Q(\mu) 
   = \frac{\alpha_s^2}{\pi^2} \,\tilde{c}_{GG}(\mu)\,\mathbbm{1} \,,
\end{align}
with 
\begin{align}
   \tilde{c}_{GG}(\mu) 
   = c_{GG} + \frac{1}{2}\,\big[ \braket{k_q(\mu)} - \braket{k_Q(\mu)} \big] \,,
\end{align}
where $\braket{\dots}$ denotes the trace over flavor indices. The flavor off-diagonal couplings are scale independent. Solving the evolution equations, one can relate the couplings defined at the low scale $\mu_\chi$ to ALP couplings at the electroweak scale. Note the important fact that at the electroweak scale flavor off-diagonal couplings $\left[k_D\right]_{12}=\left[k_D\right]_{21}^*$ are generated from one-loop diagrams involving $W$ bosons, even in models in which the ALP couplings at very high energies are flavor diagonal \cite{Bauer:2020jbp,Izaguirre:2016dfi,Gavela:2019wzg}.

The quark bilinears in the effective Lagrangian \eqref{eq:lag_alp_UV} and the ALP--gluon coupling can be written in the general form
\begin{align}
   \cL_{\mathrm{eff}}
   \ni \bar q(x) \big[  l_\mu(x)\,\gamma^\mu P_L + r_\mu(x)\,\gamma^\mu P_R 
    - s + i\gamma_5\,p \big] q(x) 
    - \frac{\alpha_s}{8\pi}\,\theta(x)\,G_{\mu\nu}^a(x)\,\tilde G^{\mu\nu,a}(x) \,,
\label{eq:lag_quark_sources}
\end{align}  
where $s=m_q$, $p=0$, and we have defined the local ALP sources
\begin{align}
\begin{aligned}
   l_\mu(x) = k_Q\,\frac{\partial_\mu a(x)}{f} \,, \qquad 
   r_\mu(x) = k_q\,\frac{\partial_\mu a(x)}{f} \,, \qquad 
   \theta(x) = - 2 c_{GG}\,\frac{a(x)}{f} \,. 
\label{eq:alp_sources}
\end{aligned}
\end{align}
The left-handed and right-handed sources can be combined into a vector and an axial-vector source according to
\begin{align}
   v_\mu = \frac{l_\mu+r_\mu}{2} = c^v\,\frac{\partial_\mu a}{2f} \,, \qquad
   a_\mu =\frac{r_\mu-l_\mu}{2} = c^a\,\frac{\partial_\mu a}{2f} \,,
\end{align}
where we have defined\footnote{We denote the diagonal entries of these matrices by $c^{v,a}_{uu}$, $c^{v,a}_{dd}$, and $c^{v,a}_{ss}$. Our parameters $c_{qq}^a$ are identical to the couplings $c_{qq}$ used in \cite{Bauer:2021mvw}.}
\begin{align}
   c^v = k_q + k_Q \,, \qquad
   c^a = k_q - k_Q \,.
\end{align}
It is well established that performing the local chiral transformation \cite{Bardeen:1986yb,Georgi:1986df,Srednicki:1985xd} 
\begin{align}
   q(x) \to e^{\frac{i}{2}\theta(x)\,\kappa_q\gamma_5}\,q(x) 
\label{eq:chiral_rotation}
\end{align}
removes the ALP coupling to gluons in \eqref{eq:lag_quark_sources} term as long as
\begin{align}
   \braket{{\kappa}_q}= \kappa_u + \kappa_d + \kappa_s = 1 \,.
\label{eq:trace_kappas}
\end{align}
Once this constraint is met, the parameters $\kappa_q$ must not affect predictions for physical quantities \cite{Bauer:2020jbp}.

\subsection[QCD chiral Lagrangian at $\mathcal{O}(p^2)$]{QCD chiral Lagrangian at $\bm{\mathcal{O}(p^2)}$}
\label{sec:QCDp2Lagrangian}

Below the scale $\mu_\chi$, the light pseudoscalar mesons $\pi$, $K$ and $\eta$ take the place of light quarks as dynamical degrees of freedom. The ALP interactions with these mesons are best described in terms of an effective chiral Lagrangian involving the meson octet and the external sources in \eqref{eq:alp_sources} \cite{Georgi:1986df,Bauer:2020jbp}. The meson octet ${\Sigma}_0$ is defined as 
\begin{align}
   \Sigma_0(x) 
   = \mathrm{exp} \left[ \frac{i\sqrt{2}}{F}\,{\Phi}(x) \right] , \qquad 
   \Phi(x) = \lambda^a\,\pi^a(x) \,.
\label{eq:meson_octet}
\end{align}
Here, $F$ is the meson  decay constant in the chiral limit (we use the normalization $F\approx F_\pi\simeq 130$\,MeV), and ${\lambda}^a$ (with $a=1,\dots,8$) are the Gell-Mann matrices. With this definition, $\Sigma_0$ satisfies 
\begin{align}
   \mathrm{det}\,\Sigma_0 = 1\,, \qquad 
   \braket{\Sigma_0\,i(D_\mu \Sigma_0)^\dagger} = 0 \,,    
\label{eq:octet_properties} 
\end{align}
where the second equality follows from the first one by taking a derivative of $\mathrm{det}\,\Sigma_0=\exp[\braket{\ln\Sigma_0}]$. An important point to notice is that in presence of a non-zero $\theta$ the relevant group becomes $G_{\chi}^\prime = U(3)_L \times U(3)_R$, and $\Sigma$ needs to be promoted to a $U(3)$ matrix with the non-trivial determinant
\begin{align}
   \mathrm{det}\,\Sigma(x) = e^{-i\,\theta(x)} \,.
\label{eq:sigma_determinant}
\end{align} 
This condition ensures the consistency with the chiral transformation \cite{Gasser:1984gg}. While there is no unique way of implementing this constraint, we choose the option that reflects exactly the rotation in \eqref{eq:chiral_rotation}, namely
\begin{align}
   \Sigma(x) = e^{-\frac{i}{2}\theta(x)\,\kappa_q}\,\Sigma_0(x)\,
    e^{-\frac{i}{2}\theta(x)\,{\kappa}_q} \,,  
\label{eq:Sigma_in_terms_of_Sigma0}
\end{align}
which implies \eqref{eq:sigma_determinant} due to the condition \eqref{eq:trace_kappas}. The matrix $\Sigma$ transforms linearly under $G_\chi^\prime$, and we assign it to the $(3_L, \bar{3}_R)$ representation, i.e.\ ${\Sigma}\to g_L\Sigma \,g_R^\dagger$. This convention differs from the one adopted in the seminal paper of Gasser and Leutwyler \cite{Gasser:1984gg}, where the meson octet is represented by $U=\Sigma^\dagger$, which transforms in the $({3}_R, \bar{3}_{L})$ representation of the chiral group. Given that $\theta$ transforms non-linearly under $G_\chi^\prime$, i.e., $\theta\to\theta+i\ln\mathrm{det}(g_L g_R^\dagger)$, it is useful to introduce a corresponding covariant derivative
\begin{align}
   D_\mu\theta 
   = \partial_\mu\theta - 2\braket{a_\mu} 
   = - 2\,\tilde c_{GG}(\mu_\chi)\,\frac{\partial_\mu a}{f} \,, 
\label{eq:D_theta}
\end{align}
which is invariant under $G_\chi^\prime$ once the transformation of the axial current is taken into account \cite{Gasser:1984gg}. In the second step we have defined 
the effective ALP--gluon coupling at the scale of chiral symmetry breaking  \cite{Bauer:2020jbp},
\begin{align}
   \tilde c_{GG}
   = c_{GG} + \frac{\braket{c^a}}{2}
   = c_{GG} + \frac{c_{uu}^a+c_{dd}^a+c_{ss}^a}{2} \,.
   \label{eq:cGGtildedefinition}
\end{align}
We now have all the necessary ingredients to construct the most general Lorentz-invariant Lagrangian at $\mathcal{O}(p^2)$ which is (formally) invariant under $G_\chi^\prime$ in the presence of the sources in \eqref{eq:alp_sources}, if we assume that the spurion fields $\chi=2B_0\,m_q$, $l_\mu$ and $r_\mu$ obey the transformation rules 
\begin{align}
\label{eq:spurion_transformations}
   \chi\to g_L\chi\,g_R^\dagger \,, \qquad
   l_\mu\to g_L\,l_\mu\,g_L^\dagger \,, \qquad
   r_\mu\to g_R\,r_\mu\,g_R^\dagger \,.
\end{align} 
Including also the free ALP Lagrangian, we obtain
\begin{align}
\begin{aligned}   
   \cL^{(p^2)}_{\rm QCD} + \cL^{(0)}_{\rm ALP} 
   &= \frac{F^2}{8} \braket{ (D_\mu\Sigma)\,(D^\mu\Sigma^\dagger)
    + \chi\Sigma^\dagger + \Sigma\chi^\dagger} 
    + \frac{F^2}{8}\,H_0 \,(D_\mu\theta)(D^\mu\theta) \\
   &\quad + \frac12\,(\partial_\mu a)(\partial^\mu a) - \frac{m_{a,0}^2}{2}\,a^2 \,, 
\label{eq:lag_p2_QCD}
\end{aligned}
\end{align}
with
\begin{align}
\begin{aligned}
   \chi &= 2 B_0\,(s+ip) = 2 B_0\,m_q \,, \\[2mm]
   D_\mu\Sigma 
   &= \partial_\mu \Sigma - i(Q e A_\mu + l_\mu)\,\Sigma + i\Sigma\,(Q e A_\mu + r_\mu) \,, \\[2mm]
   D_\mu\Sigma^\dagger 
   &= \partial_\mu \Sigma^\dagger - i(Q e A_\mu + r_\mu)\,\Sigma^\dagger 
    + i\Sigma^\dagger\,(Q e A_\mu + l_\mu) \,, 
\label{eq:QCDp2_basic_definitions}
\end{aligned}
\end{align}
where $Q=\text{diag}(\frac23,-\frac13,-\frac13)$ contains the electric charges of the quarks. In the present work we ignore QED corrections, so the photon field can be dropped for our purposes. The Lagrangian \eqref{eq:lag_p2_QCD} is equivalent to the one used in \cite{Bauer:2021wjo}: substituting \eqref{eq:Sigma_in_terms_of_Sigma0} into \eqref{eq:lag_p2_QCD} and using the properties in \eqref{eq:octet_properties} yields equation~(7) in \cite{Bauer:2021wjo}. Note that the ALP mass term breaks the shift symmetry for the ALP and thus is an explicit source of chiral symmetry breaking.

As expected, the electrically neutral states in the theory, $a$, $\pi^0$ and $\eta_8$, undergo both kinetic and mass mixing. In the following we replicate the analysis performed in \cite{Bauer:2020jbp} for the $\pi^0$\,--\,$a$ mixing, but also include the $\eta_8$. Expanding the QCD chiral Lagrangian in the presence of the ALP up to quadratic order in the fields, we obtain 
\begin{align}
   \mathcal{L}^{\mathrm{ALP}}_\chi 
   \supset \frac12 (\partial_\mu \varphi)^T Z\,(\partial_\mu \varphi) 
    - \frac12\,\varphi^T M^2\,\varphi\,, 
\end{align} 
where $\varphi^T=(\pi^0,\eta_8,a)$, and 
\begin{align}
\label{eq:mass_and_kinetic_matrices_neutral_states}
   Z &= \begin{pmatrix}
    1 & 0 & \frac{F}{f}\,\frac{\hat{c}_{uu}^a - \hat{c}_{dd}^a}{2\sqrt{2}} \\[2mm]
    0 & 1 & \frac{F}{f}\,\frac{\hat{c}_{uu}^a + \hat{c}_{dd}^a - 2\hat{c}_{ss}^a}{2\sqrt{6}} \\[2mm]
  \frac{F}{f}\,\frac{\hat{c}_{uu}^a - \hat{c}_{dd}^a}{2\sqrt{2}} &
   ~\frac{F}{f}\, \frac{\hat{c}_{uu}^a + \hat{c}_{dd}^a - 2\hat{c}_{ss}^a}{2\sqrt{6}}~ &
   ~1 + \frac{F^2}{4f^2} \left(\braket{\hat{c}^a\hat{c}^a} + H_0 \left( \braket{c^a} + 2c_{GG} \right)^2 \right) ~
   \end{pmatrix} , \\[4mm]
   M^2 &= \begin{pmatrix} 
    B_0 (m_u + m_d) & \frac{B_0 (m_u - m_d)}{\sqrt{3}} & \frac{\sqrt{2} B_0 c_{GG} F\,g_\kappa(m_u,-m_d,0)}{f} \\[2mm]
    \frac{B_0 (m_u - m_d)}{\sqrt{3}} & \frac{B_0 (4 m_s + m_u + m_d)}{3} &
    \frac{\sqrt{2} c_{GG} B_0 F\,g_\kappa(m_u,m_d,-2 m_s)}{\sqrt{3} f} \\[2mm]
   \frac{\sqrt{2} c_{GG} B_0 F\,g_\kappa(m_u,-m_d,0)}{f} & \frac{\sqrt{2} c_{GG} B_0 F\,
    g_\kappa(m_u,m_d,-2 m_s)}{\sqrt{3} f} & m_{a,0}^2 + \frac{2 c_{GG}^2 B_0 F^2}{f^2} \braket{m_q\kappa_q^2} 
  \end{pmatrix} , \notag
\end{align}
with
\begin{align}
\begin{aligned}
   \hat{c}^a &= c^a + 2 c_{GG}\,\kappa_q \,, \\
   g_\kappa(a,b,c) &= a\,\kappa_u + b\,\kappa_d + c\,\kappa_s \,.
\end{aligned}
\end{align}
Note that in the isospin-conserving limit, which corresponds to setting $m_u=m_d\equiv\hat{m}$, the states $\pi^0$ and $\eta_8$ do not mix, since $M^2_{12}=M^2_{21}=0$ in this limit.

To identify the physical mass eigenstates it is necessary to diagonalize both the kinetic and the mass term. As outlined in \cite{Bauer:2020jbp}, we first find an orthogonal matrix $U_Z$ such that $U_Z^T\,Z\,U_Z=Z_{\rm{diag}}$. By performing the rotation $\varphi\to U_Z\,\varphi$, the bilinear terms in the Lagrangian become
\begin{align}
   \cL_\chi^{\mathrm{ALP}}
   \supset \frac12 (\partial_\mu\varphi)^T Z_{\mathrm{diag}}\,(\partial_\mu\varphi)
    - \frac12\,\varphi^T (U_Z^T M^2\,U_Z)\,\varphi \,.
\end{align} 
At this stage, kinetic terms are not yet canonically normalized. To correct this, we rescale $\varphi\to Z_{\mathrm{diag}}^{-1/2}\,\varphi$, which transforms the mass matrix into
\begin{align}
   \hat{M}^2 = Z_{\mathrm{diag}}^{-1/2}\,U_Z^T\,M^2\,U_Z\,Z_{\mathrm{diag}}^{-1/2} \,.
\end{align} 
The last step is to diagonalize $\hat{M}^2$, i.e., to find an orthogonal matrix $U_M$ such that $U_M^T\hat{M}^2 U_M=M_\mathrm{diag}^2\equiv M_\mathrm{phys}^2$. With the last rotation $\varphi\to U_M\,\varphi$, we finally obtain
\begin{align}
   \cL_\chi^{\mathrm{ALP}}
   \supset \frac12 (\partial_\mu\varphi_{\mathrm{phys}})^T (\partial_\mu \varphi_{\mathrm{phys}}) 
    - \frac12\,\varphi^T_{\mathrm{phys}}\, M^2_{\mathrm{phys}}  \,\varphi_{\mathrm{phys}} \,,
\end{align} 
where we have added the pedix ``phys'' to highlight that these are the physical fields. They are related to those appearing at the beginning via
\begin{align}
   \varphi = R\,\varphi_{\mathrm{phys}} \,, \quad \text{with} \quad
   R = U_Z\,Z_{\mathrm{diag}}^{-1/2}\,U_M \,.
\end{align} 

Before finding the explicit rotations, we want to find the physical masses, i.e.\ the eigenvalues of $\hat{M}^2$. Since $\hat{M}^2$ and $M^2\,Z^{-1}$ are similar matrices they have the same eigenvalues, and hence it is sufficient to solve \cite{Bauer:2020jbp}
\begin{align}
   \mathrm{det}\left(M^2 Z^{-1} - m^2\,\mathbbm{1} \right) = 0 \,.  
\end{align} 
Keeping terms up to second order in $F/f$, and working in the isospin limit where $m_u=m_d=\hat m$, we find 
\begin{align}
\begin{aligned}
   m_{\pi^0}^2 
   &= \tilde{m}_{\pi^0}^2  \left[ 1 + \frac{F^2}{8 f^2}\,\frac{\tilde{m}_{\pi^0}^2}{\tilde{m}_{\pi^0}^2 - m_{a,0}^2}
    \left( c_{uu}^a - c_{dd}^a \right)^2 \right] , \\
   m_{\eta_8}^2 
   &= \tilde{m}_{\eta_8}^2 \left[ 1 + \frac{F^2}{24 f^2}  \frac{\tilde{m}_{\eta_8}^2}{\tilde{m}_{\eta_8}^2 - m_{a,0}^2} 
   \left(4 c_{GG}\,\frac{m_s-\hat{m}}{2m_s+\hat{m}} + c_{uu}^a + c_{dd}^a - 2 c^a_{ss} \right)^2 \right] , \\
   m_a^2 
   &= m_{a,0}^2 \left\{ 1 - \frac{F^2}{4 f^2} \left[ \Delta + H_0 \left( \braket{c^a} + 2c_{GG} \right)^2 
    + \frac{  m_{a}^2 }{2 \,(\tilde{m}_{\pi^0}^2 - m_{a,0}^2)} 
    \left( c_{uu}^a - c_{dd}^a\right)^2 \right] \right\} \\
   &\quad + c_{GG}^2\,\frac{F^2\, \tilde{m}_{\pi^0}^2}{2 f^2}
    -  \frac{F^2}{24 f^2} \frac{\left((c^a_{uu}+ c^a_{dd}-2 c^a_{ss})\, m_{a,0}^2 + 2 c_{GG} \, (m_{a,0}^2 -\tilde{m}_{\pi^0}^2)\right)^2}{ \tilde{m}_{\eta_8}^2 - m_{a,0}^2 } \,, 
\label{eq:masses_after_diagonalization}
\end{aligned}
\end{align}
where 
\begin{equation}
  \Delta  = 2 c_{GG}(c_{GG} + c^a_{uu}+c^a_{dd}) +  \braket{{c}^a {c}^a} \,.
\end{equation}
Note in particular that having set $\braket{\kappa_q}=1$ the result for the physical masses is independent of the individual $\kappa_q$ parameters.\footnote{For the cancellation of the $\kappa_q$ parameters it is important to include the $1/f^2$-suppressed terms in the expression for $Z$ in \eqref{eq:mass_and_kinetic_matrices_neutral_states}, which were omitted in \cite{Bauer:2020jbp}.} The parameters
\begin{align}
\begin{aligned}
\label{eq:meson_masses_in_terms_of_quark_masses_in_isospin_conserving_limit}
   \tilde{m}_{\pi^0}^2 
   &= \tilde{m}_{\pi^-}^2 = B_0\,(m_u+m_d) \,, \\
   \tilde{m}_{\eta_8}^2 
   &= \frac{B_0}{3}\,(m_u+m_d+4 m_s) 
\end{aligned}
\end{align}
denote the zeroth-order contributions (in $F^2/f^2$) to the pion and $\eta_8$ masses. The masses of the kaons do not receive corrections in $F/f$ and are given by
\begin{align}
\begin{aligned}
   m_{K^-}^2 &= B_0\,(m_u+m_s) \,, \\
   m_{\bar K^0}^2 &= B_0\,(m_d+m_s) \,. 
\end{aligned}
\end{align}
In our analysis of the $K^\pm\to\pi^\pm a$ decay amplitudes we work consistently to first order in $F/f$, so the above mass corrections can be ignored. We will also treat isospin-breaking effects in the quark-mass difference $(m_d-m_u)$ as a perturbation around the average value $\hat m\equiv\frac12(m_u+m_d)$. In the isospin limit one has $m_{K^-}^2=m_{\bar K^0}^2$ and 
\begin{align}
\label{eq:eta8mass}
   \tilde{m}_{\eta_8}^2 = \frac{4 m_{K^-}^2-m_{\pi^-}^2}{3} \,.
\end{align}

The physical fields are obtained from the initial ones via the transformation 
\begin{align}
   R = \begin{pmatrix}
    1 & 0 & \theta_{\pi^0 a} \\
    0 & 1 & \theta_{\eta_8 a} \\
    \theta_{a\pi^0} & \theta_{a\eta_8} & 1 
   \end{pmatrix} + \mathcal{O}\left( \frac{F^2}{f^2} \right) ,  
\label{eq:rotation}
\end{align}
where in the isospin limit the rotation angles are given by
\begin{align}
\begin{aligned}
\label{eq:rotation_explicit_angles}
   \theta_{\pi^0 a} 
   &= - \frac{M^2_{13} - m_{a,0}^2\,Z_{13}}{\tilde{m}_{\pi^0}^2 - m_{a,0}^2} 
    = \frac{F}{f}\,\frac{(\hat{c}_{uu}^a-\hat{c}_{dd}^a)\,m_{a,0}^2 
     - 2 c_{GG}\,(\kappa_u-\kappa_d)\,\tilde{m}_{\pi^0}^2}{2\sqrt{2}\,(\tilde{m}_{\pi^0}^2-m_{a,0}^2)} \,, \\ 
   \theta_{\eta_8 a} 
   &= - \frac{M^2_{23} - m_{a,0}^2\,Z_{23}}{\tilde{m}_{\eta_8}^2 - m_{a,0}^2} 
    = \frac{F}{f}\,\frac{(\hat{c}_{uu}^a+\hat{c}_{dd}^d-2\hat{c}_{ss}^a)\,m_{a,0}^2
     - 2 c_{GG}\, \tilde{m}_{\pi^0}^2 + 6 c_{GG}\,\kappa_s\,\tilde{m}_{\eta_8}^2}{2\sqrt{6}\, 
     (\tilde{m}_{\eta_8}^2-m_{a,0}^2)} \,, \\
   \theta_{a\pi^0} 
   &= \frac{M^2_{13} - \tilde{m}_{\pi^0}^2\,Z_{13}}{\tilde{m}_{\pi^0}^2 - m_{a,0}^2} 
    = - \frac{F}{f}\,\frac{(c^{a}_{uu}-c^{a}_{dd})\,\tilde{m}_{\pi^0}^2}{2\sqrt{2}\,
     (\tilde{m}_{\pi^0}^2-m_{a,0}^2)} \,, \\
   \theta_{a\eta_8 } 
   &= \frac{M^2_{23} - \tilde{m}_{\eta_8}^2 Z_{23}}{\tilde{m}_{\eta_8}^2 - m_{a,0}^2 }
    = - \frac{F}{f}\,\frac{(\hat{c}_{uu}^a+\hat{c}_{dd}^a-\hat{c}_{ss}^a+6 c_{GG}\,\kappa_s)\,\tilde{m}_{\eta_8}^2
     - 2 c_{GG}\,\tilde{m}_{\pi^0}^2}{2\sqrt{6}\,(\tilde{m}_{\eta_8}^2-m_{a,0}^2)} \,. 
\end{aligned}
\end{align}
These expressions are derived under the assumption of small mixing angles, working to first order in $F/f$, and hold as long as $|\tilde{m}_{\pi^0}^2-m_{a,0}^2| \gg\tilde{m}_{\pi^0}^2\,F/f$ and $|\tilde{m}_{\eta_8}^2-m_{a,0}^2|\gg\tilde{m}_{\eta_8}^2\,F/f$. In the opposite case, where $m_{a,0}=\tilde{m}_{\pi^0}$, one would obtain maximal mixing between the ALP and the $\pi^0$ states. Besides the fact that such a large mixing would require a fine-tuning of the mass parameters that is rather implausible, it would modify the properties of the neutral pion in a way that is incompatible with experimental findings (e.g.\ with the measured rate of the $\pi^0\to\gamma\gamma$ decay). An analogous statement holds for the case where $m_{a,0}=\tilde{m}_{\eta_8}$.

In \eqref{eq:rotation_explicit_angles} the quark masses have been written in terms of the meson masses by using the relations \eqref{eq:meson_masses_in_terms_of_quark_masses_in_isospin_conserving_limit}. Note the important fact that these mixing angles are not independent of the $\kappa_q$ parameters, and hence they are not physical quantities \cite{Bauer:2021wjo}. In the following we will apply the rotation in \eqref{eq:rotation} in order to remove mass and kinetic mixing of $a$, $\pi^0$ and $\eta_8$ from the $\mathcal{O}(p^2)$ Lagrangian. Mixing effects still arise at $\mathcal{O}(p^4)$, and we treat them as interaction vertices. Isospin-breaking corrections will be included in our LO calculation in Section~\ref{sec:amplitude_LO} at linear order in $(m_d-m_u)$ and treated as small perturbations. They are not included explicitly in \eqref{eq:rotation_explicit_angles}. 

\subsection[A comment on the $\eta^\prime$ meson]{\boldmath A comment on the $\eta^\prime$ meson}
\label{sec:eta_prime}

Because of its large mass, the $\eta^\prime$ meson appears only indirectly through the values of the low-energy constants and is usually not included as a light degree of freedom in the chiral effective theory. However, its mixing with the $\eta$ meson is known to be numerically important and it is needed to consistently perform the large-$N_c$ limit. In the present context, it might appear to be possibly even more important, since the $\eta^\prime$ is (largely) an $SU(3)$ singlet, much like the ALP. In this section we sketch briefly how the $\eta^\prime$ can be included in the chiral Lagrangian, leaving a quantitative investigation of its contributions for future work.

The extension from eight to nine pseudoscalar mesons has already been described in \cite{Gasser:1984gg} and, more recently, in \cite{Herrera-Siklody:1996tqr}. In a nutshell, the matrix $\Sigma$ is extended with a singlet field $\phi_0$ as 
\begin{align}
   \Sigma^\prime(x) = \Sigma_0(x)\,e^{i\frac{\phi_0(x)}{3}} \,,
\label{nonet}
\end{align}
with $\det\Sigma_0=1$ as in \eqref{eq:octet_properties}.
The new field $\phi_0$ is not invariant under $U(1$) axial transformations, however its transformation is precisely compensated by the transformation properties of the $\theta$ parameter, such that the combination $(\phi_0+\theta)$ is invariant under $U(3)_L\times U(3)_R$. This fact allows one to replace all low-energy coefficients in the chiral Lagrangian by arbitrary functions $V_i=V_i (\phi_0+\theta)$, i.e.\ at $\mathcal{O}(p^2)$ \cite{Gasser:1984gg,Herrera-Siklody:1996tqr} 
\begin{align}
\begin{aligned}
   \cL^{(p^2)}_{\rm QCD} 
   &= V_0(\phi_0+\theta) 
    + V_1(\phi_0+\theta) \braket{(D_\mu\Sigma')(D^\mu\Sigma^{\prime\dagger})} 
    + V_2(\phi_0+\theta) \braket{\chi\Sigma^{\prime\dagger} + \Sigma' \chi^{\dagger}} \\[2mm]
   &\quad + V_3(\phi_0+\theta) \braket{i(\chi\Sigma^{\prime\dagger} - \Sigma'\chi^{\dagger})} + V_4(\phi_0+\theta)\,(D_\mu\phi_0)(D^\mu\phi_0) \\[2mm]
   &\quad + V_5(\phi_0+\theta)\,(D_\mu\phi_0)(D^\mu\theta)
    + V_6(\phi_0+\theta)\,(D_\mu\theta)(D^\mu\theta) \,,
\label{eq:U3}
\end{aligned}
\end{align}
where $V_1(0)=V_2(0)=F^2/8$, and $V_6(0)=F^2 H_0/8$, whereas $V_3(0)=0$. Parity invariance requires that $V_3$ is an odd function of its argument, while all other potentials are even. The covariant derivative
\begin{align}
   D_\mu\phi_0 = \partial_\mu\phi_0 + 2\braket{a_\mu} 
\end{align}
is defined such that $D_\mu(\phi_0+\theta)=\partial_\mu(\phi_0+\theta)$, as appropriate for a singlet.

The potential $V_0$ can generate a mass term for the field $(\phi_0+\theta)$, which is not protected by chiral symmetry and hence can be arbitrarily heavy. We thus identify
\begin{align}
   \eta_1\equiv \frac{F}{2\sqrt{3}} \left( \phi_0+\theta \right)
\end{align}
with the heavy singlet meson $\eta_1$,\footnote{The ``heavy'' state $\eta_1$ mixes with the ``light'' state $\eta_8$ to produce the physical mesons $\eta'$ and $\eta$.} where the prefactor on the right-hand side has been chosen such that $\eta_1$ has a canonically normalized kinetic term. From \eqref{nonet}, we then obtain
\begin{align}
   \Sigma^\prime(x) 
   = \Sigma_0(x)\,e^{i\,\frac{2}{\sqrt{3} F}\,\eta_1(x)}\,e^{-i\,\frac{\theta(x)}{3}} \,.
\end{align}
Upon integrating out the heavy field $\eta_1$, this leaves us with
\begin{align}
   \Sigma^\prime(x) 
   = \Sigma_0(x)\,e^{-i\,\frac{\theta(x)}{3}} \,.
\end{align}
This way of implementing the constraint $\det\Sigma'(x)=e^{-i\,\theta(x)}$ is a particular case of \eqref{eq:Sigma_in_terms_of_Sigma0}, corresponding to the special choice $\kappa_q=\frac13\mathbbm{1}$.

Given that $m_{\eta^\prime}\gg F$, we believe that not including the $\eta^\prime$ as a propagating degree of freedom in the low-energy chiral Lagrangian is a reasonable approximation. Nevertheless, it would be interesting to study in the future the alternative approach, where $\eta_1$ is included as the ninth pseudoscalar meson. 

\subsection[Weak chiral Lagrangian at $\mathcal{O}(G_F\,p^2)$]{Weak chiral Lagrangian at $\bm{\mathcal{O}(G_F\,p^2)}$}
\label{sec:weak_Lagrangian_p2}

The decays $K^\pm\to\pi^\pm a$ involve an interplay of strong and non-leptonic weak interactions. Starting from the current-current form of the effective weak Lagrangian at energies well below the electroweak scale, it is possible to classify the relevant four-quark operators according to their transformation properties under the chiral group \cite{Pich:1985st,Pich:1990mw}. Neglecting electroweak penguin operators,\footnote{This is a very good approximation for our purposes. Including electroweak penguins, one would also obtain operators transforming as $(8_L,8_R)$ \cite{Ecker:2000zr,Cirigliano:2003gt}.} one finds operators transforming as $(8_L,1_R)$ and $(27_L,1_R)$. In kaon physics, the octet operators are dynamically enhanced and give the leading contribution to the decay amplitudes. 

It is convenient to express the operators in the chiral Lagrangian in terms of the left-handed building blocks $S$, $P$, $L_\mu$, and $W_{\mu\nu}$ defined in Table~\ref{tab:building_blocks}, all of which transform as $S\to g_L S\,g_L^\dagger$ etc.\ under chiral transformations. In the presence of the ALP, operators can also contain $D_\mu\theta$, which transforms as a singlet. The last three columns in the table show the properties of these objects under the discrete transformations $P$, $C$, and $CP$ \cite{Kambor:1989tz}, derived under the assumption that the spurion fields transform as
\begin{align}
   l_\mu \stackrel{P}{\longleftrightarrow} r^\mu \,, \qquad
   l_\mu \stackrel{C}{\longleftrightarrow} (-r_\mu)^T \,. 
\end{align}
In the lower portion of the table we introduce associated right-handed objects $S_R$, $P_R$, $R_\mu$, and $W_{\mu\nu}^R$, which transform as $S_R\to g_R\,S_R\,g_R^\dagger$ etc., and which enter in the transformation rules for parity and charge conjugation. These right-handed objects are related to the left-handed ones via
\begin{align}
\begin{aligned}
\label{eq:RHD_objects}
   S_R &= \Sigma^\dagger S\,\Sigma \,, \qquad 
    &P_R &= - \Sigma^\dagger P\,\Sigma \,, \\
   R_\mu &= - \Sigma^\dagger L_\mu\,\Sigma \,, \qquad 
    &W_{\mu\nu}^R &= \Sigma^\dagger W_{\mu\nu}\,\Sigma \,.
\end{aligned}
\end{align}

At $\mathcal{O}(p^2)$, the weak chiral Lagrangian can now be written as 
\begin{align}
\begin{aligned}
   \cL_{\rm weak}^{(p^2)} 
   &= \frac{F^4}{4}\,\Big[ G_8 \mathcal{O}_{8}+ G_8^\prime \braket{\lambda_+ S} 
    + G_{8}^\theta\,(D_\mu\theta) \braket{\lambda_+ L^\mu} \\
   &\hspace{1.4cm} + G_{27}^{1/2}\,\mathcal{O}_{27}^{1/2} 
    + G_{27}^{3/2}\,\mathcal{O}_{27}^{3/2} + \text{h.c.} \Big] \,,
\label{eq:lag_p2_weak}
\end{aligned}
\end{align}
where
\begin{align}
\mathcal{O}_{8} = \braket{\lambda_+ L_\mu L^\mu} \,,
\end{align}
$\lambda_+=\frac12(\lambda_6+i\lambda_7)$, and the objects $L_\mu$ and $S$ are the chiral representations of the left-handed current $\bar q_L\gamma_\mu q_L$ and of the scalar current $m_q (\bar q_L q_R+\bar q_R q_L)$, respectively. The terms shown explicitly in \eqref{eq:lag_p2_weak} mediate weak $s\to d$ transitions, while their hermitian conjugates describe $\bar s\to\bar d$ transitions. The first octet operator, $\mathcal{O}_{8}$, was written down as early as in 1967 by Cronin \cite{Cronin:1967jq}. The 27-plet operators with isospin change $\Delta I=\frac12$ and $\frac32$ have the form \cite{Bernard:1985wf}
\begin{align}
\begin{aligned}
   \mathcal{O}_{27}^{1/2} 
   &= L_{\mu\,32} L_{11}^\mu + L_{\mu\,31} L_{12}^\mu + 2 L_{\mu\,32} L_{22}^\mu - 3 L_{\mu\,32} L_{33}^\mu \,, \\
   \mathcal{O}_{27}^{3/2} 
   &= L_{\mu\,32} L_{11}^\mu + L_{\mu\,31} L_{12}^\mu -  L_{\mu\,32} L_{22}^\mu \,.
\end{aligned}
\end{align}
The second octet operator in \eqref{eq:lag_p2_weak}, $\braket{\lambda_+ S}$,  is known as the ``weak mass term''. In absence of external fields, this operator can be removed through field redefinitions (see Section~\ref{sec:weakmass}). However, its presence leads to non-zero effects when the ALP is present, as discussed in the next section. On the other hand, the third octet operator is a novel object, which, to the best of our knowledge, has not been previously discussed in the literature. It explicitly involves the ALP source $\theta$. 

\begin{table}[t] 
\centering 
\renewcommand{\arraystretch}{1.2} 
\scalebox{0.96}{
\begin{tabular}{|c|c||c|c|c|}
\hline
Object & Definition & $P$ & $C$ & $CP$ \\
\hline
\hline  
$S$ & $\chi\Sigma^\dagger + \Sigma\chi^\dagger$ & $S\to S_R$ & $S\to (S_R)^T$ & $S\to S^T$ \\ 
$P$ & $i(\chi\Sigma^\dagger - \Sigma\chi^\dagger)$ & $P\to P_R$ & $P\to -(P_R)^T$ & $P\to -P^T$ \\
$L_\mu$ & $\Sigma\,i(D_\mu\Sigma)^\dagger$ & $L_\mu\to R^\mu$ & $L_\mu\to -(R_\mu)^T$ & $L_\mu\to -(L^\mu)^T$ \\
$W_{\mu\nu}$ & $2(D_\mu L_\nu+D_\nu L_\mu)$ & $W_{\mu\nu}\to W^{R\,\mu\nu}$ & $W_{\mu\nu}\to -(W_{\mu\nu}^R)^T$
 & $W_{\mu\nu}\to -(W^{\mu\nu})^T$ \\
$D_\mu\theta$ & $\partial_\mu\theta-2\braket{a_\mu}$ & $D_\mu\theta\to -D^\mu\theta$ & $D_\mu\theta\to D_\mu\theta$
 & $D_\mu\theta\to -D^\mu\theta$ \\
\hline\hline
$S_R$ & $\chi^\dagger\Sigma + \Sigma^\dagger\chi$ & $S_R\to S$ & $S_R\to S^T$ & $S_R\to (S_R)^T$ \\ 
$P_R$ & $i(\chi^\dagger\Sigma - \Sigma^\dagger\chi)$ & $P_R\to P$ & $P_R\to -P^T$ & $P_R\to -(P_R)^T$ \\
$R_\mu$ & $\Sigma^\dagger i D_\mu\Sigma$ & $R_\mu\to L^\mu$ & $R_\mu\to -(L_\mu)^T$ & $R_\mu\to -(R^\mu)^T$ \\ 
$W_{\mu\nu}^R$ & $2(D_\mu R_\nu+D_\nu R_\mu)$ & $W_{\mu\nu}^R\to W^{\mu\nu}$ & $W_{\mu\nu}^R\to -(W_{\mu\nu})^T$
 & $W_{\mu\nu}^R\to -(W^{R\,\mu\nu})^T$ \\
\hline
\end{tabular}
}
\caption{Hermitian building blocks of the $\mathcal{O}(p^2)$ and $\mathcal{O}(p^4)$ chiral Lagrangians. The definition of $L_\mu$ used here coincides with the one in \cite{Bauer:2021wjo} up to an overall factor. Our definitions differ from those used in \cite{Gasser:1984gg,Kambor:1989tz} because of the transformation properties assumed for $\Sigma$ (i.e.\ $\Sigma \to g_L\Sigma\,g_R^\dagger$ here vs.\ $U\to g_R\,U g_L^\dagger$ in \cite{Gasser:1984gg,Kambor:1989tz}).}
\label{tab:building_blocks}
\end{table}

The constants $G_i$ are defined as 
\begin{align}
   G_i = - \frac{G_F}{\sqrt2}\,V_{ud}^\ast V_{us} \left( g_i
    - \frac{V_{td}^\ast\,V_{ts}}{V_{ud}^\ast V_{us}}\,g_i^t \right) ,
\end{align}
with dimensionless constants $g_i$ and $g_i^t$. In the standard definition of the CKM matrix the combination $V_{ud}^\ast V_{us}$ is real, while the ratio of CKM elements (expressed here using the Wolfenstein approximation)
\begin{align}
\label{eq:CKMelements}
   - \frac{V_{td}^\ast\,V_{ts}}{V_{ud}^\ast V_{us}}
   \approx A^2\lambda^4 \left( 1 - \rho + i\eta \right)
\end{align}
carries a non-trivial CP-odd phase. This ratio is numerically very small, of order $10^{-3}$, and we will neglect it for simplicity. Consequently, the couplings $G_i$ become real, and adding the hermitian conjugate has the effect of replacing $\lambda_+\to\lambda_6$ in the three octet operators, while in the 27-plet operators one must add the corresponding operators with the flavor indices on each current interchanged. 

To make contact with the notation used elsewhere (e.g.\ in \cite{Kambor:1989tz,Cirigliano:2003gt,Cirigliano:2011ny,Pich:2021yll}), we consider the limit of exact $SU(3)$ symmetry. In this case the relation $g_{27}^{3/2}=5 g_{27}^{1/2}$ holds and the combination of 27-plet operators simplifies to
\begin{align}
   G_{27}^{1/2}\,\mathcal{O}_{27}^{1/2} + G_{27}^{3/2}\,\mathcal{O}_{27}^{3/2} 
   = 9\,G_{27}^{1/2} \left( L_{\mu\,32} L^{\mu}_{11} + \frac23 L_{\mu\,31} L^{\mu}_{12} 
   - \frac13 L_{\mu\,32} \braket{L^\mu} \right) . 
\end{align}
This coincides with the $27$-plet as given in the cited references, since in absence of a $\theta$ source $\braket{L^\mu}=0$, see \eqref{eq:octet_properties}. In the presence of the ALP, we have instead 
\begin{align}
\label{eq:traceL}
    \braket{L^\mu} = - D^\mu\theta \,,
\end{align}
and the extra term gives a contribution to $G_8^\theta$.

The low-energy couplings are determined at the scale $\mu=m_\rho\approx 770$\,MeV through a phenomenological analysis of the $K\to\pi\pi$ decay rates at NLO in chiral perturbation theory. Including isospin-breaking corrections, their values are \cite{Cirigliano:2011ny}
\begin{align}
    g_8 = 3.61\pm 0.28 \,, \qquad
    g_{27} \equiv 9 \,g_{27}^{1/2} = 0.297\pm 0.028 \,,
    \label{eq:inputs_for_gi}
\end{align}
where the errors are completely theory-dominated, and the $SU(3)$ limit has been assumed for $g_{27}$. For the individual 27-plet couplings we then obtain $g_{27}^{1/2}\approx 0.033\pm 0.003$ and $g_{27}^{3/2}\approx 0.165\pm 0.016$. The fact that $g_8$ is about a factor 22 larger than $g_{27}^{3/2}$ is referred to as the $\Delta I=\frac12$ selection rule in non-leptonic kaon decays. Because of this pronounced hierarchy, in this work we focus solely on calculating octet contributions to the $K^-\to\pi^- a$ decay amplitude at NLO. Meanwhile, for the 27-plet contributions, we will limit ourselves to improving the LO predictions from \cite{Bauer:2021mvw} by adding isospin-breaking corrections.

Contrary to $g_8$ and $g_{27}$, the low-energy couplings $g_8^\prime$ and $g_8^\theta$, which enter \eqref{eq:lag_p2_weak} via $G_8'$ and $G_8^\theta$, are currently unknown, since they do not intervene in SM processes. If an ALP was to be found in the future, the decays $K^\pm\to\pi^\pm a$ and $K_L\to\pi^0 a$ would give access to these two couplings, which would then be promoted to physical parameters. In principle, it should be possible to calculate $g_8^\prime$ and $g_8^\theta$ using lattice QCD, by finding an appropriate representation of the external sources in terms of quark and gluon fields. 

\begin{figure}[t]
\centering
\includegraphics[width=0.5\textwidth]{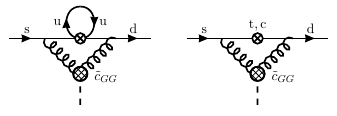}
\caption{Schematic representation of the low-energy coupling $g_8^\theta$. The crossed circle represents an effective weak vertex obtained after integrating out the electroweak gauge bosons and heavy quarks of the SM. The cross-hatched circle represents the ALP--gluon interaction governed by the coefficient $\tilde c_{GG}$ in \eqref{eq:cGGtildedefinition}. Being defined at low energy, these diagrams can receive large non-perturbative QCD corrections.}
\label{fig:aGGeff}
\end{figure}

Let us briefly comment about the new octet operator including the $\theta$ source, which is the chiral representation of the quark operator 
\begin{align}
   \frac{F^2}{4}\,(D_\mu\theta) \braket{\lambda_+ L^\mu} 
   \longleftrightarrow 2\,\tilde c_{GG}(\mu_\chi)\,\frac{(\partial_\mu a)}{f}\,
    \bar d_L\gamma^\mu s_L \,.
\end{align} 
The coupling $g_8^\theta$ thus accounts for the gluon-induced interactions of the ALP with the loop-induced flavor-changing neutral current interaction present in the SM, which in a perturbative setting would be mediated by diagrams such as those shown in Figure~\ref{fig:aGGeff}. One might expect that this interaction is a small effect, since it starts at two-loop order in perturbation theory. However, in the low-energy regime relevant here, the gluon interactions are non-perturbative and loop suppression becomes meaningless. For example, it is well known that gluon-induced low-energy interactions yield a sizable contribution to the effective ALP--photon coupling in the chiral Lagrangian, such that $c_{\gamma\gamma}^{\rm eff}\approx c_{\gamma\gamma}-(1.92\pm 0.04)\,c_{GG}$ for $m_a^2\ll m_{\pi^-}^2$ \cite{GrillidiCortona:2015jxo,Bauer:2020jbp}, even though also in this case the interaction starts at two-loop order in perturbation theory.

\subsection[QCD chiral Lagrangian at $\mathcal{O}(p^4)$]{QCD chiral Lagrangian at $\bm{\mathcal{O}(p^4)}$}

At NLO in the chiral expansion, the operators entering the QCD Lagrangian have been constructed in \cite{Gasser:1984gg} under some specific assumptions about the external sources. As long as the conditions $\theta=0$, $\braket{a_\mu}=0$ and $\braket{v_\mu}=0$ hold, the basis contains 12 operators $O_i$, whose coefficients are called $L_i$ for $i=1,\dots,10$ and $H_{1,2}$ for $i=11,12$. In the presence of an ALP these conditions are generally not fulfilled, however, since $\theta\ne 0$ and $\braket{a_\mu}\ne 0$ for generic ALP couplings to gluons and quarks. Therefore, a modification of the basis is needed. 

On the one hand, some operators in the general basis are absent for the specific sources provided by the ALP. The field strength-tensors $F_{\mu\nu}^L$ and $F_{\mu\nu}^R$ associated with the external currents $l_\mu$ and $r_\mu$ in \eqref{eq:alp_sources} vanish, and therefore $O_i=0$ for $i=9,10,11$. The definitions of the remaining operators are shown in the left portion of Table~\ref{table:p4QCD}. Note that, since $s=m_q$ and $p=0$ in our model, the operator $O_{12}=\braket{\chi^\dagger\chi}$ is a field-independent constant and can be dropped from the Lagrangian. On the other hand, the basis must be extended by additional operators containing the fields $\theta$ and $\braket{a_\mu}$ in the invariant combination $D_\mu\theta$ in \eqref{eq:D_theta}, which according to \eqref{eq:traceL} is minus the trace of $L_\mu$. An exhaustive derivation of such terms has been performed in \cite{Herrera-Siklody:1996tqr}. For the purposes of this work, we can restrict ourselves to operators linear in $D_\mu\theta$, since we consistently work to first order in $a/f$ (with $f\gg 4\pi F$ being the ALP decay constant). There are only three such operators, denoted by $O_i^\theta$, whose definition is given in the right portion of Table~\ref{table:p4QCD}. They correspond to $O_{53}$, $O_{46}$, and $O_{31}$ in the notation of \cite{Herrera-Siklody:1996tqr}. 

\begin{table}[t] 
\centering 
\renewcommand{\arraystretch}{1.2} 
\begin{tabular}{|c|c|c||c|c|c|}
\hline  
$i$ & $O_i$ & $\Gamma_i$ & $i$ & $O_i^\theta$ & $\Gamma_i^\theta$ \\
\hline
\hline
1 &$\braket{L^2}^2$ & $\frac{3}{32}$ & 1 & $-(\partial^\mu D_\mu\theta) \braket{P}$ & 0 \\
2 &$\braket{L^\mu L^\nu}\braket{L_\mu L_\nu}$ & $\frac{3}{16}$ & 2 & $ -(D_\mu\theta) \braket{L^\mu S}$ & $-\frac14$ \\
3 &$\braket{L^4}$ & 0 & 3 & $(D_\mu\theta) \braket{L^\mu L^2}$ & 0 \\
4 & $\braket{L^2}\braket{S}$ & $\frac{1}{8}$ & & & \\
5 &  $\braket{L^2 S}$ &$\frac{3}{8}$ & & & \\
6 &$\braket{S}^2$ &$\frac{11}{144}$ & & & \\
7 & $-\braket{P}^2$ & 0 & & & \\
8 & $\frac12 \braket{S^2-P^2}$ & $\frac{5}{48}$ & & & \\
12 & $\frac14 \braket{S^2+P^2}$ & $\frac{5}{24}$ & & & \\[0.5mm]
\hline
\end{tabular}
\caption{Operators in the $\mathcal{O}(p^4)$ QCD Lagrangian \eqref{eq:lag_p4_QCD} relevant for the SM extended by a light ALP. We use the short-hand notation $L^2\equiv L_\mu L^\mu$. The coefficients $\Gamma_i$ and $\Gamma_i^\theta$ are taken from \cite{Gasser:1984gg} and \cite{Herrera-Siklody:1996tqr}, respectively.}
\label{table:p4QCD}
\end{table}

Using the transformation properties shown in Table~\ref{tab:building_blocks} along with relations \eqref{eq:RHD_objects}, it is straightforward to check that the operators $O_i$ and $O_i^\theta$ are invariant under both parity and charge conjugation, as required by the symmetries of QCD.\footnote{The operators $O_{41}$, $O_{48}$, $O_{52}$ listed in \cite{Herrera-Siklody:1996tqr} are even under $C$ but odd under $P$.} 
All in all, the $\mathcal{O}(p^4)$ QCD Lagrangian then takes the form
\begin{align}
   \mathcal{L}_{\rm QCD}^{(p^4)} 
   = \sum_{i=1}^8 L_i\,O_i + \sum_{i=1}^3 L_i^\theta\,O_{i}^\theta \,.
\label{eq:lag_p4_QCD}
\end{align}
The bare low-energy constants $L_i^{(\theta)}$ are customarily written as
\begin{align}
\begin{aligned}
   L_i^{(\theta)} &= L_{i,r}^{(\theta)}(\mu) + \lambda\,\Gamma_i^{(\theta)} \,, 
\end{aligned}
\end{align} 
where $L_{i,r}^{(\theta)}(\mu)$ are the renormalized couplings, also called low-energy constants, and the quantity \begin{align}
   \lambda = \frac{\mu^{d-4}}{32\pi^2} \left( \frac{2}{d-4} - \ln 4\pi + \gamma_E - 1 \right) , 
\label{eq:lambda_factor}
\end{align}  
where $d$ denotes the number of spacetime dimensions, absorbs the UV poles in the dimensional regularization scheme. Note that the couplings $L_i$ and $L_i^\theta$ have mass dimension $[L_i^{(\theta)}]=(d-4)$, and we have added an auxiliary scale $\mu$ in the definition of $\lambda$ to account for this fact. The coefficients $\Gamma_i^{(\theta)}$ are the one-loop anomalous dimensions entering the evolution equations for the renormalized couplings. One obtains
\begin{align}
\label{eq:RGevol}
   \frac{d}{d\ln\mu}\,L_{i,r}^{(\theta)}(\mu) 
   = - \frac{\Gamma_i^{(\theta)}}{16\pi^2} \,.
\end{align}
The values of the coefficients $\Gamma_i^{(\theta)}$ as computed in \cite{Gasser:1984gg,Herrera-Siklody:1996tqr} are shown in Table~\ref{table:p4QCD}. 

\begin{table}[t] 
\centering 
\renewcommand{\arraystretch}{1.2} 
\begin{tabular}{|c|c|c|}
\hline
$10^3\,L_{i,r}$ & Ref.~\cite{Bijnens:2014lea} & Ref.~\cite{Dowdall:2013rya} \\
\hline
\hline
$10^3\,L_{4,r}(\mu)$ & $\phantom{-}0.0\pm 0.3$ & $0.09\pm 0.34$ \\
$10^3\,L_{5,r}(\mu)$ & $\phantom{-}1.2\pm 0.1$ & $1.19\pm 0.25$ \\
$10^3\,L_7$ & $-0.3\pm 0.2$ & -- \\
$10^3\,L_{8,r}(\mu)$ & $\phantom{-}0.5\pm 0.2$ & $0.55\pm 0.15$ \\
\hline
\end{tabular}
\caption{Phenomenological values of the renormalized QCD low-energy constants relevant to our calculation, evaluated at the scale $\mu=m_\rho$. $L_7$ is scale independent.}
\label{table:LECs}
\end{table}

\begin{figure}
\centering
\includegraphics[scale=.3]{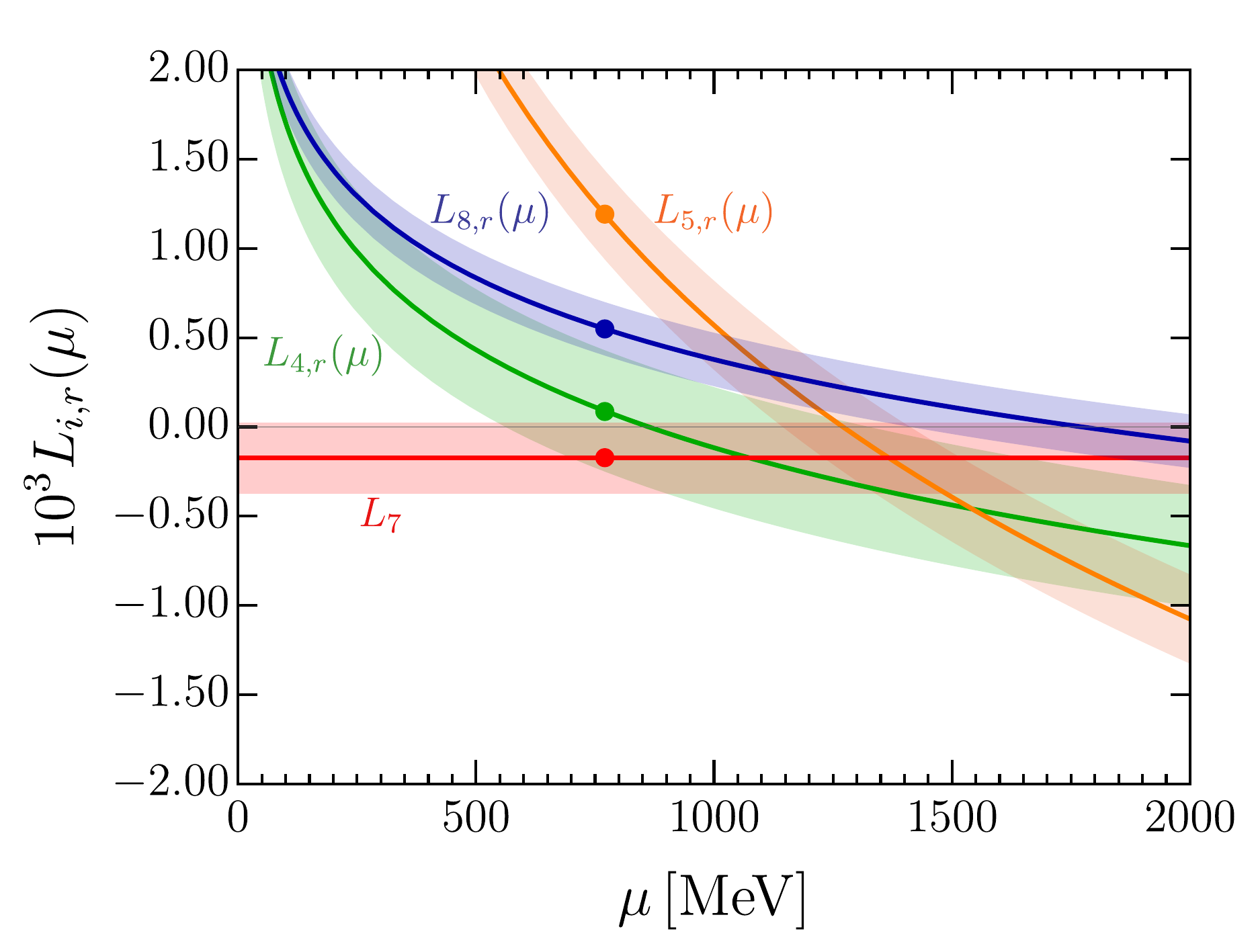}
\caption{Scale dependence and uncertainty bands of the relevant QCD low-energy constants $L_{i,r}(\mu)$. The dots refer to the values obtained in \cite{Dowdall:2013rya} at $\mu=m_\rho$.}
\label{fig:QCD_LECs}
\end{figure}

While no estimates of $L^{\theta}_{i,r}$ are available in the literature, the low-energy constants $L_{i,r}$ with $i=4,5,7,8$ relevant to our analysis have been determined -- with uncertainties ranging from about 10\% to more than 100\% -- from fits to low-energy data \cite{Bijnens:2014lea} and using lattice QCD \cite{FlavourLatticeAveragingGroupFLAG:2021npn}. For our estimates in Section~\ref{sec:Phenomenology} we employ two data sets: the values of the $\mathcal{O}(p^4)$ fit reported in Table~1 of \cite{Bijnens:2014lea}, and the lattice results obtained by the HPQCD collaboration from a simulation with $n_f=2+1+1$ dynamical quark flavors \cite{Dowdall:2013rya}. Both sets refer to the scale $\mu=m_\rho$ and are summarized in Table~\ref{table:LECs}. The value of $L_7$ is not available from the lattice. Given the good agreement of the phenomenological values and with the estimates $L_4^\infty=0$, $L_5^\infty=1.0\cdot 10^{-3}$, $L_7^\infty=-0.175\cdot 10^{-3}$, $L_8^\infty=0.375\cdot 10^{-3}$ obtained in the large-$N_c$ approximation \cite{Cirigliano:2003yq}, we use $L_7=(-0.175\pm 0.2)\cdot 10^{-3}$ with a conservative error for the second parameter set. The scale evolution of the relevant low-energy constants is shown in Figure~\ref{fig:QCD_LECs}, where the dots refer to the central values from \cite{Dowdall:2013rya} and $10^3\,L_7=-0.175$.

\subsection[Weak chiral Lagrangian at $\mathcal{O}(G_F\,p^4)$]{Weak chiral Lagrangian at $\bm{\mathcal{O}(G_F\,p^4)}$}
\label{sec:weak_Lagrangian_p4}

As discussed in Section~\ref{sec:weak_Lagrangian_p2}, the weak chiral Lagrangian involves operators transforming as $(8_L,1_R)$ or $(27_L,1_R)$ under chiral transformations, where, as is well known, the octet operators are strongly enhanced dynamically. Consequently, in this work we focus on the octet operators only. We start from the octet basis constructed in \cite{Ecker:1992de} under the assumptions $\theta=0$, $\braket{a_\mu}=0$ and $\braket{v_\mu}=0$, which consists of 37 operators $W_i^8$. This basis is the reduced version of a redundant set of operators $\{O_i^8\}$ originally presented in \cite{Kambor:1989tz}. In appendix \ref{app:A}, we discuss the reduction of these operators to the basis proposed in \cite{Ecker:1992de} in detail, focusing on the case of the SM extended by an ALP.

The vanishing of the field-strength tensors $F_{\mu\nu}^L$ and $F_{\mu\nu}^R$ associated with the external ALP currents $l_\mu$ and $r_\mu$ implies that 16 of the $W_i^8$ operators are absent, namely those with $i=14,\dots,18,25,26,27,29,\dots,35,37$. Next, the fact that the object $\chi=2 B_0\,m_q$ is a diagonal matrix implies that  $W_{36}^8=0$. Finally, for the SM extended by an ALP, and working consistently to first order in $a/f$, the operator $W_{22}^8$ is redundant and can be expressed as
\begin{align}
\label{eq:W22}
   W_{22}^8 = \frac12\,W_{23}^8 + \mathcal{O}\bigg( \frac{a^2}{f^2} \bigg) \,.
\end{align}
This leaves us with 19 operators, which we collect in the left portion of Table~\ref{table:p4weak}. The six operators containing four left-handed currents ($W_i^8$ with $i=1,\dots,4,19,28$) do not contribute to the $K^-\to\pi^- a$ decay amplitude. Note that the five operators $W_i^8$ with $i=19,20,21,23,24$ explicitly contain the left-handed ALP current and hence vanish in the SM. These operators can be rewritten in a different form using the identity $\braket{\lambda_6\,[l_\mu,B^\mu]}=\braket{[\lambda_6,l_\mu] B^\mu}$, which explains why the corresponding operators are absent for currents associated with left-handed gauge bosons, for which down and strange quarks have identical couplings, and hence $[\lambda_6,l_\mu]=0$.

\begin{table}[t]
\centering
\renewcommand{\arraystretch}{1.2} 
\scalebox{0.88}{
\begin{tabular}{|c|c|c|c||c|c|c|c|c|}
\hline 
$i$ & $W_i^8$ & $Z_i$ & $Z_i'$
 & $i$ & $W_i^{\theta\,8}$ & $Z_i^\theta$ & $Z_i^{\prime\theta}$ & $Z_i^{\theta\theta}$ \\
\hline
\hline
1 & $\braket{\lambda_6 L^2 L^2}$ & 2 & 0 
 & 1 & $(D_\mu\theta) \braket{\lambda_6 \{L^\mu,S\}}$ & \eqref{eq:Zitheta_relations} & $\frac12$
 & \eqref{eq:Zithetatheta_relations} \\
2 & $\braket{\lambda_6 L_\mu L^2 L^\mu}$ & $-\frac12$ & 0 
 & 2 & $i(D_\mu\theta) \braket{\lambda_6 [L^\mu,P]}$ & \eqref{eq:Zitheta_relations} & $\frac12$ 
 & \eqref{eq:Zithetatheta_relations} \\
3 & $\braket{\lambda_6 L_\mu L_\nu} \braket{L^\mu L^\nu}$ & 0 & 0 
 & 3 & $i(D_\mu\theta) \braket{\lambda_6 [L_\nu,W^{\mu\nu}]}$ & $Z_3^\theta$ & 0 & $Z_3^{\theta\theta}$ \\
4 & $\braket{\lambda_6 L_\mu} \braket{L^\mu L^2}$ & 1 & 0 
 & 4 & $(D_\mu\theta) \braket{\lambda_6 L^\mu} \braket{S}$ & $-\frac34$ & 0 & $\frac12$ \\
5 & $\braket{\lambda_6 \{S,L^2\}}$ & $\frac32$ & $\frac34$
 & 5 & $(\partial^\mu D_\mu\theta) \braket{\lambda_6 P}$ & $Z_5^\theta$ & 0 & $Z_5^{\theta\theta}$ \\
6 & $\braket{\lambda_6 L_\mu} \braket{S L^\mu}$ & $-\frac14$ & 0
 & 6 & $(D_\mu\theta) \braket{\lambda_6 \{L^\mu,L^2\}}$ & -- & 0 & -- \\
7 & $\braket{\lambda_6 S} \braket{L^2}$ & $-\frac98$ & $\frac12$
 & 7 & $(D_\mu\theta) \braket{\lambda_6 L^\mu} \braket{L^2}$ & -- & 0 & -- \\
8 & $\braket{\lambda_6 L^2} \braket{S}$ & $-\frac12$ & 0
 & 8 & $(D_\mu\theta) \braket{\lambda_6 L_\nu} \braket{L^\mu L^\nu}$ & -- & 0 & -- \\
9 & $i\braket{\lambda_6 [P,L^2]}$ & $\frac34$ & $-\frac34$
 & 9 & $i\epsilon_{\mu\nu\rho\sigma} (D^\mu\theta) \braket{\lambda_6 L^\nu L^\rho L^\sigma}$ & -- & 0 & -- \\
10 & $\braket{\lambda_6 S^2}$ & $\frac23$ & $\frac{5}{12}$ & & & & & \\
11 & $\braket{\lambda_6 S} \braket{S}$ & $-\frac{13}{18}$ & $\frac{11}{18}$ &&&&& \\
12 & $-\braket{\lambda_6 P^2}$ & $-\frac{5}{12}$ & $\frac{5}{12}$ &&&&& \\
13 & $-\braket{\lambda_6 P} \braket{P}$ & 0 & 0 &&&&& \\
19 & $\braket{\lambda_6\left[ l_\mu, [L^2,L^\mu] \right]}$ & $-\frac54$ & 0 &&&&& \\
20 & $\frac{i}{2} \braket{\lambda_6\left[ l_\mu, \{L_\nu,W^{\mu\nu}\} \right]}$ & $\frac34$ & 0 &&&&& \\
21 & $-\braket{\lambda_6\left[ l_\mu, [S,L^\mu] \right]}$ & $\frac56$ & 0 &&&&& \\
23 & $-i\braket{\lambda_6\left[ l_\mu, \{P,L^\mu\} \right]}$ & $\frac{5}{12}$ & 0 &&&&& \\
24 & $-i\braket{\lambda_6\left[ l_\mu,L^\mu \right]} \braket{P}$ & 0 & 0 &&&&& \\
28 & $i\epsilon_{\mu\nu\rho\sigma} \braket{\lambda_6 L^\mu} \braket{L^\nu L^\rho L^\sigma}$ & 0 & 0 &&&&& \\
\hline
\end{tabular}
}
\caption{CP-invariant operators $W_i^8$ (left) and $W_i^{\theta\,8}$ (right) entering the $\mathcal{O}(p^4)$ weak octet chiral Lagrangian \eqref{eq:lag_p4_weak} for the SM extended by an ALP. The coefficients $Z_i$ are taken from \cite{Ecker:1992de}. For the operator $W_{23}^8$ there is an additional contribution from the redundant operator $W_{22}^8$, see \eqref{eq:W22}. The calculation of the coefficients $Z_i'$ is described in Section~\ref{sec:weakmass}, whereas consistency conditions on the coefficients $Z_i^\theta$ and $Z_i^{\theta\theta}$ are derived in Section~\ref{sec:amplitude_NLO}. Entries with a dash remain unconstrained.
}
\label{table:p4weak}
\end{table}

To the best of our knowledge, the weak octet basis in the general case where $\theta$, $\braket{a_\mu}$ and $\braket{v_\mu}$ can be non-zero has not yet been derived in the literature. Working again to linear order in $D_\mu\theta$, we find that there are 12 such operators, which we can organize by the number of insertions of the left-handed current $L_\mu$. We find 
\begin{align}
\begin{aligned}
\label{eq:Wi_theta_set}
   & 1) && (\partial^\mu D_\mu\theta) \braket{\lambda_6 P} , \qquad
   & \phantom{1} 7) \quad & (\Box D_\mu\theta) \braket{\lambda_6 L^\mu} , \\
   & 2) && (\partial_\nu D_\mu\theta) \braket{\lambda_6 W^{\mu\nu}} , \qquad
   & \phantom{1} 8) \quad & (D_\mu\theta) \braket{\lambda_6 \{L^\mu,L^2\}} , \\
   & 3) && (D_\mu\theta) \braket{\lambda_6 \{L^\mu,S\}} , \qquad
   & \phantom{1} 9) \quad & (D_\mu\theta) \braket{\lambda_6 L_\nu L^\mu L^\nu} , \\
   & 4) && i(D_\mu\theta) \braket{\lambda_6 [L^\mu,P]} , \qquad
   & 10) \quad & (D_\mu\theta) \braket{\lambda_6 L^\mu} \braket{L^2} , \\
   & 5) && i(D_\mu\theta) \braket{\lambda_6 [L_\nu,W^{\mu\nu}]} , \qquad
   & 11) \quad & (D_\mu\theta) \braket{\lambda_6 L_\nu} \braket{L^\mu L^\nu} , \\
   & 6) && (D_\mu\theta) \braket{\lambda_6 L^\mu} \braket{S} , \qquad
   & 12) \quad & i\epsilon_{\mu\nu\rho\sigma} (D^\mu\theta) \braket{\lambda_6 L^\nu L^\rho L^\sigma} .
\end{aligned}
\end{align}
We have used that $\partial_\nu D_\mu\theta=\partial_\mu D_\nu\theta$, and that the equation of motion following from the $\mathcal{O}(p^2)$ chiral Lagrangian \cite{Gasser:1984gg}, namely
\begin{align}
\label{eq:EOM}
   W_\mu^{\,\mu} + 2 P - \frac23 \braket{P} \mathbbm{1} + \frac43\,\partial^\mu D_\mu\theta\,\mathbbm{1} 
   = \mathcal{O}(G_F) \,,
\end{align}
can be used to eliminate $W_\mu^{\,\mu}$. Here the $\theta$-dependent term ensures that
\begin{align}
   \braket{W_\mu^{\,\mu}} = 4 \braket{D_\mu L^\mu}
   = 4 \partial_\mu \braket{L^\mu} = - 4 \partial_\mu D^\mu\theta 
\end{align}
in accordance with \eqref{eq:traceL}. There is no need to introduce operators containing derivatives on the building blocks $S$, $P$, and $W^{\mu\nu}$ in \eqref{eq:Wi_theta_set}, since these derivatives can be moved over to the $\theta$ field using an integration by parts (up to higher-order terms in $a/f$). The above list of operators is still redundant. The Cayley--Hamilton theorem can be employed to eliminate the operator 9) using 
\begin{align}
\begin{aligned}
   (D_\mu\theta) \braket{\lambda_6 L_\nu L^\mu L^\nu}
   &= - (D_\mu\theta) \braket{\lambda_6 \{L^\mu,L^2\}} 
    + (D_\mu\theta) \braket{\lambda_6 L_\nu} \braket{L^\mu L^\nu} \\
   &\quad + \frac12 (D_\mu\theta) \braket{\lambda_6 L^\mu} \braket{L^2} + \mathcal{O}\bigg(\frac{a^2}{f^2}\bigg) \,.
\end{aligned}
\end{align}
Next, using an integration by parts and the equation of motion, and working to first order in $a/f$, we obtain
\begin{align}
\begin{aligned}
   (\Box D_\mu\theta) \braket{\lambda_6 L^\mu}
   &= \frac12\,(\partial^\mu D_\mu\theta) \braket{\lambda_6 P} \,, \\[1mm]
   (\partial_\nu D_\mu\theta) \braket{\lambda_6 W^{\mu\nu}}
   &= - 4 (\Box D_\mu\theta) \braket{\lambda_6 L^\mu} 
    = - 2 (\partial^\mu D_\mu\theta) \braket{\lambda_6 P} \,,
\end{aligned}
\end{align}
which eliminates two additional operators. This leaves us with the nine operators shown in the right portion of Table~\ref{table:p4weak}. Operators containing three factors of $L_\mu$ ($i=6,\dots,9$) do not contribute to the $K^\pm\to\pi^\pm a$ decay amplitudes.

The bare ALP mass term in the effective Lagrangian \eqref{eq:lag_p2_QCD} provides an additional low-energy scale besides $F$, $H_0$, and $B_0\,m_q$. This term is absent for the QCD axion, whose mass is generated dynamically through the potential following from the chiral Lagrangian \cite{GrillidiCortona:2015jxo}. For a general ALP, however, the presence of a non-zero $m_{a,0}^2$ allows one, in principle, to construct additional operators in the effective chiral Lagrangian starting at $\mathcal{O}(p^4)$. The fact that such operators must only contribute to processes involving an ALP eliminates the $\mathcal{O}(p^4)$ operators $m_{a,0}^2\braket{L^2}$ and $m_{a,0}^2\braket{S}$ in the QCD chiral Lagrangian, as well as $m_{a,0}^2\braket{\lambda_6 L^2}$ and $m_{a,0}^2\braket{\lambda_6 S}$ in the weak chiral Lagrangian. {\em A priori}, this argument does not eliminate the weak-interaction operator
\begin{align}
   m_{a,0}^2 \left( D_\mu\theta \right) \braket{\lambda_6 L^\mu} .
\end{align}
However, to first order in $a/f$ the ALP equation of motion implies
\begin{align}
   m_{a,0}^2 \left( D_\mu\theta \right) \braket{\lambda_6 L^\mu} 
    = - (\Box D_\mu\theta) \braket{\lambda_6 L^\mu}
    = - \frac12\,(\partial^\mu D_\mu\theta) \braket{\lambda_6 P} \,,
\end{align}
and hence there is no need to include this operator in the basis.

The complete $\mathcal{O}(p^4)$ weak octet Lagrangian for the SM extended by an ALP thus reads 
\begin{align}
   \mathcal{L}_{\rm weak}^{(p^4)} 
   = \frac{G_8 F^2}{2} \left( \sum_{i\in \mathcal{S}} N_i\,W_i^8 + \sum_{i=1}^9 N_i^\theta\,W_i^{\theta\,8} \right),
\label{eq:lag_p4_weak}
\end{align}
where the set $\mathcal{S}$ is defined as $\mathcal{S}=\{1,\dots,13,19,20,21,23,24,28\}$, and we follow the conventions of \cite{Ecker:1992de}. Under renormalization, the operators $W_i^8$ present in the SM in general require the new-physics operators $W_i^{\theta\,8}$ as counterterms to remove UV poles in Green's functions involving an external ALP. On the other hand, the new-physics operators only mix among themselves. For the structure of the $1/(d-4)$ pole terms in the corresponding couplings $N_i$ and $N_i^\theta$, this implies
\begin{align}
\begin{aligned}
\label{eq:Nilambda}
   N_i 
   &= N_{i,r}(\mu) + \lambda \left( Z_i + \frac{G_8'}{G_8}\,Z_i' \right) ; \qquad
    & i &\in S \,, \\[1mm]
   N_i^\theta 
   &= N_{i,r}^\theta(\mu) + \lambda \left( Z_i^\theta + \frac{G_8'}{G_8}\,Z_i^{\prime\,\theta}  
    + \frac{G_8^\theta}{G_8}\,Z_i^{\theta\theta} \right) ; \quad 
    & i &= 1,\dots,9 \,.
\end{aligned}
\end{align}
Here $G_8$, $G_8'$ and $G_8^\theta$ denote the coefficients of the three octet operators in the $\mathcal{O}(p^2)$ weak chiral Lagrangian \eqref{eq:lag_p2_weak}. The $Z_i$ factors have been computed in \cite{Ecker:1992de}, whereas the remaining renormalization coefficients are considered here for the first time. The coefficients $Z_i'$ and $Z_i^{\prime\,\theta}$ will be calculated in the following section, while consistency relations for the coefficients $Z_i^\theta$ and $Z_i^{\theta\theta}$ will be determined in Section~\ref{sec:amplitude_NLO} from the requirement that the $1/(d-4)$ pole terms cancel in the result for the $K^\pm\to\pi^\pm a$ decay amplitudes. We find that
\begin{align}
\label{eq:Zitheta_relations}
   Z_1^\theta = \frac12 + \frac12\,Z_5^\theta \,, \qquad
   Z_2^\theta = - \frac12 + 2 Z_3^\theta + \frac12\,Z_5^\theta\,,
\end{align}
and
\begin{align}
\label{eq:Zithetatheta_relations}
   Z_1^{\theta\theta} = \frac34 + \frac12\,Z_5^{\theta\theta} \,, \qquad
   Z_2^{\theta\theta} = 2 Z_3^{\theta\theta} + \frac12\,Z_5^{\theta\theta} \,.
\end{align}

In analogy with \eqref{eq:RGevol}, the relations \eqref{eq:Nilambda} imply the evolution equations
\begin{align}
\begin{aligned}
\label{eq:RGevolNi}
   \frac{d}{d\ln\mu}\,N_{i,r}(\mu) 
   &= - \frac{1}{16\pi^2} \left( Z_i + \frac{G_8'}{G_8}\,Z_i' \right) , \\
   \frac{d}{d\ln\mu}\,N_{i,r}^\theta(\mu) 
   &= - \frac{1}{16\pi^2} \left(  Z_i^\theta + \frac{G_8'}{G_8}\,Z_i^{\prime\,\theta}  
    + \frac{ G_8^\theta}{G_8}\,Z_i^{\theta\theta} \right) .
\end{aligned}
\end{align}
Note that the contributions proportional to $G_8'$ in the first equation modify the scale dependence of the renormalized low-energy coefficients $N_{i,r}(\mu)$, which seems to imply that the weak mass term has observable consequences even in the SM. However, as we will discuss in the next section, this effect can be removed by a redefinition of the low-energy constants $N_{i,r}$, in accordance with the findings of \cite{Kambor:1989tz}.

Regarding the renormalized weak low-energy constants $N_{i,r}$, our knowledge is limited to the three combinations that can be extracted from $K \to 3 \pi$ data, namely \cite{Bijnens:2004ai,Bijnens:2002vr,Kambor:1991ah}
\begin{align}
\begin{aligned}
   \tilde{K}_1(\mu) 
   &= g_8 \big[ N_{5,r}(\mu) - 2 N_{7,r}(\mu)  + 2 N_{8,r}(\mu) + N_{9,r}(\mu) \big] \,,  \\
   \tilde{K}_2(\mu) 
   &= g_8 \big[ N_{1,r}(\mu) + N_{2,r}(\mu) \big] \,, \\
   \tilde{K}_3
   &= g_8\,N_3 \,.
\end{aligned}
\end{align}
To the best of our knowledge, the study in \cite{Bijnens:2004ai} is the most recent extraction of these combinations.
Given that the operators $W^{8}_{1,2,3,7}$ do not contribute to $K^\pm\to\pi^\pm a$ decays, none of these combinations is directly relevant to our analysis. The only potentially useful estimate is the one of $\tilde{K}_1(\mu)$, since $W^8_{5,8,9}$ do contribute to $K^\pm\to \pi^\pm a$. However, in the fits performed in \cite{Bijnens:2004ai,Bijnens:2002vr} this combination has an $\mathcal{O}(1)$ uncertainty or is set to 0 by hand.
The uncertainty on $\tilde{K}_1(\mu)$ in the precedent analysis of \cite{Kambor:1991ah} is even larger ($\mathcal{O}(10)$). Given these substantial uncertainties and considering that the $K^\pm\to\pi^\pm a$ amplitudes depend on several other combinations, this estimate holds limited practical utility.

In this paper we focus on CP-even weak-interaction operators, because as explained in Section~\eqref{sec:weak_Lagrangian_p2} this is sufficient for our purposes. We mention for completeness that a basis of CP-odd operators can be obtained by simply replacing $\lambda_6\to\lambda_7$ in the basis operators \cite{Kambor:1989tz}.

\subsection{Weak mass term in the presence of external sources}
\label{sec:weakmass}

In \eqref{eq:lag_p2_weak} we have introduced the weak mass term as an octet operator of the form $\braket{\lambda_+ S}$ or, equivalently, $\braket{\lambda_6 S}$. When external sources are absent, there are several ways to show that this operator has no physical significance at lowest order in the chiral expansion (see e.g.\ \cite{Crewther:1985zt,Bernard:1985wf,Leurer:1987ih}). Moreover, at $\mathcal{O}(p^4)$ the effects of this operator can be absorbed into a redefinition of the couplings of the octet operators $W_i^8$ in the weak chiral Lagrangian \cite{Kambor:1989tz}. As first suggested by Crewther \cite{Crewther:1985zt}, the reparametrization invariance of the low-energy Lagrangian under chiral transformations plays a key role in eliminating the weak mass term from the LO chiral Lagrangian in the absence of external sources (up to linear order in $G_F$). We discuss the argument in some detail, because the presence of an ALP introduces some significant changes. 

In presence of only the weak mass term, the $\mathcal{O}(p^2)$ chiral Lagrangian can be expressed as
\begin{align}
\label{chiralOp2}
   \cL^{(p^2)} 
   = \frac{F^2}{8} \braket{(D_\mu\Sigma) (D^\mu\Sigma^\dagger) + \chi^D \Sigma^\dagger + \Sigma\,\chi^{D\dagger}}\,,
\end{align}
where
\begin{align}
   \chi^D = \left( 1 + 2 F^2 G_8^\prime\,\lambda_6 \right) \chi   \,,
\end{align}
contains linear (tadpole) terms in the meson fields, indicating that $\Sigma=1$ is no longer the correct ground-state of the theory. The linear terms can be removed, and the scalar potential be minimized, by means of an infinitesimal chiral transformation $\Sigma\to\Sigma^\prime=g_L\Sigma\,g_R^\dagger$ with
\begin{align}
\label{eq:weak_mass_term_field_redefintions}
   g_L = 1 + i\alpha_L \,, \qquad  g_R = 1 + i\alpha_R \,,
\end{align}
and $\alpha_{L,R}=\mathcal{O}(G_8')$. Under this transformation,  $\Sigma$ transforms as $\Sigma\to\Sigma+\delta\Sigma$, with
\begin{align}
   \delta\Sigma = i\alpha_L\,\Sigma - i\Sigma\,\alpha_R \,.
\label{eq:variation}
\end{align}
In the absence of external sources the kinetic term is unaffected by this transformation, and the condition
\begin{align}
   g_L\chi^D g_R^\dagger = \chi + \mathcal{O}(G_F^2)
\label{eq:condition}
\end{align}
ensures that after the transformation the ground-state is indeed at $\Sigma=1$. This condition is met by choosing the parameters $\alpha_L$ and $\alpha_R$ as
\begin{align}
\begin{aligned}
\label{eq:alphaLRsol}
   \alpha_L &= G_8^\prime F^2 \left( \frac{m_s+m_d}{m_s-m_d} + \frac{m_s-m_d}{m_s+m_d} \right) \lambda_7 \,, \\
   \alpha_R &= G_8^\prime F^2 \left( \frac{m_s+m_d}{m_s-m_d} - \frac{m_s-m_d}{m_s+m_d} \right) \lambda_7 \,.
\end{aligned}
\end{align}
In the absence of external sources, this choice removes the weak mass term from the LO chiral Lagrangian, indicating that it is a redundant operator at this order. Indeed, when computing the $K\to\pi\pi$ matrix elements of the weak mass term, one finds that there are two tree-level diagrams which exactly cancel each other. Beyond the LO the situation becomes more subtle. Since the weak mass term is redundant at $\mathcal{O}(p^2)$, one would find that its one-loop matrix elements also vanish, since they are computed using the leading-order strong and weak chiral Lagrangians.\footnote{To the best of our knowledge, this calculation has not yet been done.} However, tree-level diagrams involving the weak mass term along with an insertion of an operator $O_i$ from the $\mathcal{O}(p^4)$ QCD chiral Lagrangian \eqref{eq:lag_p4_QCD} do {\em not\/} vanish. The reason is that some of the operators $O_i$ are not invariant under the field redefinition in \eqref{eq:variation} \cite{Kambor:1989tz}. Hence the weak mass term is not a redundant operator at $\mathcal{O}(p^4)$ and beyond.

If external sources are present, the weak mass term is in general not redundant even at LO. In our case, the chiral Lagrangian contains the spurion fields $\chi=2B_0\,m_q$ along with $l_\mu$ and $r_\mu$, and it is formally invariant under chiral transformations only if we impose the hypothetical transformation rules \eqref{eq:spurion_transformations}. The fact that in reality the spurion $\chi$ is not invariant under a chiral transformation was the key to eliminating the weak mass term in the first place, alongside with the invariance of the kinetic term. However, in the presence of the ALP currents, this invariance no longer holds. In particular, the covariant derivative transforms as
\begin{align}
   D_\mu\Sigma \to D_\mu\Sigma
    + i \left( \alpha_L D_\mu\Sigma - D_\mu\Sigma\,\alpha_R\right)
    - [\alpha_L,l_\mu]\,\Sigma + \Sigma\,[\alpha_R,r_\mu] \,.
\end{align} 
The variation of the kinetic term in \eqref{chiralOp2} then yields an extra contribution, so that after the field redefinition we obtain
\begin{align}
\label{eq:weak_mass_term_after_field_redefinition}
   \cL^{(p^2)} 
   = \frac{F^2}{8} \braket{(D_\mu\Sigma) (D^\mu\Sigma)^\dagger 
    + \chi \Sigma^\dagger + \Sigma\,\chi^\dagger}
    + \frac{F^2}{4} \braket{[i\alpha_L,l_\mu] L^\mu + [i\alpha_R,r_\mu] R^\mu} \,,
\end{align}
with $\alpha_{L,R}$ as given in \eqref{eq:alphaLRsol}. This term is non-zero if there are sources $l_\mu$ and/or $r_\mu$ that do not commute with $\lambda_7$. In the SM, the only possible sources $l_\mu$ and $r_\mu$ at low energies arise from the photon, whose interactions are flavor-universal and thus commute with $\lambda_7$. In the ALP model, however, we have
\begin{align}
\begin{aligned}
   [i\lambda_7,l_\mu] 
   &= \frac{\partial_\mu a}{f} \left[ \big( [k_Q]_{33} - [k_Q]_{22} \big)\,\lambda_6 +  \text{Re}\,[k_Q]_{23} \,(\sqrt3\,\lambda_8 - \lambda_3) \right] , \\
   [i\lambda_7,r_\mu] 
   &= \frac{\partial_\mu a}{f} \left[ \big( [k_q]_{33} - [k_q]_{22} \big)\,\lambda_6 +  \text{Re}\,[k_q]_{23} \,(\sqrt3\,\lambda_8 - \lambda_3) \right] , 
\end{aligned} 
\end{align} 
which in general can be non-zero. Phenomenological bounds impose tight constraints on flavor off-diagonal ALP couplings in the down-quark sector, especially for the quarks of the first two generations \cite{Bauer:2021mvw}. Since the ALP couplings in the mass basis are obtained from the couplings in the interaction basis via a CKM-like rotation, a simple way to satisfy these bounds is to impose the conditions 
\begin{align}
   [k_Q]_{33} = [k_Q]_{22} \,, \qquad 
   [k_q]_{33} = [k_q]_{22} \,, \qquad 
   [k_Q]_{23} = [k_q]_{23} = 0
\end{align}
in the UV theory far above the electroweak scale, i.e., on the couplings in \eqref{eq:UVcouplings}. Corrections to these relations are generated by one-loop matching contributions at the electroweak scale through the weak interactions of the SM \cite{Bauer:2021mvw,Izaguirre:2016dfi,Gavela:2019wzg}. Since the weak mass term itself is of first order in $G_F$, these effects are then of second order in weak interactions and can be neglected. We will use the assumption of a ``flavor-universal ALP'' in some of our phenomenological analysis later in this work. However, for the purposes of the following discussion we treat the more general case.

For the calculation of the $K^\pm\to\pi^\pm a$ decay amplitudes at NLO in the chiral expansion in Section~\ref{sec:decayamplitude} we are going to use the weak mass term in the transformed form \eqref{eq:weak_mass_term_after_field_redefinition}, which has the advantage of not introducing tadpole vertices at $\mathcal{O}(p^2)$. However, this has non-trivial consequences, because we need to perform the chiral transformation in \eqref{eq:weak_mass_term_field_redefintions} also in the $\mathcal{O}(p^4)$ Lagrangian. Given that the field redefinition is of order $G_F$, we only need to perform it in the QCD chiral Lagrangian \eqref{eq:lag_p4_QCD}. In general, the basis operators in tis Lagrangian transform as 
\begin{align}
   O_i^{(\theta)} \to O_i^{(\theta)} + \delta O_i^{(\theta)} \,,
\end{align}
with variations $\delta O_i^{(\theta)}$ that are proportional to $G_8'$. To work out the explicit forms of these variations, we first consider the transformations of the left-handed building blocks, for which we define
\begin{align}
\begin{aligned}
   L_\mu &\to g_L\,L_\mu\,g_L^\dagger + \delta L_\mu \,, \\
   S &\to g_L\,S\,g_L^\dagger + \delta S \,, \\
   P &\to g_L\,P\,g_L^\dagger + \delta P \,,
\end{aligned}    
\end{align}
where the extra terms arise from the presence of the spurion fields. We find
\begin{align}
\begin{aligned}
   \delta L_\mu &= [i\alpha_L,l_\mu] - \Sigma\,[i\alpha_R,r_\mu]\,\Sigma^\dagger \,, \\
   \delta S &= - G_8' F^2 \big( \{ \lambda_6, S \} - i\,[\lambda_6, P] \big) \,, \\
   \delta P &= - G_8' F^2 \big( \{ \lambda_6, P \} + i\,[\lambda_6, S] \big) \,.
\end{aligned}    
\end{align}
The latter two relations have first been derived in \cite{Kambor:1989tz}. Using these results, we obtain
\begin{align}
\begin{aligned}
\label{eq:deltaOi}
   \delta O_1 & = 4 \braket{[i\alpha_L,l_\mu] L^\mu + [i\alpha_R,r_\mu] R^\mu} \braket{L^2} , \\
   \delta O_2 &= 4 \braket{[i\alpha_L,l_\mu] L_\nu + [i\alpha_R,r_\mu] R_\nu} \braket{L^\mu L^\nu} , \\
   \delta O_3 &= 2 \braket{[i\alpha_L,l_\mu] \{ L^\mu,L^2 \} + [i\alpha_R,r_\mu] \{ R^\mu,R^2 \}} , \\
   \delta O_4 &= - 2 G_8^\prime F^2 \braket{\lambda_6 S} \braket{L^2} \\
   &\quad + 2 \braket{[i\alpha_L,l_\mu] L^\mu + [i\alpha_R,r_\mu] R^\mu} \braket{S} , \\
   \delta O_5 &= - G_8^\prime F^2 \braket{\lambda_6 \left( \{ S,L^2 \} - i\,[ P,L^2 ] \right) } \\
   &\quad + \braket{[i\alpha_L,l_\mu] \{ L^\mu,S\} + [i\alpha_R,r_\mu] \{ R^\mu,S_R\}} , \\  
   \delta O_6 &= - 4 G_8^\prime F^2 \braket{\lambda_6 S} \braket{S} , \\
   \delta O_7 &= \phantom{+} 4 G_8^\prime F^2 \braket{\lambda_6 P} \braket{P} , \\
   \delta O_8 &= - 2 G_8^\prime F^2 \braket{\lambda_6 (S^2-P^2)} , 
\end{aligned}    
\end{align}
and 
\begin{align}
\begin{aligned}
\label{eq:deltaOitheta}
   \delta O_1^\theta &= 2 G_8' F^2 \left( \partial^\mu D_\mu\theta \right) \braket{\lambda_6 P} , \\
   \delta O_2^\theta &= G_8' F^2 \left( D_\mu\theta \right) 
    \braket{\lambda_6 \left( \{ L^\mu,S \} + i\,[ L^\mu,P ] \right)} \\
   &\quad - \left( D_\mu\theta \right) \braket{[i\alpha_L,l_\mu] S + [i\alpha_R,r_\mu] S_R} , \\
   \delta O_3^\theta &= \left( D_\mu\theta \right) \braket{[i\alpha_L,l_\mu] L^2 + [i\alpha_R,r_\mu] R^2} \\
   &\quad + \left( D_\mu\theta \right) \braket{[i\alpha_L,l_\nu] \{ L^\mu,L^\nu \}
    + [i\alpha_R,r_\nu] \{ R^\mu,R^\nu \}} .
\end{aligned}    
\end{align}
where we have used the relations \eqref{eq:RHD_objects}. Note that the terms involving $l_\mu$ and $r_\mu$ in the relations \eqref{eq:deltaOitheta} are of $\mathcal{O}(a^2/f^2)$ and can therefore be dropped. On the other hand, these contributions must be kept in the relations in \eqref{eq:deltaOi}, which generalize corresponding results derived in \cite{Kambor:1989tz} in the absence of the ALP sources. Our expressions for the $\delta O_i$ involve operators that are already present in \eqref{eq:lag_p4_weak}, specifically $W_{5,7,9,\dots,13}^8$ and $W_{1,2,5}^{\theta\,8}$, as well as new structures involving explicitly the external currents $l_\mu$ and $r_\mu$.

In the SM model the extra contributions involving the ALP currents are absent, and the entire contribution of the weak mass term to amplitudes for processes such as $K\to\pi\pi$ ands $K\to\pi\pi\pi$ is contained in the shifts $\delta O_i$, which in this case can all be expressed in terms of the weak-interaction basis operators $W_i^8$. This generates additional contributions to some of the Wilson coefficients $N_i$ proportional to the Gasser--Leutwyler coefficients $L_i$. Effectively,
\begin{equation}
N_i \to  N_i^\prime   \,, 
\end{equation}
with
\begin{align}
\begin{aligned}
\label{eq:combi1}
   N_5^\prime &= N_5 - 2\,\frac{G_8'}{G_8}\,L_5 \,, \qquad && N_{11}^\prime = N_{11} - 8\,\frac{G_8'}{G_8}\,L_6 \,, \\
   N_7^\prime &= N_7 - 4\,\frac{G_8'}{G_8}\,L_4 \,, \qquad && N_{12}^\prime= N_{12} - 4\,\frac{G_8'}{G_8}\,L_8 \,, \\
   N_9^\prime &= N_9 + 2\,\frac{G_8'}{G_8}\,L_5 \,, \qquad && N_{13}^\prime = N_{13} - 8\,\frac{G_8'}{G_8}\,L_7 \,, \\
   N_{10}^\prime &= N_{10} - 4\,\frac{G_8'}{G_8}\,L_8 \,. &&
\end{aligned}    
\end{align}
For all other coefficients $N_i^\prime\equiv N_i$. In the SM there are no additional loop contributions from the weak mass term at $\mathcal{O}(p^4)$, and hence the $1/(d-4)$ poles contained in the bare coefficients $L_i$ must be absorbed by the coefficients $N_i$. This determines the coefficients $Z_i'$ in Table~\ref{table:p4weak}. 

In an analogous way, in the presence of an ALP the variations $\delta O_i^\theta$ in \eqref{eq:deltaOitheta} yield contributions to some of the operators $W_i^{\theta\,8}$ in \eqref{eq:lag_p4_weak}, giving rise to the combinations 
\begin{align}
\label{eq:combi2}
   N_1^{\theta \prime} &= N_1^\theta + 2\,\frac{G_8'}{G_8}\,L_2^\theta \,, \notag\\
   N_2^{\theta \prime} &= N_2^\theta + 2\,\frac{G_8'}{G_8}\, L_2^\theta \,, \\
   N_5^{\theta \prime} &= N_5^\theta + 4\,\frac{G_8'}{G_8}\, L_1^\theta \,, \notag
\end{align}
and $N_i^{\theta \prime}\equiv N_i^\theta$ in all other cases. From the cancellation of the $1/(d-4)$ poles within these combinations, we have derived the factors $Z_i^{\prime\theta}$ in Table~\ref{table:p4weak}. The remaining terms involving the ALP currents $l_\mu$ and $r_\mu$ in \eqref{eq:deltaOi} cannot be expressed in terms of the operators $W_i^8$ or $W_i^{8\,\theta}$. They serve as counterterms to remove the $1/(d-4)$ poles in one-loop diagrams containing the weak mass term, which contribute to the $K^\pm\to\pi^\pm a$ decay amplitudes. The coefficients of these counterterms are uniquely determined in terms of the $L_i$ coefficients.

From \eqref{eq:RGevolNi}, it follows that for the primed coefficients shown in \eqref{eq:combi1} and \eqref{eq:combi2} the terms proportional to $G_8'$ cancel in the evolution equations, i.e.\
\begin{align}
\begin{aligned}
   \frac{d}{d\ln\mu}\, N_{5,r}^\prime(\mu) 
   &= - \frac{Z_5}{16\pi^2} \,, \\
   \frac{d}{d\ln\mu}\, N_{1,r}^{\theta \prime}(\mu) 
   &= - \frac{1}{16\pi^2} \left(  Z_i^\theta + \frac{G_8^\theta}{G_8}\,Z_i^{\theta\theta} \right) ,
\end{aligned}
\end{align}
and similarly for all other combinations.

\section{\boldmath Calculation of the $K^\pm\to\pi^\pm a$ decay amplitudes}
\label{sec:decayamplitude}

\subsection{Structure of the chiral expansion}

The standard definitions and notations employed in the literature on chiral perturbation theory obscure, to some extent, the systematics of the chiral expansion. Consider the QCD chiral Lagrangian (without an ALP) as an example. From \eqref{eq:lag_p2_QCD} and \eqref{eq:lag_p4_QCD} we have schematically
\begin{align}
   \cL_{\rm QCD} 
   = \frac{F^2}{8}\,\sum_i O_i^{(p^2)} + \sum_i L_i^{(p^4)}\,O_i^{(p^4)} 
    + \frac{1}{F^2} \sum_i L_i^{(p^6)}\,O_i^{(p^6)} + \dots
\end{align}
with dimensionless low-energy constants $L_i^{(p^{2n})}$. Operators at $\mathcal{O}(p^{2n})$ produce matrix elements containing $2n$ powers of the light meson masses, so naively the expansion of scattering amplitudes generated by this Lagrangian looks like
\begin{align}
   \mathcal{M} \propto F^2\,m_{K,\pi}^2 + L_i^{(p^4)} m_{K,\pi}^4 + \frac{1}{F^2}\,L_i^{(p^6)} m_{K,\pi}^6 + \dots \,.
\end{align}
Given that $F\approx m_\pi$, the series would not converge unless the dimensionless coefficients $L_i^{(p^{2n})}$ are strongly suppressed.

The physical scale governing this suppression is the scale of chiral symmetry breaking, $\mu_\chi=4\pi F\approx 1.6$\,GeV, which is significantly larger than the light meson masses. In fact, with three light quark flavors the ratio
\begin{align}
   \frac{m_K^2}{\left(4\pi F\right)^2} \approx 0.1
\end{align}
serves as a decent expansion parameter of chiral perturbation theory. In order to make this fact explicit, it is convenient to reorganize the effective Lagrangian in the form
\begin{align}
   \cL_{\rm QCD} 
   = \frac{F^2}{8} \left[ O^{(p^2)} + \frac{1}{\left(4\pi F\right)^2} \sum_i \hat L_i^{(p^4)}\,O_i^{(p^4)}  
    + \frac{1}{\left(4\pi F\right)^4} \sum_i \hat L_i^{(p^6)}\,O_i^{(p^6)} + \dots \right] ,
\end{align}
where we have rescaled the low-energy constants such that 
\begin{align}
\label{eq:elli_rescale}
   L_i^{(p^{2n})} \equiv \frac{\hat L_i^{(p^{2n})}}{8\left(4\pi\right)^{2n-2}} \,.
\end{align}
If the new parameters $\hat L_i^{(p^{2n})}$ are $\mathcal{O}(1)$ parameters, one obtains a systematic chiral expansion, in which one-loop contributions calculated from the $\mathcal{O}(p^2)$ Lagrangian are of the same order as tree-level contributions from the $\mathcal{O}(p^4)$ Lagrangian, two-loop contributions calculated from the $\mathcal{O}(p^2)$ Lagrangian are of the same order as one-loop contributions from the $\mathcal{O}(p^4)$ Lagrangian and tree-level contributions form the $\mathcal{O}(p^6)$ Lagrangian, and so on.

For the relevant low-energy constants collected in Table~\ref{table:LECs}, the values obtained after performing the scaling in \eqref{eq:elli_rescale} are indeed of $\mathcal{O}(1)$. For the first parameter set, e.g., we obtain $\hat L_{4,r}(\mu)=0.00\pm 0.38$, $\hat L_{5,r}(\mu)=1.52\pm 0.13$, $\hat L_7=-0.38\pm 0.25$, and $\hat L_{8,r}(\mu)=0.63\pm 0.25$, all at the scale $\mu=m_\rho$. From \eqref{eq:RGevol}, it follows that
\begin{align}
   \frac{d}{d\ln\mu}\,\hat L_{i,r}^{(\theta)}(\mu) 
   = - 8 \Gamma_i^{(\theta)} \,,
\end{align}
which according to Table~\ref{table:p4QCD} yields $\mathcal{O}(1)$ coefficients on the right-hand side. In our analysis below, we will perform the rescaling \eqref{eq:elli_rescale} for all low-energy constants in the QCD chiral Lagrangian at $\mathcal{O}(p^4)$, and based on the prefactors in \eqref{eq:lag_p2_weak} and \eqref{eq:lag_p4_weak} we will perform a similar rescaling of the low-energy constants arising in the weak chiral Lagrangian at $\mathcal{O}(p^4)$. Concretely, we define
\begin{align}
\label{eq:elli_ni_rescale}
   L_i^{(\theta)} \equiv \frac{\hat L_i^{(\theta)}}{8\left(4\pi\right)^2} \,, \qquad
   N_i^{(\theta)} \equiv \frac{\hat N_i^{(\theta)}}{2\left(4\pi\right)^2} \,.
\end{align}
In predictions for observables such as the $K^-\to\pi^- a$ decay amplitude, the dependence on the renormalization scale $\mu$ cancels between the logarithms arising from the renormalized loop diagrams and the low-energy constants $\hat L_{i,r}^{(\theta)}(\mu)$ and $\hat N_{i,r}^{\prime(\theta)}(\mu)$.

\subsection{Decay amplitudes at leading order}
\label{sec:amplitude_LO}

With the general chiral Lagrangian described in the previous section at hand, we can now move forward to the calculation of the physical amplitude for the process of interest, $K^-\to\pi^- a$. The amplitude for the CP-conjugate process $K^+\to\pi^+ a$ is then obtained by complex conjugation. We work at first order in $a/f$, where $f$ is the decay constant of the ALP, and at first order in $G_F$. This setup features two sources of flavor violation: the flavor-violating ALP couplings between strange and down quarks, and the SM weak interactions. We neglect contributions from the product of the two. We thus write the amplitude as
\begin{align}
   \mathcal{A} 
   = \mathcal{A}^{\rm FV} + \mathcal{A}^{\rm FC} \,,
\end{align}
where in the second term the superscript ``FC'' refers to flavor-conserving ALP couplings. In this contribution the flavor-changing transition comes from the SM weak interactions. Both terms receive a LO and a NLO contribution. In order to make the dependence from the ALP couplings explicit, we parameterize these contributions as
\begin{align}
\begin{aligned}
   \mathcal{A}^{\rm FV} 
   &=- (m_{K^-}^2-m_{\pi^-}^2)\,\frac{ \left[ k_d + k_D \right]_{12}}{2f}\,A^{\rm FV} \,, \\
   \mathcal{A}^{\rm FC} 
   &= \sum_{c_{\mathrm{ALP}},\,G}\, 
    \frac{G\,F_{\pi^-}^2\,m_{K^-}^2\, c_{\mathrm{ALP}}}{2f}\,A^{G,\,c_{\mathrm{ALP}}} \,,
\label{eq:amplitudeLOandNLO}
\end{aligned}
\end{align}  
where
\begin{align}
\begin{aligned}
   G &\in \big\{ G_8, G_8^\theta, G_8', G_{27}^{1/2}, G_{27}^{3/2} \big\} \,, \\[1mm]
   c_{\text{ALP}} &\in \big\{ \tilde c_{GG}, c_{uu}^a, (c_{dd}^a+c_{ss}^a), (c_{dd}^a-c_{ss}^a), (c_{dd}^v-c_{ss}^v) \big\} \,.
\end{aligned}
\end{align}
The normalization in \eqref{eq:amplitudeLOandNLO} is chosen such that the reduced amplitudes $A^{\rm FV}$ and $A^{G,\, c_{\mathrm{ALP}}}$ are dimensionless and independent of the ALP couplings. In particular, the amplitude $i A^{\rm FV}$ coincides with the $K^-\to\pi^-$ form factor $F_0^{K\to\pi}(q^2)$ evaluated at $q^2=m_a^2$, i.e.\ \cite{Bauer:2021mvw}
\begin{align}
  i A^{\rm FV} = F_0^{K\to\pi}(m_a^2) \,.
\end{align}
Note that we express all quantities in terms of the physical meson masses $m_{K^-}$, $m_{\pi^-}$ and the physical pion decay constant $F_{\pi^-}$ (see below). For simplicity, we will simply write $m_K$, $m_\pi$ and $F_\pi$ from now on. It is often stated that the flavor-diagonal vector couplings $c_{ff}^v$ of the ALP to the SM fermions are unobservable because the vector currents $\bar f\gamma^\mu f$ are conserved. However, as pointed out in \cite{Bauer:2021mvw}, this is no longer true when flavor-changing neutral-current processes are considered. In $s\to d$ weak transitions, the down-quark and strange-quark flavor numbers each change by one unit. Consequently, only the sum $(c_{dd}^v+c_{ss}^v)$ is unobservable, but the difference $(c_{dd}^v-c_{ss}^v)$ becomes an observable parameter.

\begin{figure}[t]
\centering
\includegraphics[scale=1.2]
{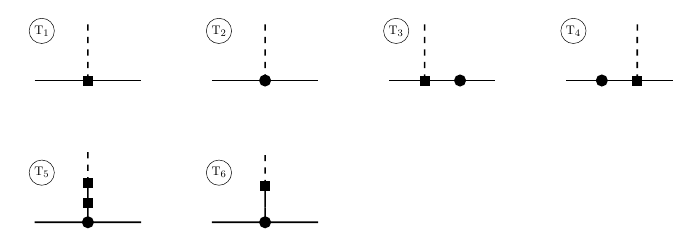}
\caption{Feynman graphs contributing to the $K^-\to\pi^- a$ decay amplitude at tree level. The ALP is represented by a dashed line. The black square $\rule[-.2ex]{1.5ex}{1.5ex}$ denotes an insertion of a vertex from the $\mathcal{O}(p^2)$ QCD Lagrangian in \eqref{eq:lag_p2_QCD}, while the black dot {\Large $\bullet$} refers to a vertex from the $\mathcal{O}(p^2)$ weak Lagrangian in \eqref{eq:lag_p2_weak}. Diagrams $T_5$ and $T_6$ account for $\pi^0$\,--\,$\eta_8$, $\pi^0$\,--\,$a$ and $\eta_8$\,--\,$a$ mixing contributions proportional to $\varepsilon^{(2)}$, which are not removed by the rotations in \eqref{eq:rotation}.}
\label{fig:tree_graphs}
\end{figure}

At LO in the chiral expansion, summing all tree-level Feynman diagrams is sufficient. These are shown in Figure~\ref{fig:tree_graphs}. In diagram $T_1$ the flavor-changing $s\to d$ transition occurs via a flavor-changing ALP--quark interaction, while in $T_{2,\dots,5}$ the flavor change originates from the CKM matrix. Since the rotations in \eqref{eq:rotation} diagonalize the bilinear terms in the Lagrangian in the isospin-conserving limit only, diagrams $T_5$ and $T_6$ are needed to account for the residual mixing between $\pi^0$, $\eta_8$ and $a$ arising at first order in the isospin-breaking parameter
\begin{align}
    \varepsilon^{(2)} 
   = \frac{m_d-m_u}{m_s-\hat{m}} \approx 0.028 \,,  \qquad 
   \hat{m} = \frac{m_u+m_d}{2} \,.
\label{eq:isospin_breaking_param}
\end{align} 
Expressed in terms of the mass ratios 
\begin{align}
   x = \frac{m_a^2}{m_K^2} \,, \qquad y = \frac{m_\pi^2}{m_K^2} \,,
   \label{eq:massratios}
\end{align}
the contribution involving flavor-changing ALP couplings reads
\begin{align}
   i A^{\rm FV}_{\rm LO} = 1 \,,
\label{eq:ALO_FV}
\end{align}
in agreement with the Ademollo--Gatto theorem. For the contributions from the three octet weak-interaction operators we obtain
\begin{align}  
   i A^{G_8,\,\tilde{c}_{GG}}_{\rm LO} 
   &= \frac{8\,(1-x)(1-y)}{4-y-3\,x} \left[ 1 - \varepsilon^{(2)}\,\frac{2(1-y)}{4-y-3\,x} \right] , \notag\\
   i A^{G_8,\,c_{uu}^a}_{\rm LO} 
   &= - \frac{y (1-y)}{4-y-3 x} \left[ 1 - \varepsilon^{(2)}\,\frac{ (1-y) \left( 8\,y-x\,(4+7\,y)+3\,x^2 \right)}{y\,(y-x) (4-y-3\,x)} \right] , \notag \\
   i A^{G_8,\,(c^a_{dd}+c^a_{ss})}_{\rm LO} 
   &= - \frac{(1-y)(4+y-6\,x)}{2\,(4-y-3\,x)} \notag \\
   &\quad \times \left[ 1 - \varepsilon^{(2)}\,\frac{(1-y) \left(16\,y-x \, (20+ 17\,y) +21\,x^2 \right)}{(y-x)(4+y-6\,x) (4-y-3\,x)} \right] , \\
   i A^{G_8,\,(c^a_{dd}-c^a_{ss})}_{\rm LO} 
   &= \frac{x\,(1+2\,y-3\,x)}{2\,(4-y-3\,x)} \notag\\
   &\quad \times \left[ 1 - \varepsilon^{(2)}\,\frac{(1-y) \left(4+13\,y+y^2-3\,x \,(7+5\,y)+18\, x^2\right)}{(y-x)(1+2\,y-3\,x)(4-y-3 \,x)} \right] , \notag \\[1mm]
   i A^{G_8,\,(c^v_{dd}-c^v_{ss})}_{\rm LO} 
   &= \frac{1-x+y}{2} \,, \notag 
\end{align}
as well as
\begin{align} 
   i A^{G_8^\theta,\,\tilde{c}_{GG}}_{\rm LO} 
   = -2 \, (1-y) \,,
\end{align}
and
\begin{align} 
\begin{aligned}
   i A^{G_8^\prime,\,(c^a_{dd}-c^a_{ss})}_{\rm LO} 
   &= (1-y)^2 \left[ 1 - \varepsilon^{(2)}\,(1-y) \right] , \\
   i A^{G_8^\prime,\,(c^v_{dd}-c^v_{ss})}_{\rm LO} 
   &= - \left[ 1 + \varepsilon^{(2)}\,(1-y) \right] . 
\end{aligned}
\end{align}
The remaining coefficients vanish. Finally, the contributions from the two 27-plet operators take the form 
\begin{align} 
\begin{aligned} 
   i A^{G_{27}^{1/2}, \, \tilde{c}_{GG}}_{\rm LO}  
   &= \frac{4\, (1-y) (4-3\, y-x)}{4-y-3\,x} 
    \left[ 1 - \varepsilon^{(2)}\,\frac{(1-y) (4-5 \, y+x)}{(4 - y-3 \,x) (4-3\,y - x)} \right] , \\
   i A^{G_{27}^{1/2},\, c_{uu}^a}_{\rm LO} 
   &= \frac{4\,y\, (1-y) }{4-y-3\, x}  
    \left[ 1 + \varepsilon^{(2)}\, \frac{(1-y) \left(y\, (4-5\,y)+ x\, (8-y)-6 x^2\right)}{2 \, y\, (y-x) (4-y-3\, x)}\right] , \\
   i A^{G_{27}^{1/2},\,(c^a_{dd}+c^a_{ss})}_{\rm LO} 
   &= - \frac{(1-y) (12-7 \,y-3 \,x)}{4 - y - 3\, x } \\
   &\quad\times \left[1 - \varepsilon^{(2)}\,\frac{(1-y) \left(2 \,y\, (4-5 \, y)-x \, (20-19 \, y)+3 \, x^2\right)}{(y-x) (4-y-3 \,x) (12-7 \, y- 3 \, x)} \right] , \\
   i A^{G_{27}^{1/2},\,(c^a_{dd}-c^a_{ss})}_{\rm LO} 
   &= \frac{1}{2\,(4-y-3\, x)}\,\bigg[ 20-25 \, y+5 \,y^2+x +2\, x \, y-3\, x^2 \\
   &\quad - \varepsilon^{(2)}\,\frac{6 \,x (1-y) \left(4+3 \,y-4 \,y^2-x \, (11-5 \,y)+ 3 \,x^2\right)}{(y-x) (4-y-3  x)} \bigg] \,, \\ 
   i\,A^{G_{27}^{1/2},\,(c^v_{dd}-c^v_{ss})}_{\rm LO} 
   &= \frac{1}{2}\,(1-x+y) \,, 
\end{aligned}
\end{align}
and
\begin{align} 
\begin{aligned} 
   i A^{G_{27}^{3/2},\,\tilde{c}_{GG}}_{\rm LO} 
   &= \frac{4\, (1-x) (1-y)}{4 - y -3\, x} \\ 
   &\quad\times \left[ 1  +  \varepsilon^{(2)}\,\frac{(1-y) \left( 12-13 \,y+9 \,x \, y-1\,  x + x^2 +2 \, y^2\right)}{(1-x) (y-x) (4 - y -3 \, x)} \right] , \\
   i A^{G_{27}^{3/2},\,c_{uu}^a}_{\rm LO} 
   &= \frac{2 \,y \, (1-y)  (3-y-2 \, x)}{(y-x) (4-y-3  x)} \\
   &\quad\times \left[1 - \varepsilon^{(2)}\,\frac{(1-y) }{y \, (3-y-2 \, x)} \left( \vphantom{\frac12} 5 - 2 \,y - 4 \,x - \frac{2 (4 - 7 \,x + 3 \,x^2) }{4 - y - 3\, x}\right) \right] , \\
   i A^{G_{27}^{3/2},\,(c^a_{dd}+c^a_{ss})}_{\rm LO} 
   &= - \frac{(1-y)}{(y-x) (4 - y -3\, x)}\,\bigg[\,  y \, (6-y)-x \, (3+5\, y)+3 \,x^2 \\
   &\quad + \varepsilon^{(2)}\,\frac{1-y}{4-y-3\,x} \left( 24 -26 \, y + 4 \, y^2- x \, (16 - 17  \, y) - 3 \, x^2 \right) \bigg] \,, \\
   i A^{G_{27}^{3/2},\,(c^a_{dd}-c^a_{ss})}_{\rm LO}
   &= - \frac{1}{2 (y-x) (4-y-3  x)} \\
   &\quad\times \bigg[\, y \left(4-5 \, y+\,y^2\right)+x \left(2-8 \,  \,y+3 \, y^2\right)  +x^2 \,(1+ 5\, y)-3 \,x^3 \\
   &\hspace{1.1cm} - \varepsilon^{(2)}\,\frac{6 \,x \, (1-y) \left(2-6 \,y+y^2 +x \, (2+ 4 \,y)-3 \,x^2\right)}{4-y-3 \,x} \bigg] \,, \\
   i\, A^{G_{27}^{3/2},\,(c^v_{dd}-c^v_{ss})}_{\rm LO} 
   &= \frac12\, (1 - x + y) \,. 
\end{aligned}
\end{align}
In these expressions all quark masses have been expressed in terms of the charged meson masses $m_{K^-}^2$ and $m_{\pi^-}^2$ plus, where necessary, an isospin-breaking correction proportional to $\varepsilon^{(2)}$. In the isospin-conserving limit $m_u=m_d$, the contributions to the amplitude mediated by $\mathcal{O}_8$ have already been calculated in \cite{Bauer:2021wjo}, while those mediated by $\mathcal{O}_{27}^{1/2}$ and $\mathcal{O}_{27}^{3/2}$ have been derived in \cite{Bauer:2021mvw}. We have included first-order corrections in the isospin-breaking parameter $\varepsilon^{(2)} $ defined in \eqref{eq:isospin_breaking_param}. As correctly anticipated in \cite{Bauer:2021wjo}, in the contributions proportional to $G_8$ and  $G_{27}^{1/2}$ these isospin-breaking effects are suppressed by $(m_d-m_u)/m_s$ and hence yield contributions to the amplitude at the $\mathcal{O}(1\%)$ level. This is not the case in the $G_{27}^{3/2}$ term, where the isospin-breaking contribution is of the same size as the isospin-conserving part of the amplitude. Indeed, in this case isospin-breaking effects scale as $(m_d-m_u)/\hat{m}$. In the results above, we have already imposed the condition \eqref{eq:trace_kappas} on the parameters entering the chiral rotation. As already noted in \cite{Bauer:2021wjo}, the results do not exhibit any residual dependence on the $\kappa_q$ parameters, as expected for consistency reasons. This applies also to the contributions computed here for the first time. 

\begin{figure}
\centering
\includegraphics[scale=1.3]{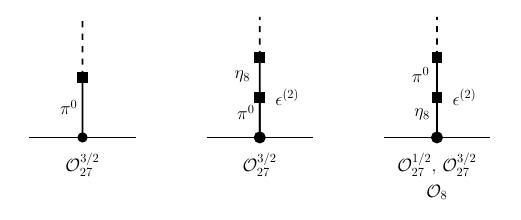}
\caption{Representation of Feynman diagrams that generate a divergent behavior for $m_a\to m_{\pi^0}$. For illustration purposes, we draw these graphs in the unrotated basis. The ALP is represented by the dashed line and generated via the insertion of a mixing with the $\pi^0$ or the $\eta_8$.}
\label{fig:divergence_diagrams}
\end{figure}

Some contributions to the LO amplitudes diverge in the limit $x\to y$, corresponding to $m_a\to m_{\pi^0}$. This (unphysical) behavior is due to the fact that the mixing between the ALP and the neutral pion is included only linearly in $F/f$ in our diagonalization procedure, see \eqref{eq:rotation_explicit_angles}. As can be seen from the explicit expressions shown above, at LO this effect arises only in the isospin-breaking contribution mediated by the $\mathcal{O}_{8}$ and  $\mathcal{O}_{27}^{1/2}$ operators, while for $\mathcal{O}_{27}^{3/2}$ it shows up also in the isospin-conserving contributions. 
The reason is that $\mathcal{O}_8$ and $\mathcal{O}_{27}^{1/2}$ cannot mediate the decay $K^-\to\pi^-\pi^0$ on shell, as they describe $\Delta I=1/2$ transitions only, and hence the corresponding amplitudes are proportional to $(\tilde m_{\pi^0}^2-p_{\pi^0}^2)$. When the neutral pion mixes into the ALP, this factor cancels the pole $1/(\tilde m_{\pi^0}^2-m_a^2)$ in the expressions for $\theta_{\pi^0 a}$ in \eqref{eq:rotation_explicit_angles} (or the $\pi^0$ propagator in the unrotated basis). Instead, $\mathcal{O}_{27}^{3/2}$ provides the necessary isospin change for the $K^{-} \to \pi^{-} \pi^0$ decay to be allowed on shell (first diagram in Figure~\ref{fig:divergence_diagrams}). This amplitude is therefore not proportional to $m_{\pi^0}^2 - m_a^2$, so the resulting contribution to $K^- \to \pi^- a$ diverges for $m_a = m_{\pi^0}$. 
The same behavior is generated when the neutral pion first mixes into the $\eta_8$ via an isospin-violating interaction, followed by a $\eta_8$-$a$ mixing (second diagram in Figure~\ref{fig:divergence_diagrams}). Similarly, $\mathcal{O}_{8}$ and $\mathcal{O}_{27}^{1/2}$ can only exhibit a divergent behavior in the isospin-violating case (see third diagram in Figure~\ref{fig:divergence_diagrams}). Thus, the pole from the propagator term is only generated at $\mathcal{O}(\epsilon^{(2)})$ for $\mathcal{O}_{8}$ and  $\mathcal{O}_{27}^{1/2}$ operators.  

As mentioned, the divergence is an artifact of the linearization in $F/f$. In a full, all-order diagonalization, the mixing angle $\theta_{\pi^0 a}$ does not contain a pole for $m_a =  m_{\pi^0}$. 
Still, one would find that the amplitude tends to different (finite) values for $m_a \to m_{\pi^0}^+$ and $m_a \to m_{\pi^0}^-$ (see Appendix \ref{app:C} for a detailed explanation).  This discontinuity is a physical effect: in this limit, the neutral pion and the ALP become essentially indistinguishable, hence the amplitude to the orthogonal state should also be computed to get a physical result. Indeed, the sum of the two amplitudes would be continuous over the entire domain. However, as discussed above, the case where the ALP and the neutral pion are degenerate -- corresponding to maximal mixing -- is experimentally excluded, hence we do not consider it here.

\subsection{Decay amplitudes at next-to-leading order}
\label{sec:amplitude_NLO}

In this section, we discuss the dominant corrections to the $K^-\to\pi^- a$ decay amplitude arising at NLO in the chiral expansion. For the reasons explained in Section~\ref{sec:weak_Lagrangian_p2}, we do not consider $\mathcal{O}(p^4)$ corrections for the 27-plets. Additionally, due to the negligible impact of isospin-breaking corrections on the LO octet result, we compute all $\mathcal{O}(p^4)$ contributions in the isospin-conserving limit. 

\begin{figure}
\centering
\includegraphics[scale=1.3]{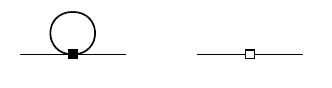}
\caption{Feynman graphs contributing to the mesons self-energy at NLO in the chiral expansion. The black square $\rule[-.2ex]{1.5ex}{1.5ex}$ denotes the insertion of a vertex from the $\mathcal{O}(p^2)$ QCD Lagrangian in \eqref{eq:lag_p2_QCD}, while the empty square $\square$ refers to a vertex from the $\mathcal{O}(p^4)$ QCD Lagrangian in \eqref{eq:lag_p4_QCD}.}
\label{fig:self_energy_mesons}
\end{figure}

We include one-loop diagrams with zero or one insertion of $\mathcal{O}(G_F\,p^2)$, along with arbitrary insertions from the $\mathcal{O}(p^2)$ QCD Lagrangian. Furthermore, we  consider tree-level graphs with either a $\mathcal{O}(G_F\,p^4)$ insertion plus an arbitrary number of $\mathcal{O}(p^2)$ QCD insertions or, alternatively, one insertion from the $\mathcal{O}(p^4)$ QCD Lagrangian and the $\mathcal{O}(G_F\,p^2)$ weak Lagrangian, accompanied by an arbitrary number of  insertions from the $\mathcal{O}(p^2)$ QCD Lagrangian. Furthermore, we account for external-leg corrections and incorporate NLO corrections to the LO meson masses and to the pion decay constant. Therefore, we write the total amplitude in the form 
\begin{align}
   \mathcal{A}_{\rm LO}+ \mathcal{A}_{\rm NLO} 
   = \sqrt{Z_{\pi} Z_{K}} \,\mathcal{A}_{\rm LO}^\prime + \mathcal{A}_{\Delta m_i^2,\,\Delta F_\pi}^{(p^4)} 
    + \mathcal{A}_{\text{1-loop}}^{(p^4)} + \mathcal{A}_{\rm tree}^{(p^4)} \,.
\label{eq:nlo_result}
\end{align}
Now we describe each of these contributions in detail, starting from the factor $\sqrt{Z_{\pi} Z_{K}}$, which represents the corrections from wave-function renormalization. The factors $Z_{\pi,K}$ are defined as 
\begin{align}
   Z_i^{-1} = 1 - \left. \frac{d\Sigma_i(p^2)}{d p^2} \right|_{p^2=m_i^2} \,, 
\end{align}
where $\Sigma_i(p^2)$ is the self-energy of the $i$-th meson, obtained by adding the one-loop 1-particle irreducible contribution with an insertion of a $O(p^2)$ QCD vertex, plus the tree-level contribution with an $O(p^4)$ QCD vertex, see Figure~\ref{fig:self_energy_mesons}.  We find 
\begin{align}
   Z_K &= 1 - \frac{1}{\left(4\pi F\right)^2}\,\bigg[ 
    \left( 4 m_K^{d-2} + \frac12\,m_\pi^{d-2} - \frac12\,m_{\eta_8}^{d-2} \right)
    \left( \frac{2}{d-4} - \ln 4\pi + \gamma_E - 1 \right) \notag \\
   &\hspace{2.88cm} + 5 m_K^2 \ln\frac{\mu^2}{m_K^2} + m_\pi^2 \ln\frac{\mu^2}{m_\pi^2} \notag \\
   &\hspace{2.88cm} + 2 m_K^2 \left( 2 \hat L_{4,r}(\mu) + \hat L_{5,r}(\mu) \right) 
    + 2 m_\pi^2\,\hat L_{4,r}(\mu) \bigg] \,, \\
   Z_\pi &= 1 - \frac{1}{\left(4\pi F\right)^2}\,\bigg[
    \left( \frac43\,m_K^{d-2} + \frac83\,m_\pi^{d-2} \right)
    \left( \frac{2}{d-4} - \ln 4\pi + \gamma_E - 1 \right) \notag  \\
   &\hspace{2.88cm} + 2 m_K^2 \ln\frac{\mu^2}{m_K^2} + 4 m_\pi^2 \ln\frac{\mu^2}{m_\pi^2} \notag \\
   &\hspace{2.88cm} + 4 m_K^2\,\hat L_{4,r}(\mu)
    + 2 m_\pi^2 \left( \hat L_{4,r}(\mu) + \hat L_{5,r}(\mu) \right) \bigg] \,, \notag 
\end{align}
where in the first relation $m_{\eta_8}^2$ can be eliminated using \eqref{eq:eta8mass}. The second term in \eqref{eq:nlo_result} is finite and comprises corrections in the masses and the decay constant. The relation between the meson masses at LO and NLO is given by 
\begin{align}
   m_{i,\rm{NLO}}^2 = m_i^2 + \Delta m_i^2 \,,
\label{eq:mass_shift}
\end{align}
where $\Delta m_i^2=\Sigma_i(m_i^2)$. The mass shift affects only the meson masses originating from Lagrangian parameters (specifically the quark masses, which appear both in vertices and in propagators), not those stemming from products of external momenta. Since we have expressed all quark masses in terms of the charged kaon and pion masses, it is sufficient to compute these two mass shifts. We find
\begin{align}
\begin{aligned}
   \Delta m_{\pi^-}^2 
   &= \frac{m_\pi^2}{\left(4\pi F\right)^2}\,\bigg[ 
    - m_\pi^2 \ln\frac{\mu^2}{m_\pi^2}
    + \frac{m_{\eta_8}^2}{3}\,\ln\frac{\mu^2}{m_{\eta_8}^2} \\
   &\quad + (4 m_K^2 + 2 m_\pi^2) \left( 2 \hat L_{6,r}(\mu) - \hat L_{4,r}(\mu) \right)
    + 2 m_\pi^2 \left( 2\hat L_{8,r}(\mu) - \hat L_{5,r}(\mu) \right) \bigg] \,, \\[1mm]
   \Delta m_{K^-}^2  
   &= \frac{m_K^2}{\left(4\pi F\right)^2}\,\bigg[ 
    - \frac23\,m_{\eta_8}^2 \ln\frac{\mu^2}{m_{\eta_8}^2} \\
   &\quad + (4 m_K^2 + 2 m_\pi^2) \left( 2\hat L_{6,r}(\mu) - \hat L_{4,r}(\mu) \right)
    + 2 m_K^2 \left( 2\hat L_{8,r}(\mu) - \hat L_{5,r}(\mu) \right) \bigg] \,, 
\end{aligned}
\end{align}
in agreement with \cite{Gasser:1984gg}. In the LO result we thus replace $m_i^2=m_{i,\rm{NLO}}^2-\Delta m_i^2$ for $i=\pi,K$. Also the decay constant $F$ can be expressed in terms of the physical pion decay constant $F_\pi$ as 
\begin{align}
   F_\pi =  F+ \Delta F_\pi \,,
\label{eq:Fshift}
\end{align}
where at NLO \cite{Gasser:1984gg}
\begin{align}
   \Delta F_{\pi} 
   = \frac{F_\pi}{\left(4\pi F\right)^2}\,\bigg[ m_K^2 \ln\frac{\mu^2}{m_K^2} 
    + 2 m_\pi^2 \ln\frac{\mu^2}{m_\pi^2} 
   + 2 m_K^2\,\hat L_{4,r}(\mu) 
    + m_\pi^2 \left( \hat L_{4,r}(\mu) + \hat L_{5,r}(\mu) \right) \bigg] \,.\notag\\
\label{eq:FshiftExplicit}
\end{align}
Note that this is a finite renormalization, and the scale dependence on the right-hand side cancels between the logarithms and the renormalized low-energy constants. 

\begin{figure}[t]
\centering
\includegraphics[scale=1.3]{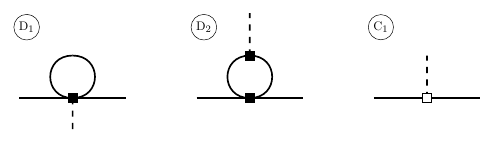}
\caption{Feynman graphs with a flavor-violating ALP coupling, contributing to the $K^-\to\pi^- a$ decay amplitude at NLO in the chiral expansion. The black square $\rule[-.2ex]{1.5ex}{1.5ex}$ denotes the insertion of a vertex from the $\mathcal{O}(p^2)$ QCD Lagrangian in \eqref{eq:lag_p2_QCD}, while the empty square $\square$ refers to a vertex from the $\mathcal{O}(p^4)$ QCD Lagrangian in \eqref{eq:lag_p4_QCD}.}
\label{fig:QCD_loops}
\end{figure}

\begin{figure}[t]
\centering
\includegraphics[width = \textwidth]{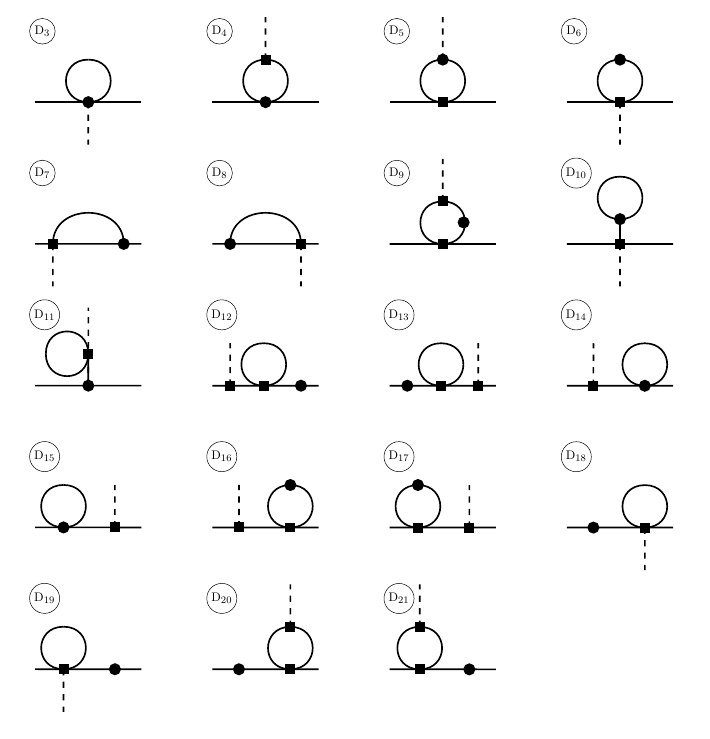}
\caption{Feynman graphs contributing to the $K^-\to\pi^- a$ decay amplitude at NLO in the chiral expansion, and generated by flavor-conserving ALP interactions. The flavor change is mediated by an insertion of a vertex from the $\mathcal{O}(p^2)$ weak chiral Lagrangian in \eqref{eq:lag_p2_weak} and shown by a black dot {\Large $\bullet$}. The black square $\rule[-.2ex]{1.5ex}{1.5ex}$ refers to a vertex from the $\mathcal{O}(p^2)$ QCD chiral Lagrangian in \eqref{eq:lag_p2_QCD}.}
\label{fig:weak_loops}
\end{figure}

By substituting \eqref{eq:mass_shift} and \eqref{eq:Fshift} in the LO amplitude, effectively we split it up in two pieces,
\begin{align}
   \mathcal{A}_{\rm LO} (m_i^2,F)  
   = \mathcal{A}_{\rm LO} (m_{i,\rm{NLO}}^2,F_\pi) + \mathcal{A}_{\Delta m_i^2,\,\Delta F_\pi}^{(p^4)} 
   \equiv \mathcal{A}_{\rm LO}^\prime + \mathcal{A}_{\Delta m_i^2,\,\Delta F_\pi}^{(p^4)}  \,.
\label{mass-shift}
\end{align}
The first one is formally of $\mathcal{O}(p^2)$ and is strictly speaking the ``true'' LO result (expressed in terms of physical parameters), while the second one is part of the $\mathcal{O}(p^4)$ correction. Both these quantities are trivially independent of the $\kappa_q$ parameters. In the following we will drop the prime when referring to the LO result. 

\begin{figure}[t]
\centering
\includegraphics[scale=1.4]{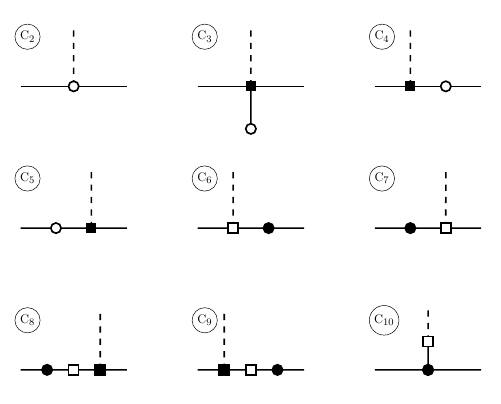}
\caption{Feynman graphs contributing to the $K^-\to\pi^- a$ decay amplitude at NLO in the chiral expansion, generated by flavor-conserving ALP interactions involving one insertion of the $\mathcal{O}(p^4)$ QCD  Lagrangian in \eqref{eq:lag_p4_QCD} (empty square $\square$) or $\mathcal{O}(p^4)$ weak Lagrangian in \eqref{eq:lag_p4_QCD} (empty dot $\Circle$). The black square $\rule[-.2ex]{1.5ex}{1.5ex}$ and circle {\Large $\bullet$} are as explained in Figure \ref{fig:weak_loops}.}
\label{fig:weak_counterterms}
\end{figure}

The third term in \eqref{eq:nlo_result}, $\mathcal{A}_{\text{1-loop}}^{(p^4)}$, comprises all 1-particle irreducible one-loop graphs built out of $\mathcal{O}(p^2)$ vertices and including at most one weak vertex. This leads to 21 distinct topologies, displayed in Figure~\ref{fig:QCD_loops} ($D_{1,2}$) and Figure~\ref{fig:weak_loops} ($D_{3,\dots,21}$), each with various internal states, for a total of 89 distinct graphs. Finally, the last term in \eqref{eq:nlo_result}, $\mathcal{A}_{\rm tree}^{(p^4)}$, contains tree-level diagrams with one $\mathcal{O}(p^4)$ insertion and arbitrarily many $\mathcal{O}(p^2)$ insertions. The corresponding topologies are $C_1$ in Figure~\ref{fig:QCD_loops} and $C_{2,\dots,10}$ in Figure~\ref{fig:weak_counterterms}. Only $C_{10}$ has two distinct possible virtual states ($\pi^0$ and $\eta_8$), hence in total we have 11 distinct graphs. Note that even though we have removed the ALP mixing with $\pi^0$ and $\eta_8$ at LO, both are generated at NLO. Instead of performing a NLO diagonalization of the mass and kinetic terms, we choose to keep these terms explicitly and treat them as interactions. This is shown in graphs $D_{11}$ and $C_{10}$.

We have checked explicitly that the sums $\sum_{i=1}^{21} D_i$ and $\sum_{i=1}^{10} C_i$ are (separately) independent of the $\kappa_q$ parameters. This is an important consistency check of the correctness of our NLO calculation. Another crucial check is the cancellation of UV divergences. We have confirmed that the UV divergences cancel in the contribution involving the flavor-changing couplings of the ALP, as well as in the contribution proportional to $G_8$ in the limit where $\braket{a_\mu}=0$ and $\theta=0$ (i.e.\ $\braket{c^a}=0$ and $c_{GG}=0$). This requires knowledge of the coefficients $\Gamma_i$, $\Gamma^\theta_i$ and $Z_i$ collected in Tables~\ref{table:p4QCD} and \ref{table:p4weak}. The cancellation of the remaining poles, proportional to $G_8\,(2c_{GG}+\braket{c^a})$ and $G_8^\theta\,(2c_{GG}+\braket{c^a})$, is achieved via the operators $W_{1,2,3}^{\theta\,8}$ when choosing the anomalous dimensions as shown in Table~\ref{table:p4weak}. The remaining (finite) result $\mathcal{A}_{\rm LO}+\mathcal{A}_{\rm NLO}$ retains an explicit $\ln\mu$ dependence, which is canceled by the scale dependence of the various low-energy constants. Explicit expressions for the NLO contributions -- parametrized as in  \eqref{eq:amplitudeLOandNLO} -- in terms of the mass ratios in \eqref{eq:massratios} and the low-energy constants $\hat{L}_{i,r} (\mu)$, $\hat{L}_{i,r}^{\theta} (\mu)$, $\hat{N}_{i,r}^{\prime} (\mu)$, and $\hat{N}_{i,r}^{\theta \prime}(\mu)$ are collected in Appendix~\ref{app:B}. We also provide diagram by diagram expressions 
in a \texttt{Mathematica} notebook attached to the arXiv version of this article as an ancillary file.

An unexpected aspect of our result is the complete absence of an absorptive part of the $K^-\to\pi^- a$ decay amplitude. This is in stark contrast to the case of the $K_S\to\pi^+\pi^-$ decay, in which case the one-loop diagrams in chiral perturbation theory generate a significant strong rescattering phase \cite{Buras:2016fys}. In principle, non-zero imaginary parts could arise from the loop diagrams $D_4$, $D_8$, $D_9$, $D_{20}$, and $D_{21}$ in Figure~\ref{fig:weak_loops}, when the initial-state kaon in graph $L_8$ decays into two pions, or when the final-state ALP couples to a two-pion state (remaining graphs). However, the vertex factors in these diagrams ensure that all relevant cuts vanish. In diagram $D_8$, the Feynman rule for the $K^-\,\pi^-\,\pi^0$ vertex (black circle) is proportional to the difference of the two pion propagators, yielding a sum of two tadpole integrals that are real. Similarly, in diagrams $D_4$, $D_{20}$, and $D_{21}$ the Feynman rule for the $a\,\pi\,\pi$ vertex is proportional to the difference of the two pion propagators, yielding a sum of two tadpole integrals in all cases. In diagram $D_9$ a contribution remains with two propagators in the loop, which arises when the upper right pion propagator is cancelled by the numerator structure. In this case, however, the loop contains a $K\,\pi$ intermediate state, which cannot be on-shell, since $m_a<m_K-m_\pi$ for the decay $K^-\to\pi^- a$ to be kinematically allowed.

\section{Phenomenological consequences}
\label{sec:Phenomenology}

We now provide numerical estimates of the NLO effects calculated above. For our study we use PDG values for the meson masses and the kaon width, i.e.\ $m_{K^-}=493.7$\,MeV, $m_{\pi^-}=139.6 $\,MeV and $\Gamma_{K^-}=5.31\cdot 10^{-14}$\,MeV \cite{ParticleDataGroup:2022pth}, along with the FLAG average for the pion decay constant, $F_{\pi^-}=(130.2\pm 0.8)$\,MeV \cite{FlavourLatticeAveragingGroupFLAG:2021npn}.

\subsection{Numerical size of higher-order effects}

\begin{figure}[t]
\centering
\includegraphics[width=0.55\textwidth]{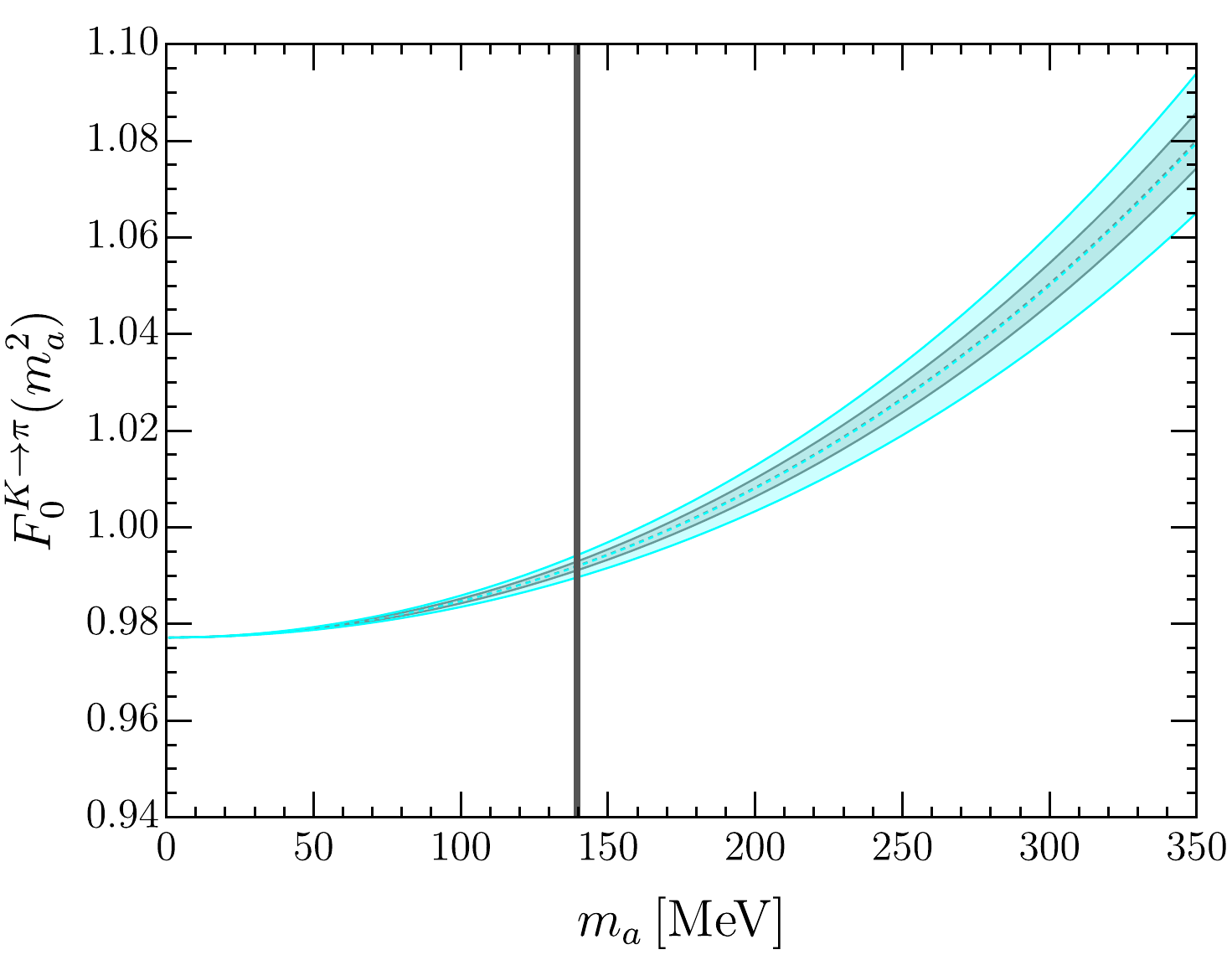}
\caption{Total (LO\,+\,NLO) contribution of flavor-violating ALP couplings to the $K^- \to \pi^- a$ decay amplitude. The  gray (light blue) band corresponds to the $1\sigma$  uncertainty estimated using the inputs of the first (second) column of Table~\ref{table:LECs} for the QCD low-energy constant $L_{5,r}$. The region $m_a \approx m_{\pi^0}$ is excluded for the reason explained below  \eqref{eq:rotation_explicit_angles}.} 
\label{fig:pheno_deltaFV}
\end{figure}

\begin{figure}[t]
    \centering
 \includegraphics[width=.48\textwidth]{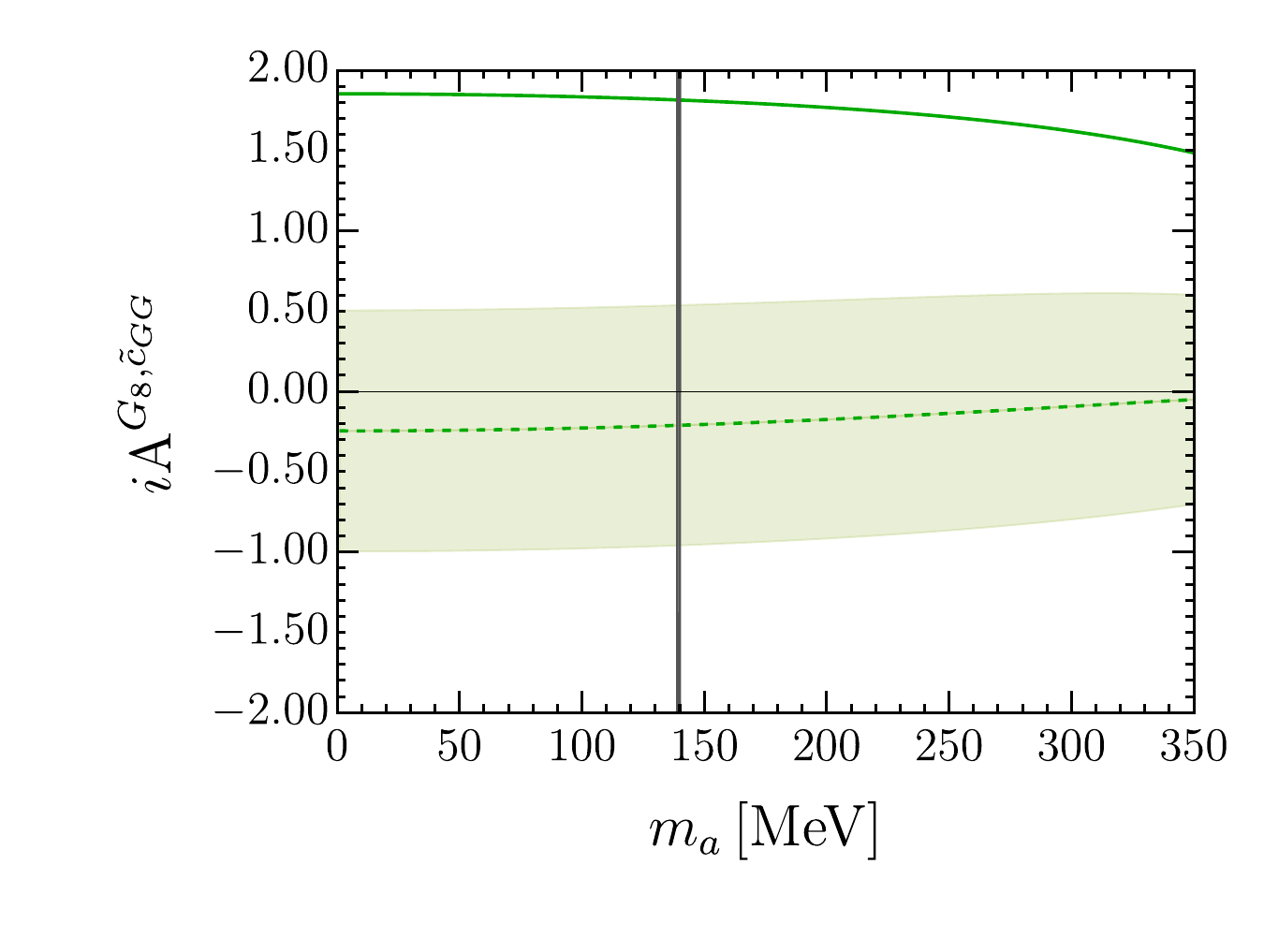}
\includegraphics[width=.48\textwidth]{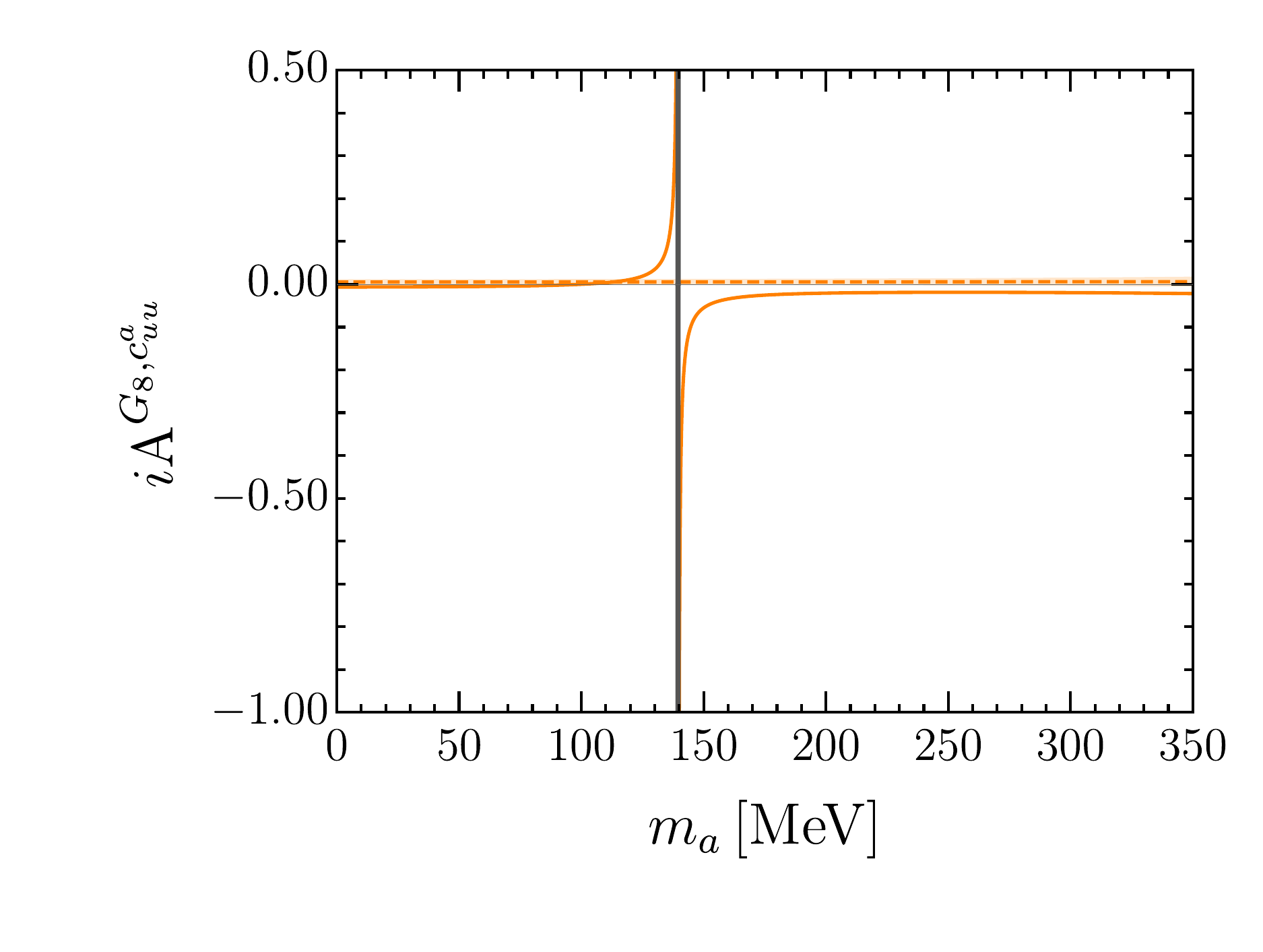}
\\[-2mm]
\includegraphics[width=.48\textwidth]{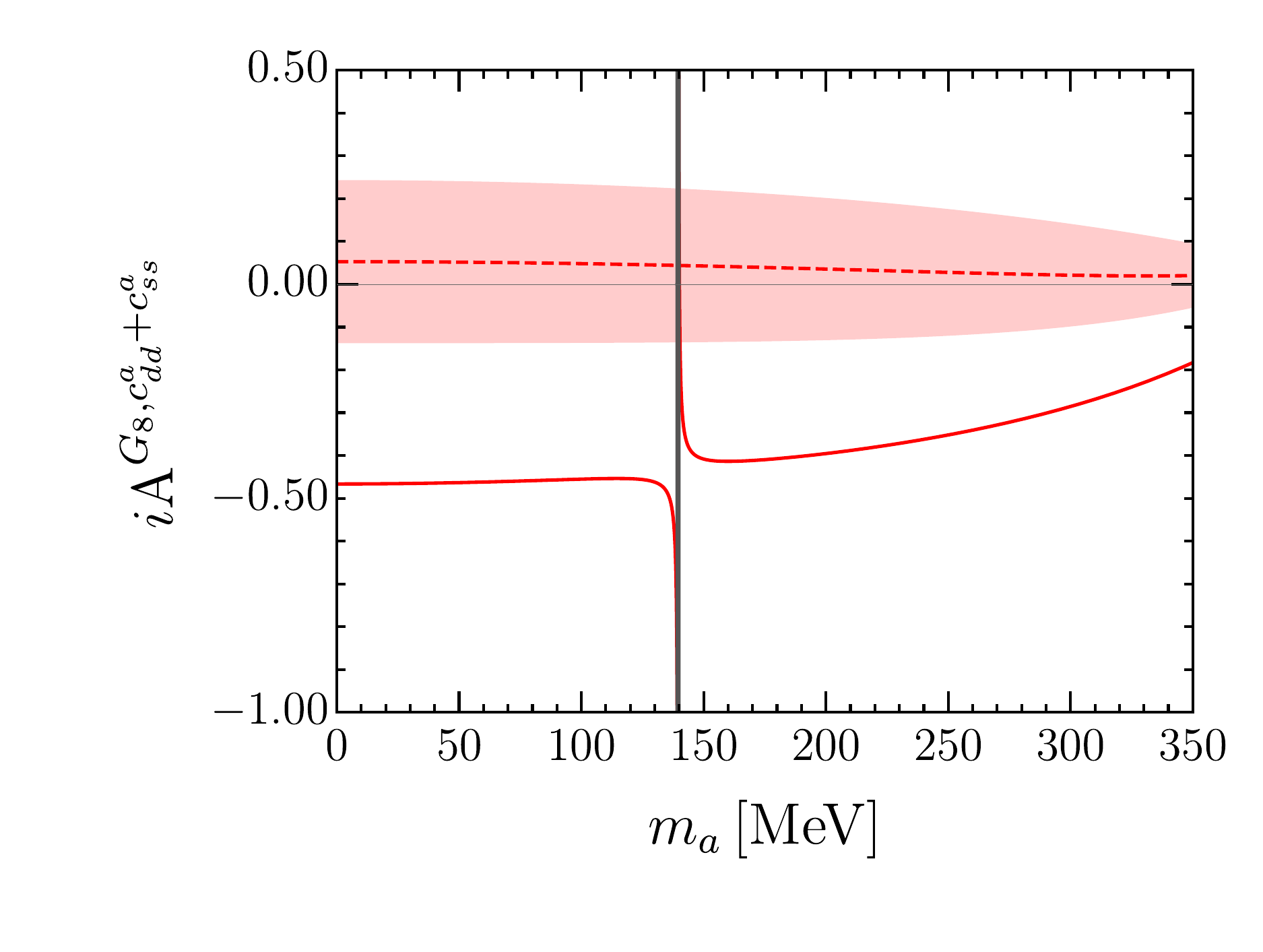}
\includegraphics[width=.48\textwidth]{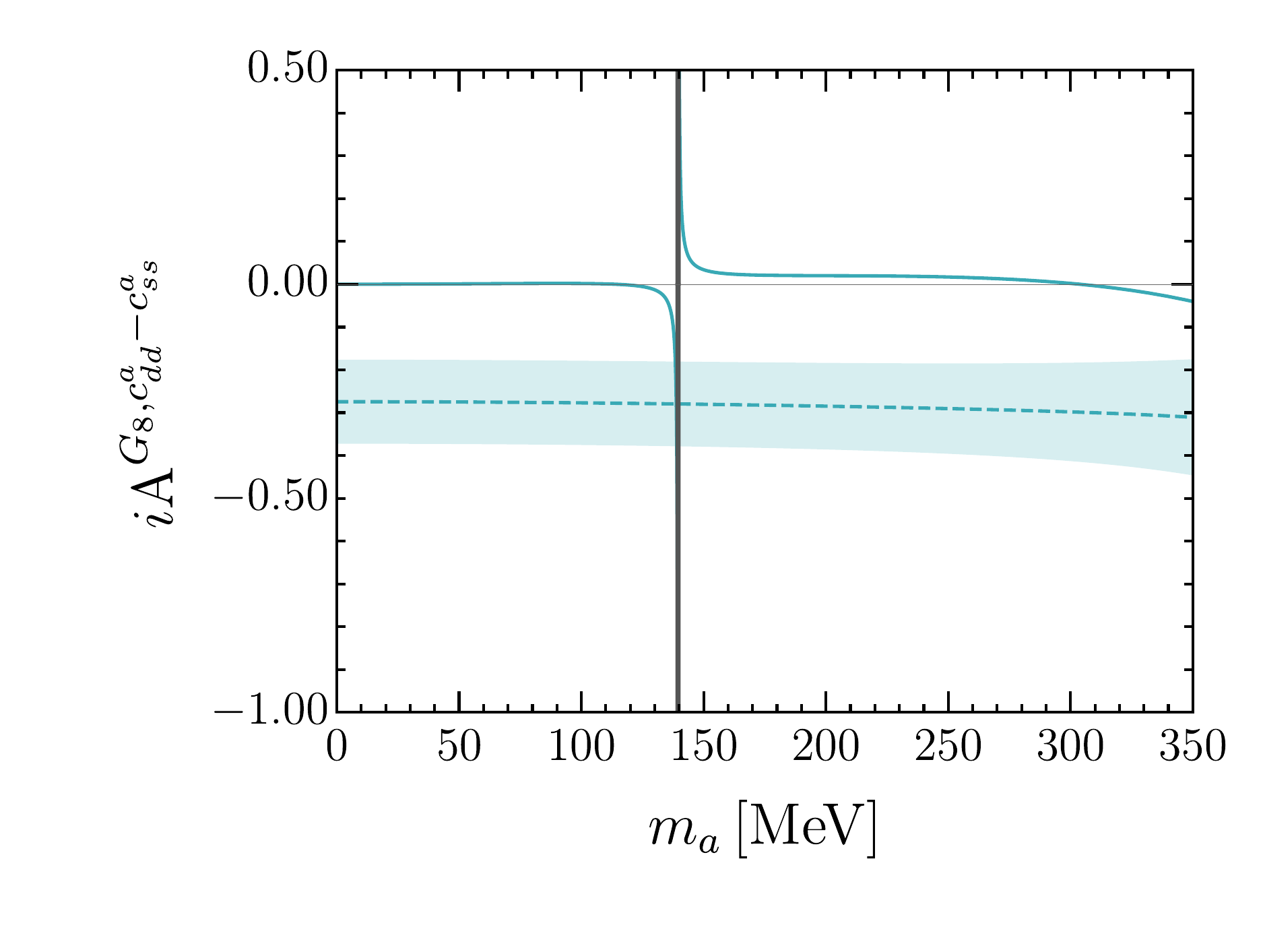} \\[-2mm]    
\includegraphics[width=.48\textwidth]{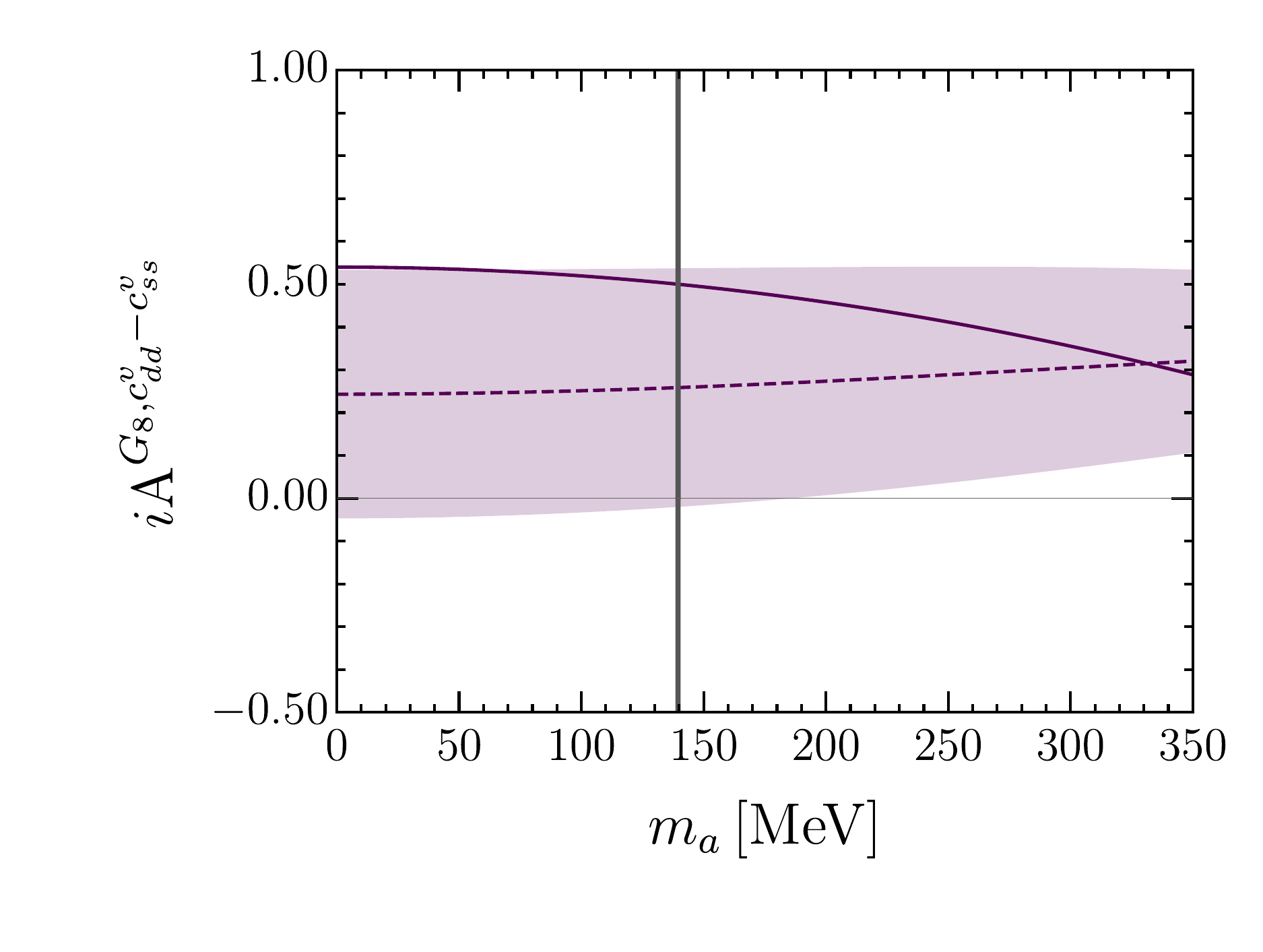}
\vspace{-2mm}
\caption{LO (solid) and NLO (dashed lines) contributions proportional to $G_8$ to the $K^-\to\pi^- a$ decay amplitude, as defined in \eqref{eq:amplitudeLOandNLO}. The bands show the combined $1\sigma$ uncertainties estimated using the inputs of the second column of Table~\ref{table:LECs} for the QCD low-energy constants and varying the scale $\mu_0$ by a factor of $\sqrt{2}$ around the default choice $\mu_0=1.4$\,GeV. The gray vertical line in the region $m_a \approx m_{\pi^0}$ is excluded for the reasons explained below \eqref{eq:rotation_explicit_angles}. Note that for the second plot we have used a different scale on the vertical axis.
}
\label{fig:Amp_Plots}
\end{figure}

\begin{figure}[t]
\centering
\includegraphics[width=.49\textwidth]{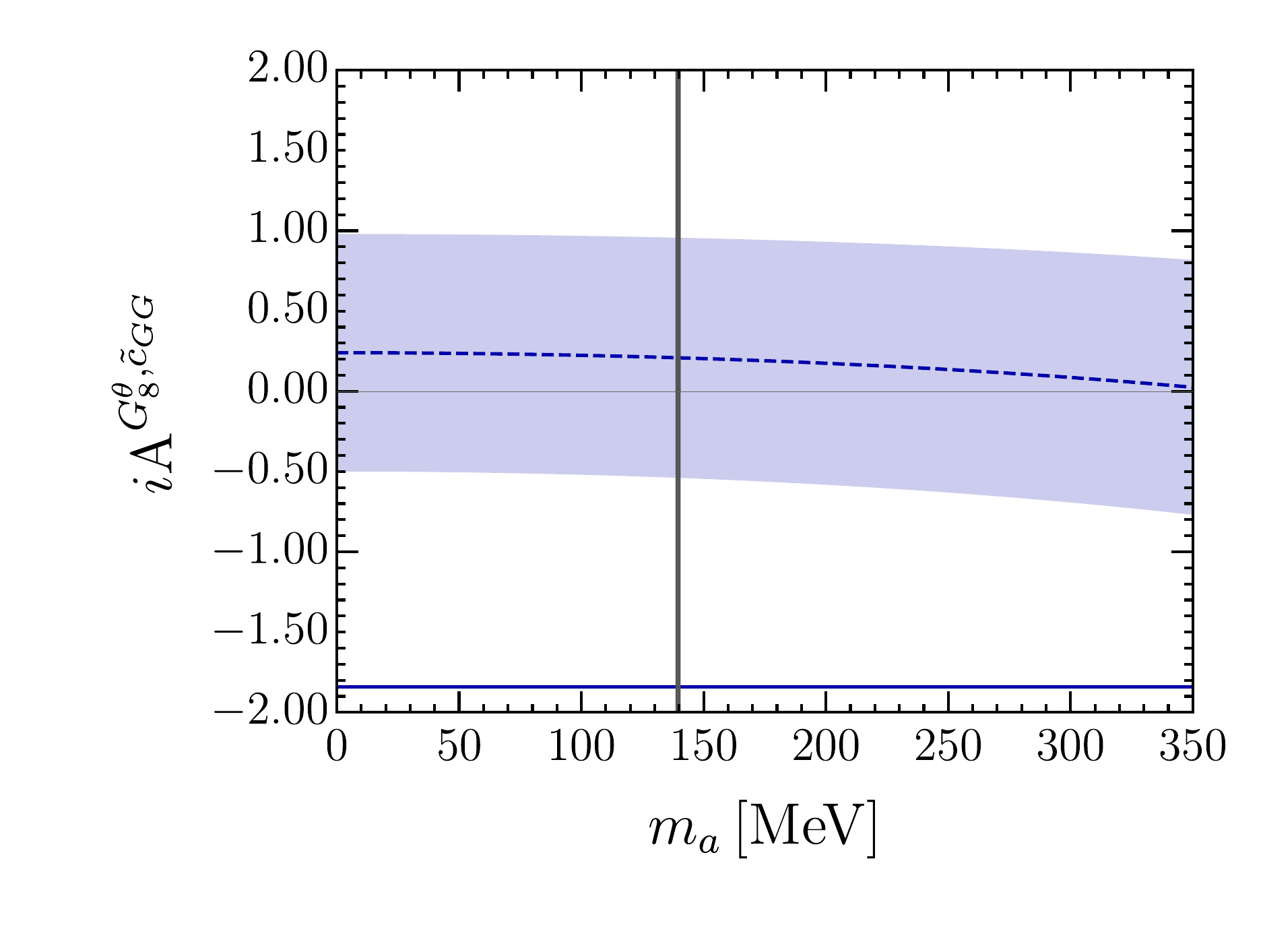} \\
\includegraphics[width=.485\textwidth]{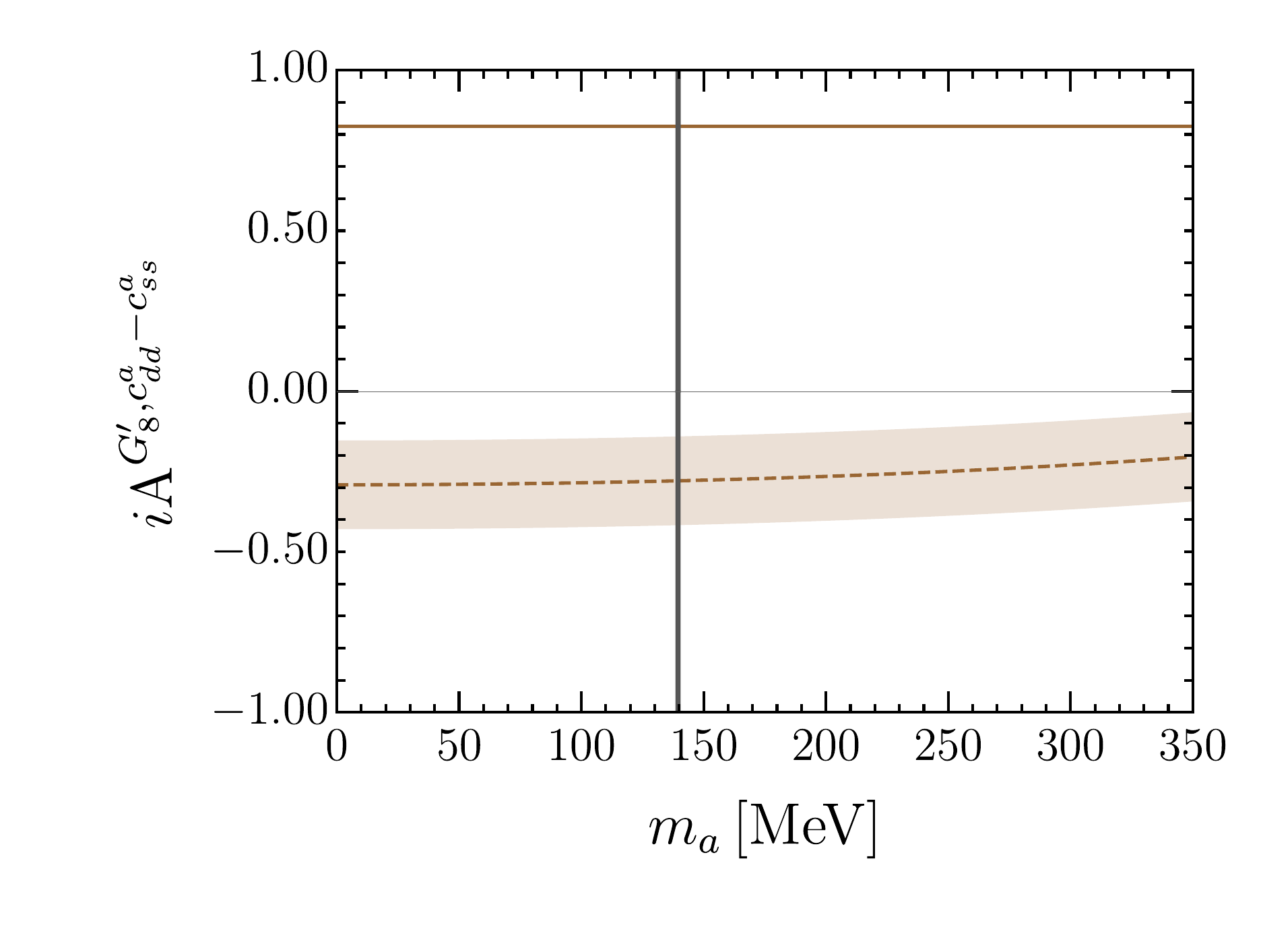}
\includegraphics[width=.485\textwidth]{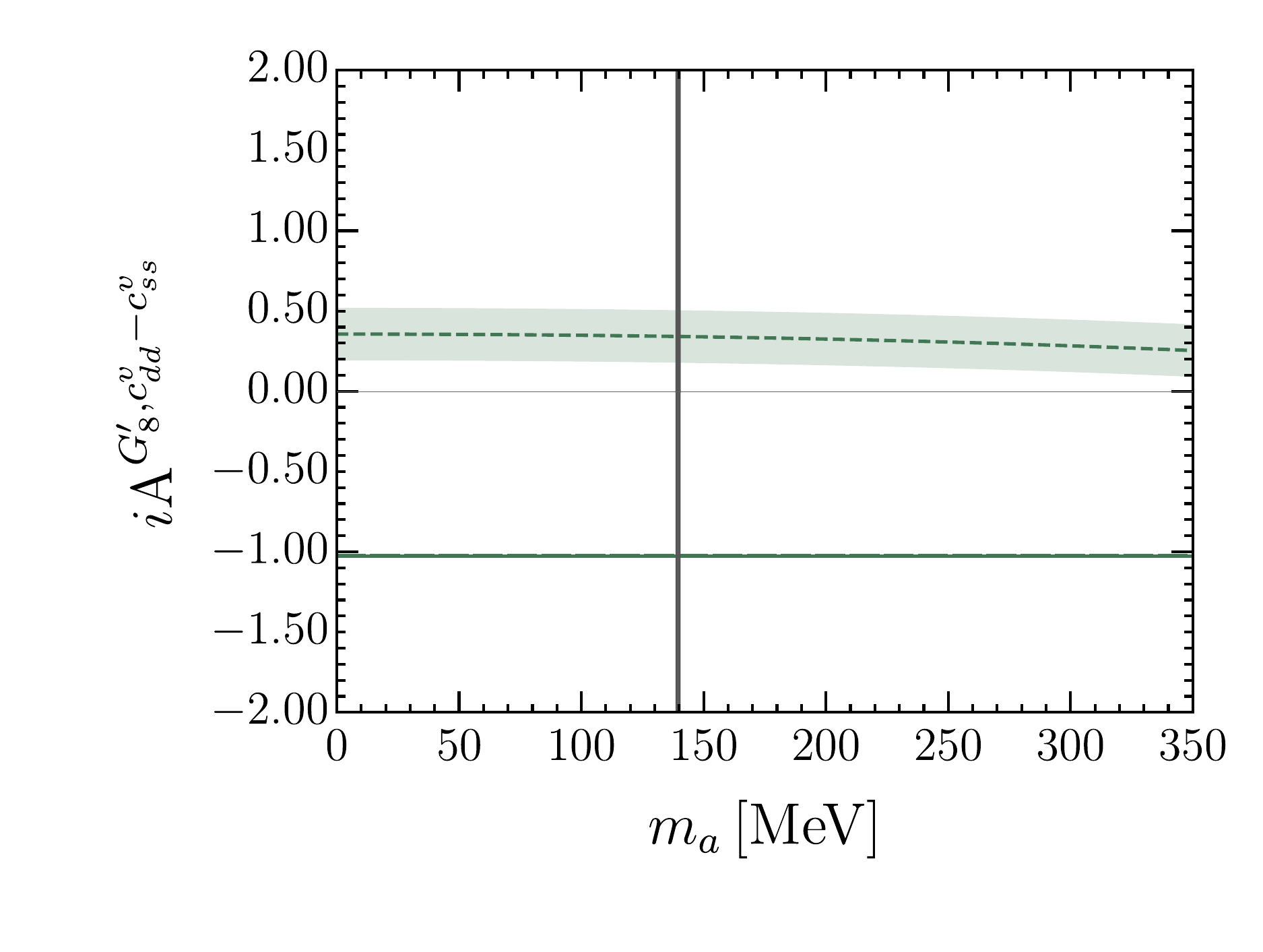}   
 \caption{LO (solid) and NLO (dashed lines) contributions proportional to $G^\theta_8$ (top) and $G_8^\prime$ (bottom) to the $K^-\to\pi^- a$ decay amplitude. The meaning of the bands and vertical line is the same as in Figure~\ref{fig:Amp_Plots}.}
 \label{fig:Amp_Plots_WMT}
\end{figure}

To illustrate the magnitude of the corrections, we plot the LO and NLO contributions to the decay amplitude for $K^-\to \pi^- a$, defined as in \eqref{eq:amplitudeLOandNLO}, as a function of the ALP mass. The contribution proportional to the flavor-violating ALP coupling is illustrated in Figure~\ref{fig:pheno_deltaFV}, while those proportional to the flavor-conserving ALP couplings are shown in Figures~\ref{fig:Amp_Plots} and \ref{fig:Amp_Plots_WMT}. The same plots apply for the CP-conjugate amplitude for $K^+\to\pi^+ a$ decay. (Recall that we are neglecting the very small CP-violating effects in the SM, see the discussion after \eqref{eq:CKMelements}.)

Keeping in mind that at LO $i A^{\rm{FV}} \equiv F_0^{K\to\pi}(m_a^2)=1$, Figure~\ref{fig:pheno_deltaFV} shows that the NLO corrections to the contribution involving flavor-violating ALP couplings are relatively small over the entire mass range, reaching a maximum value of about 8\% at the largest kinematically allowed value of $m_a$. As can be seen from the explicit expressions in Appendix \ref{app:B}, $i A^{\rm{FV}}_{\rm NLO}$ depends on a single low-energy constant,  $L_{5,r}$. Using \eqref{eq:RGevol} together with the corresponding value from Table~\ref{table:LECs} yields a fully scale independent expression for $i A^{\rm{FV}}$. The only source of uncertainty, illustrated by the light-blue band, is the error on $L_{5,r}$. As a check of the correctness of our result we have verified that, for $m_a=0$, $i A_{\rm NLO}^{\rm FV}$ coincides with the NLO correction  to the $K^-\to\pi^-$ form factor at $q^2 =0$, usually denoted by $f_2$, which amounts to $f_2\approx -0.023$ \cite{Leutwyler:1984je}. 

As to flavor-conserving ALP couplings, the plots in Figures  \ref{fig:Amp_Plots} and \ref{fig:Amp_Plots_WMT} separately show their LO (solid line) and NLO (dashed line) contributions to the amplitude, together with the uncertainty affecting the latter. Indeed, as can be seen from the explicit expressions of the various $i A^{G, \mathrm{c_{ALP}}}_{\rm NLO}$ in Appendix \ref{app:B}, the NLO contributions from flavor-conserving ALP couplings depend on the ALP mass $m_a$, some known quantities (the meson masses, $F_\pi$, and the QCD low-energy constants $L_{4,r}$, $L_{5,r}$,  $L_7$, and $L_{8,r}$), but also on 18 unknown $p^4$ low-energy constants that appear in 15 independent combinations. The fact that so many low-energy constants are currently unknown significantly limits the predictive power of our results. Our strategy will be to assume that at some suitably chosen scale $\mu=\mu_0$ the values of these yet unknown low-energy constants are small, so that they can be neglected to a reasonable approximation. On theoretical grounds, one expects the most reasonable choice for $\mu_0$ to be the scale of chiral symmetry breaking, $\mu_\chi=4\pi F_\pi\approx 1.6$\,GeV, at which the low-energy constants (the Wilson coefficients of the chiral effective theory) are free of large logarithms. Figure~\ref{fig:QCD_LECs} indeed suggests that the relevant low-energy constants of the $\mathcal{O}(p^4)$ QCD chiral Lagrangian are compatible with zero in the vicinity of $\mu_0\approx 1.4$\,GeV, in contrast to their values at the lower scale $\mu_0=m_\rho$. Assuming that a similar behavior is exhibited by the unknown low-energy constants $\hat L_{i,r}^\theta(\mu)$, $\hat N_{i,r}'(\mu)$ and $\hat N_{i,r}^{\prime\theta}(\mu)$, we set all the unknown low-energy constants to zero at $\mu_0 = 1.4\,\text{GeV}$. To estimate the uncertainty inherent in this model assumption, we vary this default scale by a factor of $\sqrt{2}$ up and down, yielding a scale variation between approximately 1 and 2\,GeV.\footnote{In chiral perturbation theory it is conventional to define the scale $\mu$ in the Gasser--Leutwyler scheme, where the subtraction parameter $\lambda$ in \eqref{eq:lambda_factor} contains a constant $(-1)$ that is absent in the $\overline{\rm MS}$ scheme. Consequently, the scale $\mu$ used in our results is related to the scale of the $\overline{\rm MS}$ scheme by $\mu_{\overline{\rm MS}}=\mu/\sqrt{e}$, so that our scale interval corresponds to $0.6\,\text{GeV}<\mu_{\overline{\rm MS}}<1.2$\,GeV.} 
The uncertainty obtained in this way is then added in quadrature to the error of the low-energy constants given in Table~\ref{table:LECs} and yields the colored bands around the dashed lines in Figures  \ref{fig:Amp_Plots} and \ref{fig:Amp_Plots_WMT}. 

All in all, the NLO contribution varies from a few percent to about 60\% of the LO for most couplings and ALP masses and only significantly exceeds the LO in the case of  $i A^{G_8, c^a_{dd}-c^a_{ss}}$, which is, however, close to zero at LO almost everywhere in the entire allowed mass-range for the ALP.

To directly compare the contributions of various couplings, we now give numerical results for the amplitudes for the case $m_a=0$. For the contribution proportional to flavor-violating ALP couplings, we obtain 
\begin{align}
	i\mathcal{A}^{\text{FV}}_{\text{LO}+\text{NLO}} = - (m_K^2 - m_\pi^2)\, \frac{ [k_d +k_D]_{12} }{2f}\, (1_{\,\rm LO} - 0.023_{\,\rm NLO}) \,,
\label{eq:FV_amplitude}
\end{align}
with negligible uncertainty. For the contributions proportional to the octet couplings $G_8$, $G_8^\theta$ and $G_8^\prime$, we have 
\begin{align}
   i\mathcal{A}^{G_8}_{\text{LO}} 
   &= \frac{G_8 F_\pi^2\,m_K^2}{2f}\,\Big[ \left(1.88-0.88\,\varepsilon^{(2)}\right) \tilde{c}_{GG} -\left(0.02-0.44\,\varepsilon^{(2)}\right) c^a_{uu}\nonumber \\
   &\hspace{2.6cm} - \left(0.48- 0.44 \,\varepsilon^{(2)} \right)(c^a_{dd}+c^a_{ss})+  0.54\,(c^v_{dd}-c^v_{ss} ) \Big] \,, \notag \\
   i\mathcal{A}^{G_8}_{\text{NLO}} 
   &= \frac{G_8 F_\pi^2\,m_K^2}{2f}\,\Big[ \left(-0.25 \pm 0.43\pm 0.61 \right) \tilde{c}_{GG} + \left(5.21\pm 1.03\pm 6.52 \right) \cdot 10^{-3}\,c^a_{uu} \nonumber \\
   &\hspace{2.6cm} + \left(0.06 \pm 0.11\pm 0.16\right)(c^a_{dd}+c^a_{ss}) - \left( 0.27 \pm 0.10  \pm 0 \right)(c^a_{dd}-c^a_{ss}) \nonumber \\[2mm]
   &\hspace{2.6cm} + \left(0.24\pm 0.23 \pm 0.18 \right)(c^v_{dd}-c^v_{ss} ) \Big] \,,  
\label{eq:G8_numerical_result_lowscale}
\end{align}
as well as
\begin{align}
   i\mathcal{A}^{G^\theta_8}_{\text{LO}+\text{NLO}} 
   = \frac{G^\theta_8\,F_\pi^2\,m_K^2}{2f}\,\big[ -1.84_{\rm LO}  + \left( 0.25 \pm 0.43 \pm 0.60 \right)_{\rm NLO} \big]\, \tilde{c}_{GG}\,,
\end{align} 
and
\begin{align}
\begin{aligned}  
   i\mathcal{A}^{G^\prime_8}_{\text{LO}} 
   &= \frac{G^\prime_8\,F_\pi^2\,m_K^2}{2f}\,\big[ \left(0.85- 0.78\,\varepsilon^{(2)} \right) (c^a_{dd}-c^a_{ss}) 
    - \left(1 + 0.92\,\varepsilon^{(2)}\right)(c^v_{dd}-c^v_{ss} )\big]\,, \\
   i\mathcal{A}^{G^\prime_8}_{\text{NLO}} 
   &= \frac{G^\prime_8 F_\pi^2\,m_K^2}{2f}\,\big[ \left( -0.29 \pm 0.14 \right) (c^a_{dd}-c^a_{ss}) 
    + \left( 0.35 \pm 0.16 \right) (c^v_{dd}-c^v_{ss} ) \big] \,. 
\end{aligned}
\end{align}
In $i\mathcal{A}^{G_8}_{\text{NLO}}$ and $i\mathcal{A}^{G_8^\theta}_{\text{NLO}}$ we distinguish explicitly between the uncertainty coming from the unknown low-energy constants, which we write first,  and the one stemming from the known QCD low-energy constants introduced in (\ref{eq:lag_p4_QCD}). In $i\mathcal{A}^{G_8^\prime}_{\text{NLO}}$, instead, the sole source of uncertainty is the known QCD low-energy constants, as this contribution is scale-independent.

Finally, the contributions proportional to $G^{1/2}_{27}$ and $G^{3/2}_{27}$ read
\begin{align}
\begin{aligned}
   i\mathcal{A}^{G^{1/2}_{27}}_{\text{LO}} 
   &=\frac{G^{1/2}_{27} F_\pi^2\,m_K^2}{2f}\,\Big[\left(3.53 -0.79\,\varepsilon^{(2)} \right)\,\tilde{c}_{GG} + \left(0.08+0.40 \, \varepsilon^{(2)}\right)\,c^a_{uu} \\
   &\hspace{3.05cm} - \left(2.7 - 0.4 \, \varepsilon^{(2)}\right) (c^a_{dd}+c^a_{ss}) + 2.31 \, (c^a_{dd}-c^a_{ss}) \\[1mm]
   &\hspace{3.05cm} + 0.54\,(c^v_{dd}-c^v_{ss} )\Big] \,, \\
   i\mathcal{A}^{G^{3/2}_{27}}_{\text{LO}} 
   &= \frac{G^{3/2}_{27} F_\pi^2\,m_K^2}{2f}\,\Big[\left(0.9 + 30.2\,\varepsilon^{(2)}\right) \tilde{c}_{GG}
   + \left(1.4-15.1 \,\varepsilon^{(2)} \right) c^a_{uu} \\
   &\hspace{3.05cm} - \left(1.4 + 15.1 \,\varepsilon^{(2)}\right) (c^a_{dd}+ c^a_{ss})- 0.46 \,  (c^a_{dd} -c^a_{ss}) \\[1mm]
   &\hspace{3.05cm} + 0.54\,(c^v_{dd}-c^v_{ss}) \Big] \,.
\label{eq:27plet_numerical_result_lowscale}\end{aligned}
\end{align}
In these expressions, all ALP couplings are evaluated at the scale $\mu_{\chi}=1.6 $  GeV in the $\overline{\rm MS}$ scheme. The full amplitude can be easily obtained by adding up the various contributions in \eqref{eq:G8_numerical_result_lowscale}--\eqref{eq:27plet_numerical_result_lowscale}.

\begin{table}[t]
\centering
\renewcommand{\arraystretch}{1.2} 
\begin{tabular}[b]{|c|c|c|}
\hline
\multirow{3}{*}{$c_i(\mu_\chi)$} & \multicolumn{2}{c|}{$\Lambda_{c_i}^{\text{eff}}$ [TeV]} \\
& $m_a=0$ & $m_a=200$\,MeV \\
\hline
\hline
$ [k_D + k_d]_{12}$ & $2.9 \cdot 10^8$  & $6.5 \cdot 10^8$  \\
\hline
$\tilde{c}_{GG} \vphantom{1}^{(\ast)}$ & $43$ & $84$\\
\hline
$c^a_{uu}$&  $1.5$ &  $4.3 $ \\
\hline
$c^a_{dd}+c^a_{ss}$&   $15$ & $19$ \\
\hline
$c^a_{dd}-c^a_{ss} \vphantom{1}^{(\ast\ast)}$ & $8$ &  $8$\\
\hline
$c^v_{dd}-c^v_{ss} \vphantom{1}^{(\ast\ast)}$ & $23$ &  $47$ \\
\hline
\end{tabular}
\caption{$90\%$ CL lower bounds on the effective scales $\Lambda_{c_i}^{\text{eff}}\equiv f/|c_i|$ of the ALP couplings $c_i$ at the scale $\mu_\chi$ for two different ALP masses, derived using the $90$\% CL upper bound on the $K^+\to\pi^+ X$ branching ratio provided by NA62 \cite{NA62:2021zjw}, as explained in the main text. These bounds are independent of the specific UV structure of the ALP couplings.  The bounds for couplings denoted by $^{(\ast)}$ depend on the value of $g_8^\theta$, while those for couplings denoted with $^{(\ast \ast)}$ depend on $g_8^\prime$. Here we set $g_8^\theta=g_{8}^{\prime}=0$.} 
\label{table:bounds_low_energy}
\end{table}  

The measurement posing the strongest constraints on the magnitude of the effective couplings $|c_i|/f$  appearing in the $K^\pm\to\pi^\pm a$ decay amplitudes -- or, equivalently, on the effective scale $\Lambda_{c_i}^{\text{eff}} \equiv f/|c_i|$ -- is the NA62 upper bound on the branching ratio for the decay $K^+\to\pi^+ X$, where $X$ is a scalar or pseudoscalar particle decaying to invisibles and/or outside the detector. For $m_X\in [0,110]$\,MeV, NA62 obtains upper limits in the range $\mathcal{B}(K^+\to \pi^+  X) < \text{(3\,--\,6)}\times 10^{-11}$ ($90\%$ CL). For $m_{X} \in [160,260]$ MeV the bound tightens to $\mathcal{B}(K^+\to \pi^+  X) < 1 \times 10^{-11}$ ($90\%$ CL) \cite{NA62:2021zjw}. We can express these limits as upper bounds on the absolute value of the amplitude $|\mathcal{A}(K^+\to\pi^+ X)|$ and use them to derive lower bounds on the effective scales $\Lambda_{c_i}^{\text{eff}}$ by switching on one ALP coupling at a time. In Table~\ref{table:bounds_low_energy} we report these bounds for two representative cases, $m_a=0$ and $m_a=200$\,MeV, obtained by using the value $g_8=3.61\pm 0.28$ from \eqref{eq:inputs_for_gi} and setting $g_8^\theta=g_{8}^{\prime}=0$. Due to the dependence of the amplitude on the unknown couplings $g_8^\theta$ and $g_{8}^{\prime}$, we need to make assumptions about their values to derive specific bounds. In particular, the value of $g_8^\theta$ impacts the bound for $\tilde{c}_{GG}$, and the value of $g_{8}^{\prime}$ affects the bounds for $({c}^a_{dd}-{c}^a_{ss})$ and $({c}^v_{dd}-{c}^v_{ss})$. Concretely, the bound on $\tilde{c}_{GG}$ is weakened (strengthened) for positive (negative) values of $g_8^\theta$. For instance, for a massless ALP we find $\Lambda_{\tilde{c}_{GG}}^{\rm eff} \gtrsim 10$ TeV for $g_8^\theta=1$ and $\Lambda_{\tilde{c}_{GG}}^{\rm eff} \gtrsim 63$ TeV for $g_8^\theta=-1$.  This underlines the importance of determining these couplings from first principles to establish more reliable bounds. 

It is important to emphasize that the bounds obtained from $K^\pm\to\pi^\pm a$ decays on all ALP couplings entering these processes represent the strongest particle-physics constraints on such couplings in the entire kinematically allowed mass range, i.e.\ for $m_a\lesssim 350$\,MeV. The constraint is particularly strong for the flavor-violating ALP coupling to strange and down quarks, for which our analysis of $K^\pm\to\pi^\pm a$ decays excludes effective scales as high as $10^{8}$\,TeV. These bounds are interesting because, in the absence of a flavor symmetry, they would imply bounds on the ALP decay constant of order $f \gtrsim 10^{11}\, \mathrm{GeV}$, which are in tension with  recent (model-dependent) cosmological constraints \cite{Gorghetto:2020qws}.  Perhaps more surprisingly, $K^\pm \to \pi^\pm a$ decays currently also pose the strongest particle-physics constraints, of $\mathcal{O}(10\, \mathrm{TeV})$, also for the flavor-conserving ALP couplings to gluons and light quarks, which enter the process through diagrams involving the SM weak interactions \cite{Bauer:2021wjo}.

\subsection{Decay amplitudes in terms of UV couplings}

To apply our results to specific scenarios of new physics, it is convenient to express the $K^-\to \pi^-\,a $ amplitude in terms of ALP couplings defined at the new-physics scale $\Lambda=4\pi f\gg\mu_w\gg\mu_\chi$, where $\mu_{w}$ is  the scale of electroweak symmetry breaking. To this end, we recall that above $\mu_{w}$ the most general Lagrangian for an ALP coupling to the Standard Model at dimension-5 order is given by \cite{Georgi:1986df}
\begin{align}
	\mathcal{L}^{D=5}_{\text{SM+ALP}} &=\,c_{GG} \,\frac{\alpha_s}{4\pi} \frac{a}{f}\,G^a_{\mu\nu}\,\tilde{G}^{\mu\nu, a} +  c_{WW} \,\frac{\alpha_2}{4\pi} \frac{a}{f}\,W^I_{\mu\nu}\,\tilde{W}^{\mu\nu, I} + c_{BB} \,\frac{\alpha_1}{4\pi} \frac{a}{f}\,B_{\mu\nu}\,\tilde{B}^{\mu\nu} \notag\\
	&+\frac{\partial^\mu a}{f}\,\sum_F \bar{\psi}_F\, \boldsymbol{c}_F\, \gamma_\mu \psi_F + c_{\phi} \frac{\partial^\mu a}{f} ( \phi^\dagger i \overleftrightarrow{D}_{\mu} \phi ) \,.
\label{eq:UV_Lagrangian}
\end{align}
In this expression $W_{\mu\nu}$ and $B_{\mu\nu}$ are the field strength tensors of $SU(2)_L$ and $U(1)_Y$, with coupling parameters $\alpha_{1,2}=g_{1,2}^2/(4\pi)$, and $\phi$ is the Higgs doublet. The index $F$ runs over all chiral fermions of the Standard Model, $F \in \{Q,\,L,\,u,\,d,\,e\}$, and the quantities $\boldsymbol{c}_F$ are hermitian $3 \times 3$ matrices in flavor space. As noted already in \cite{Georgi:1986df}, not all these parameters are physical. Indeed, one can exploit the $U(1)$ symmetries of the SM (hypercharge, individual lepton numbers, and baryon number) to remove five of them. 

The renormalization-group (RG) evolution equations for the couplings in \eqref{eq:UV_Lagrangian} have been computed in \cite{Bauer:2020jbp,Chala:2020wvs}. Couplings to gauge bosons are found to be scale independent (at least up to two-loop order), with no matching corrections arising at the electroweak scale. The couplings to fermions, in contrast, are scale dependent. Notably, the RG evolution generates contributions to flavor off-diagonal couplings through loop diagrams involving the flavor-changing interactions of the $W$ bosons. As such couplings are strongly constrained by experiments (see e.g.~Table~1 in \cite{Bauer:2021mvw}), a motivated choice is to consider an ALP that is flavor universal -- hence, in particular, also flavor diagonal -- at the scale $\Lambda$:
\begin{equation}
    \boldsymbol{c}_F   (\Lambda) = c_F  (\Lambda) \,\mathbbm{1} \,; \qquad F \in \{u,d,Q,e,L\} \,. 
\end{equation}
This assumption leaves in principle nine independent ALP couplings, namely the bosonic couplings $c_{GG}$, $c_{WW}$, $c_{BB}$ and $c_{\phi}(\Lambda)$, and the fermionic couplings $c_u(\Lambda)$, $c_d(\Lambda)$, $c_Q(\Lambda)$, $c_e(\Lambda)$ and $c_L(\Lambda)$. However, thanks to the symmetries mentioned above, only six of them are physical. They can be chosen as \cite{Bauer:2020jbp, Bauer:2021mvw} 
\begin{align}
\begin{aligned}
   \tilde{c}_{GG}(\Lambda) &= {c}_{GG} + \frac32 c_u(\Lambda) + \frac32 c_d(\Lambda) - 3 c_Q(\Lambda) \,, \\     
   \tilde{c}_{WW}(\Lambda) &= {c}_{WW} - \frac92 c_Q(\Lambda) - \frac32 c_L (\Lambda) \,, \\
   \tilde{c}_{BB}(\Lambda) &={c}_{BB} + 4 c_u(\Lambda) + c_d(\Lambda) - \frac12 c_Q(\Lambda) + 3 c_e(\Lambda) - \frac32 c_L(\Lambda) \,, \\[1mm]
   \tilde{c}_u(\Lambda) &= c_u(\Lambda) - c_Q(\Lambda) - c_{\phi}(\Lambda) \,, \\[2mm]
   \tilde{c}_d(\Lambda) &= c_d(\Lambda) - c_Q(\Lambda)+  c_{\phi}(\Lambda) \,, \\[2mm]
   \tilde{c}_e(\Lambda) &= c_e(\Lambda) - c_L (\Lambda)  +  c_{\phi}(\Lambda) \,. 
\end{aligned} 
\end{align}
Below, we will express our results in terms of these physical couplings.

As mentioned above, RG effects generate flavor-violating couplings to left-handed fermions at scales below $\Lambda$. For example,
setting $f=1\,\text{TeV}$ in the RG evolution (with $\Lambda=4\pi f$) one finds \cite{Bauer:2021mvw} 
\begin{align}
\begin{aligned}
   [k_D(m_t)]^{\text{univ}}_{ij} 
   &\simeq \,10^{-5}\,V^*_{ti} V_{tj}\,\big[ -6.1 \,\tilde{c}_{GG}(\Lambda) - 2.8 \tilde{c}\,_{WW}(\Lambda) \\
   &\hspace{2.75 cm} - 0.02 \,\tilde{c}_{BB}(\Lambda) + 1.9\cdot 10^3\,\tilde{c}_u(\Lambda) \big]\,, \\[2mm]
   [k_d(m_t)]^{\text{univ}}_{ij} &= 0 \,,
\label{eq:FV_running_induced}
\end{aligned}
\end{align}
where the relevant CKM suppression factor for the $s\to d$ transition in $K^-\to\pi^- a$ is given by $V^*_{td} V_{ts}\approx -(3.0 +  1.3 \, i)\times 10^{-4}$. Neglecting the Yukawa couplings of the light quarks, the flavor off-diagonal couplings do not RG-evolve further below $m_t$, so that $[k_D(\mu_\chi)]_{ij}=[k_D(m_t)]_{ij}$. For the axial ALP couplings to quarks at $\mu_{\chi}$, we have \cite{Bauer:2021mvw}   
\begin{align}
   [c^a_{uu}(\mu_{\chi})]^\text{univ}
   \simeq & 0.90\,\tilde{c}_u(\Lambda) + 0.008\,\tilde{c}_d(\Lambda) - 0.042\,\tilde{c}_{GG}(\Lambda)\\
   & -10^{-4}\,\big[ 2.1\,\tilde{c}_{WW} (\Lambda) + 0.34\,\tilde{c}_{BB}(\Lambda)  \big] \,,\notag \\ 
   [c^a_{dd,ss}(\mu_{\chi})]^\text{univ}
   \simeq & 0.13\,\tilde{c}_u(\Lambda) + 1.00\,\tilde{c}_d(\Lambda) 
   - 0.042\,\tilde{c}_{GG}(\Lambda) \label{eq:cqqmuchi}\\
   &  - 10^{-4}\,\big[  2.3\,\tilde{c}_{WW} (\Lambda)  + 0.10\,\tilde{c}_{BB} (\Lambda) \big]	\,.	 \notag
\end{align} 
Finally, for the vector couplings, we have \cite{Bauer:2021mvw} 
\begin{align}
   [c^v_{dd}(\mu_{\chi})]^\text{univ} =[c^v_{ss}(\mu_{\chi})]^\text{univ} \,.
\end{align} 
In this scenario, the $K^-\to\pi^- a$ decay amplitude is only sensitive to the axial couplings of the ALP to quarks, since the vector couplings only enter in the combination $(c^v_{dd}-c^v_{ss})$, which vanishes in the flavor-universal ALP model. As a consequence, the weak mass term does not contribute. 

\begin{table}[t]
\renewcommand{\arraystretch}{1.2} 
\centering
\begin{tabular}[b]{|c|c|c|}
\hline
\multirow{3}{*}{$c_i(\Lambda)$} & \multicolumn{2}{c|}{$\Lambda_{c_i}^{\text{eff}}$ [TeV]} \\
& $m_a=0$ & $m_a=200$\,MeV \\
\hline
\hline
$\tilde{c}_{GG}(\Lambda)$ & 49 & 97 \\
\hline
$\tilde{c}_{WW}(\Lambda)$ & 2.6 & 6 \\
\hline
$\tilde{c}_{BB}(\Lambda)$ & 0.02 & 0.04  \\
\hline
$\tilde{c}_u(\Lambda)$ & $1.9\cdot 10^3$ &   $4.2\cdot 10^3$  \\
\hline
$\tilde{c}_d(\Lambda)$ & 51 &  78 \\
\hline
\end{tabular}
\caption{$90\%$ CL lower bounds on the effective scales $\Lambda_{c_i}^{\text{eff}}\equiv f/|c_i|$ of the ALP couplings $c_i$ at the high scale $\Lambda$ in the flavor-universal ALP model for the cases $m_a=0$ and $m_a=200$\,MeV. The bounds are derived by setting $\Lambda=4\pi f$ with $f=1$\,TeV, and using the 90\% CL upper bound on the $K^+\to\pi^+ X$ branching ratio  provided by NA62 \cite{NA62:2021zjw}. The bounds shown here are obtained by setting $g_8^\theta=0$.}
\label{table:bounds}
\end{table}  

With these ingredients at hand, and setting $g_8^\theta=0$,  the full $K^-\to\pi^- a$ decay amplitude for a flavor-universal massless ALP can be written as 
\begin{align}
\begin{aligned}
   i\mathcal{A}^{\rm univ} 
   &= - 10^{-11}\,\mathrm{GeV} \left[ \frac{1\,\mathrm{TeV}}{f} \right] \Big[ (2.4  \pm 1.0 +0.1 \,i )\,\tilde{c}_{GG}(\Lambda) \\
   &\hspace{4.55cm} + (9.37 \pm 0.02  +3.97\,i)\cdot 10^{-2}\,\tilde{c}_{WW}(\Lambda) \\[2mm]
   &\hspace{4.55cm} + (0.57 \pm 0.02  +0.26\,i)\cdot 10^{-3}\,\tilde{c}_{BB}(\Lambda) \\[2mm]
   &\hspace{4.55cm} - (68\pm 1 +28\,i)\,\tilde{c}_u(\Lambda) - (2.5\pm 1.0)\,\tilde{c}_d(\Lambda) \big] \,,
\end{aligned}
\end{align}
where we have combined in quadrature the uncertainties stemming from the known QCD low-energy constants with those coming from the scale variation. The numerical coefficients in this result are obtained for $f=1$\,TeV. The amplitude scales approximately like $1/f$ since it is linear in the ALP couplings, but in addition the coefficients carry a weak logarithmic dependence on $f$ (see Figure~\ref{fig:Lambda_eff} below). Interestingly, the amplitude is sensitive to five out of the six physical parameters characterizing the flavor-universal ALP scenario.  Although this expression is valid for $m_a = 0$, the coefficients vary by less than $10\%$ across the entire allowed mass range. The values for $g_8$, $g_{27}^{1/2}$, and $g_{27}^{3/2}$ used in this equation are those provided in \eqref{eq:inputs_for_gi}. We have omitted the uncertainties associated with these values, as they are consistently subdominant compared to those arising from the low-energy constants.

\begin{figure}[t]
\centering
\includegraphics[width=.6\textwidth]{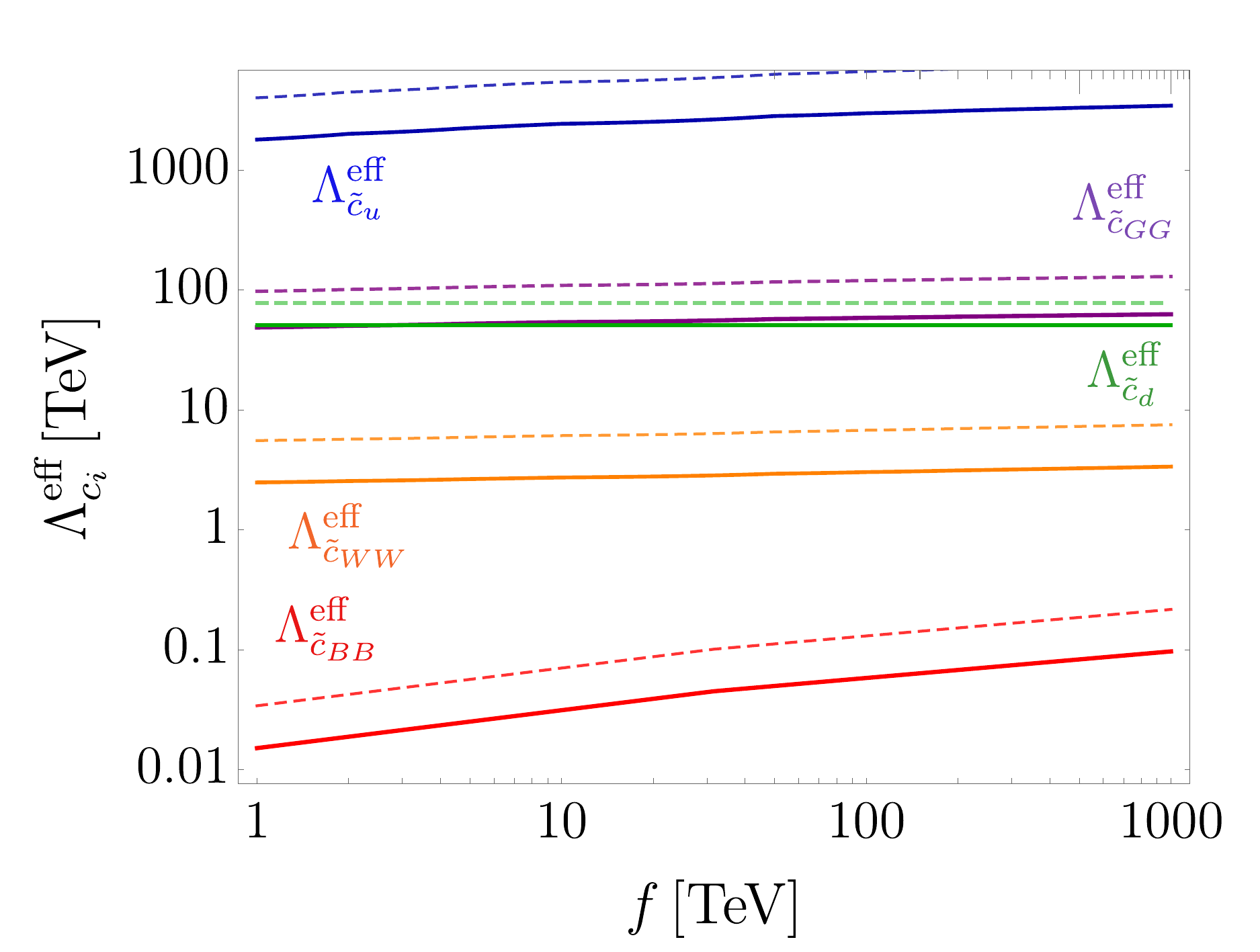}  
\caption{$90\%$ CL lower bounds on the effective scales $\Lambda_{c_i}^{\text{eff}}\equiv f/|c_i|$ of the ALP couplings $c_i(\Lambda)$ for a flavor-universal ALP as a function of the scale $f$, for $g_8^{\theta} = 0$. The solid (dashed) lines correspond to $m_a=0$ ($m_a=200$\,MeV).}
\label{fig:Lambda_eff}
\end{figure}

\begin{figure}[t]
\centering
\includegraphics[width=.6\textwidth]{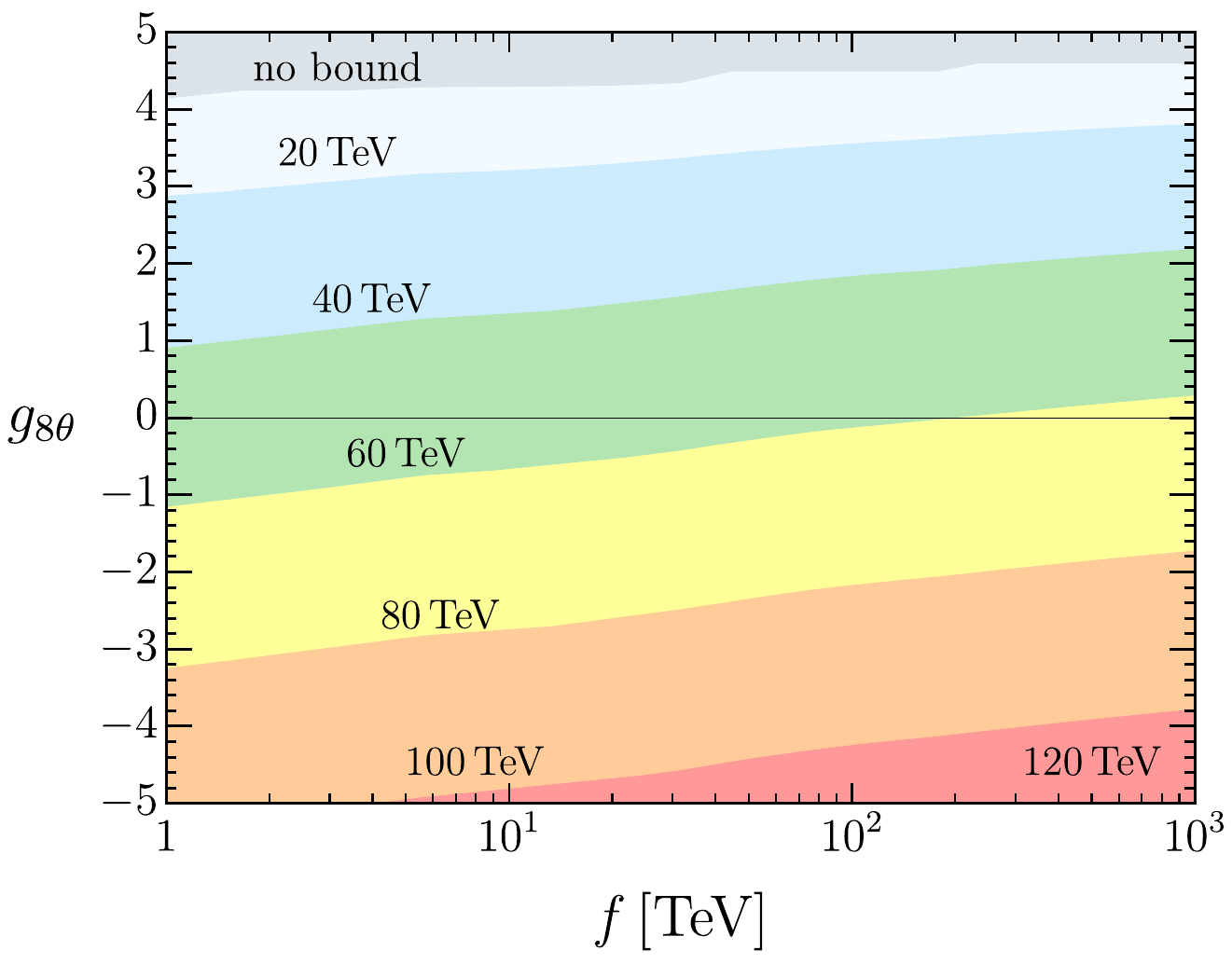} 
\caption{$90\%$ CL lower bounds on the effective scales $\Lambda_{\tilde{c}_{GG}}^{\text{eff}}$ associated to the ALP coupling $\tilde{c}_{GG}(\Lambda)$ for a flavor-universal ALP as a function of the scale $f$ and the low-energy constant $g_8^{\theta}$ for $m_a = 0$}.
 \label{fig:Lambda_eff_cGG}
\end{figure}

As in the previous section, we can use the NA62 constraint on the branching ratio for the decay $K^+\to\pi^+ X$ to derive lower bounds on the effective scales $\Lambda_{c_i}^{\text{eff}}$ associated with the five physical couplings appearing in the amplitude. In Table~\ref{table:bounds} we report these bounds for two representative cases, $m_a=0$ and $m_a=200$\,MeV, setting $g_8^\theta=0$ for simplicity. A graphical representation of these bounds is provided in Figure~\ref{fig:Lambda_eff}, where we also show how they change when choosing different values of $f$ in the RG evolution from $\Lambda=4\pi f$ to the low scale $\mu_\chi$. Overall, all bounds get stronger for larger values of $f$ due to large logarithms in the evolution equations.  Finally, we note that the choice $g_8^\theta=0$ has an appreciable impact only on the size of the effective scales associated to $\tilde{c}_{GG}(\Lambda)$ and $\tilde{c}_d(\Lambda)$. In Figure~\ref{fig:Lambda_eff_cGG} we show how $\Lambda_{\tilde{c}_{GG}}^{\text{eff}}$ changes as a function of $f$ and $g_8^\theta \in [-5,5]$. For $\tilde{c}_d(\Lambda)$ we find that  $\Lambda_{\tilde{c}_{d}}^{\text{eff}} $ varies between 25 and 75\,TeV in the same $g_8^\theta$ range, with no dependence on $f$. 

For $\tilde{c}_u(\Lambda)$, $\tilde{c}_{WW}(\Lambda)$ and $\tilde{c}_{BB}(\Lambda)$, the dominant contribution to the $K^-\to \pi^- a$ decay amplitude originates from the RG-induced flavor-changing ALP coupling to strange and down quarks, see \eqref{eq:FV_running_induced}. As demonstrated in the previous section, this contribution can be accurately determined at NLO in the chiral expansion. Consequently, the constraints on these couplings are directly dictated by the size of their contribution to the amplitude. The strongest bound is found for the combination $\tilde{c}_u(\Lambda)$. The bound on $\tilde{c}_{BB}(\Lambda)$ instead is very loose, since it stems from a two-loop effect. The bounds on these three couplings are essentially the same one would obtain by employing the LO result, since the NLO correction to $i\mathcal{A}_{\rm FV}^{\rm univ}$ is small, see \eqref{eq:FV_amplitude}. For $\tilde{c}_{GG}(\Lambda)$ and $\tilde{c}_d(\Lambda)$, the primary contribution to the $K^-\to \pi^- a$ decay amplitude stems from $i\mathcal{A}_{\rm FC}^{\rm univ}$, which, as we have shown, is subject to significant theoretical uncertainties. The resulting constraints are therefore 30\,--\,40\% weaker than those found using the LO result. In particular, by employing the LO result one would deduce $\Lambda_{\tilde{c}_{GG}}^{\rm eff} \gtrsim 73$\,TeV and $\Lambda_{\tilde{c}_{d}}^{\rm eff}\gtrsim 76$\,TeV for a massless ALP, and $\Lambda_{\tilde{c}_{GG}}^{\rm eff} \gtrsim 165$\,TeV and $\Lambda_{\tilde{c}_{d}}^{\rm eff}\gtrsim 150$\,TeV for $m_a=200$\,MeV. 

Once again, these bounds showcase the enormous power of the decays $K^\pm\to\pi^\pm a$ in constraining the couplings for ALP masses below $\sim 350$ MeV. While our NLO analysis of this process slightly weakens the bounds with respect to the LO estimates of \cite{Bauer:2021mvw} for some of the couplings, the resulting constraints are still the most stringent ones in the ALP mass range under consideration as far as particle-physics probes are concerned. And importantly, the NLO calculation allows us to assign theoretical uncertainties to bounds previously derived.

\subsection{ALP couplings to nucleons}

At low energies, the effective couplings of axions and ALPs to protons and neutrons can be described by the Lagrangian \cite{AxionLimits} 
\begin{align}
   \mathcal{L}_{aN} 
   = \frac{g_{an}}{2 m_n}\,(\partial_\mu a)\,
    \bar{\psi}_n\gamma^\mu\gamma_5\psi_n
    + \frac{g_{ap}}{2 m_p}\,(\partial_\mu a)\,
    \bar{\psi}_p\gamma^\mu\gamma_5\psi_p \,,
\end{align}
where $\psi_n$ and $\psi_p$ describe the neutron and the proton, respectively, and $m_n$ and $m_p$ denote the corresponding masses. These ALP--nucleon couplings are related to the ALP couplings to light quarks and gluons via \cite{Bauer:2021mvw}\footnote{The effective couplings $g_{Na}$ defined in \cite{Bauer:2021mvw} are related to the ones used in this paper by $g_{aN}=\frac{g_{Na}^{\rm [27]}}{2f}\,m_N$, for $N=n,p$.} 
\begin{align}
\label{eq:AL_neutron}
    g_{an} &= \frac{m_n}{2}\,\left( g_0\bigg[ \frac{c^a_{uu}}{f} + \frac{c^a_{dd}}{f}  + 2\,\frac{c_{GG}}{f} \bigg] - g_A \,\frac{m_\pi^2}{m_\pi^2-m_a^2} \bigg[\frac{c^a_{uu}}{f} - \frac{c^a_{dd}}{f} + 2 \frac{c_{GG}}{f} \,\frac{m_u-m_d}{m_u+m_d}\bigg]\right) ,\notag\\
     g_{ap} &= \frac{m_p}{2}\,\left( g_0\bigg[ \frac{c^a_{uu}}{f} + \frac{c^a_{dd}}{f}  + 2\,\frac{c_{GG}}{f} \bigg] + g_A \,\frac{m_\pi^2}{m_\pi^2-m_a^2} \bigg[\frac{c^a_{uu}}{f} - \frac{c^a_{dd}}{f} + 2 \frac{c_{GG}}{f} \,\frac{m_u-m_d}{m_u+m_d}\bigg]\right) ,
\end{align}
with dimensionless constants $g_0=0.440(44)$ and $g_A=1.2754(13)$ \cite{10.1093/ptep/ptaa104}. We can translate the bounds obtained from the process $K^+\to \pi^+ a$ into bounds on the couplings of the ALP to neutrons and protons. Importantly, these bounds hold not only for an ALP but also for the standard QCD axion, whose mass is determined by the relation \cite{GrillidiCortona:2015jxo} 
\begin{align}
    m_a = 5.7\,\mu\text{eV} \left| 2 c_{GG}\,\frac{10^{12}\,\text{GeV}}{f} \right| 
    = 11.4\,\mu\text{eV} \left| \frac{10^9\,\text{TeV}}{\Lambda_{c_{GG}}^{\rm eff}} \right| .
\end{align}
For the mass range $m_a<10^{-2}$\,eV considered below, the effective scale $\Lambda_{c_{GG}}^{\rm eff}>1.1\cdot 10^6$\,TeV is so large that it exceeds by far the bound on $\Lambda_{\tilde c_{GG}}^{\rm eff}$ shown in Table~\ref{table:bounds_low_energy}. It is then no longer justified to treat $\tilde c_{GG}$ as being independent of the quark couplings $c_{qq}^a$. As a result, the bounds on the ALP couplings to down-type quarks are significantly relaxed and, without any assumption on the flavor structure of the various couplings, the uncertainty due to the NLO contributions does not allow us to put a competitive bound on $|g_{an}|$ and $|g_{ap}|$.

\begin{figure}[t]
\centering
\includegraphics[width=0.47\textwidth]{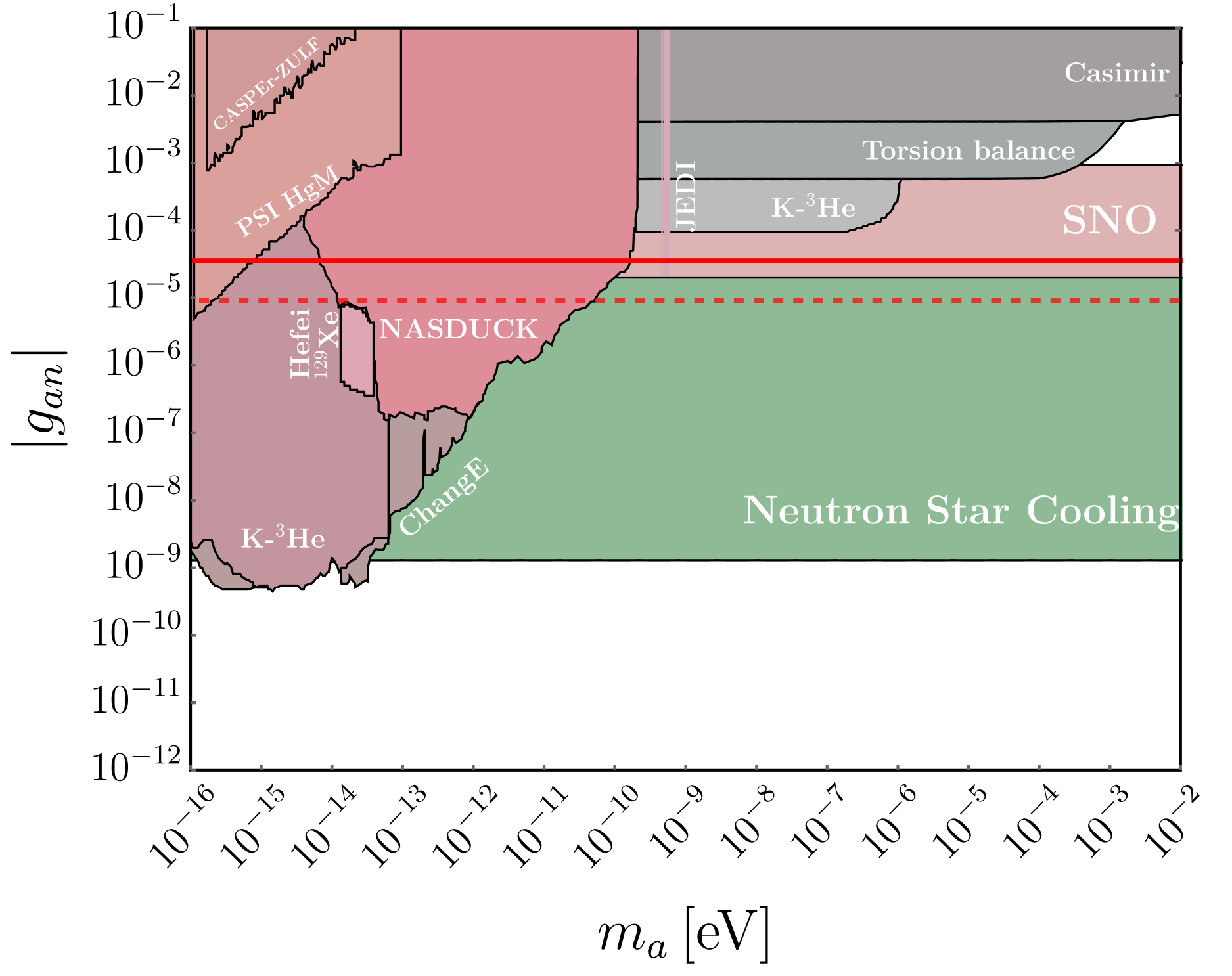} 
\quad \includegraphics[width=.47\textwidth]{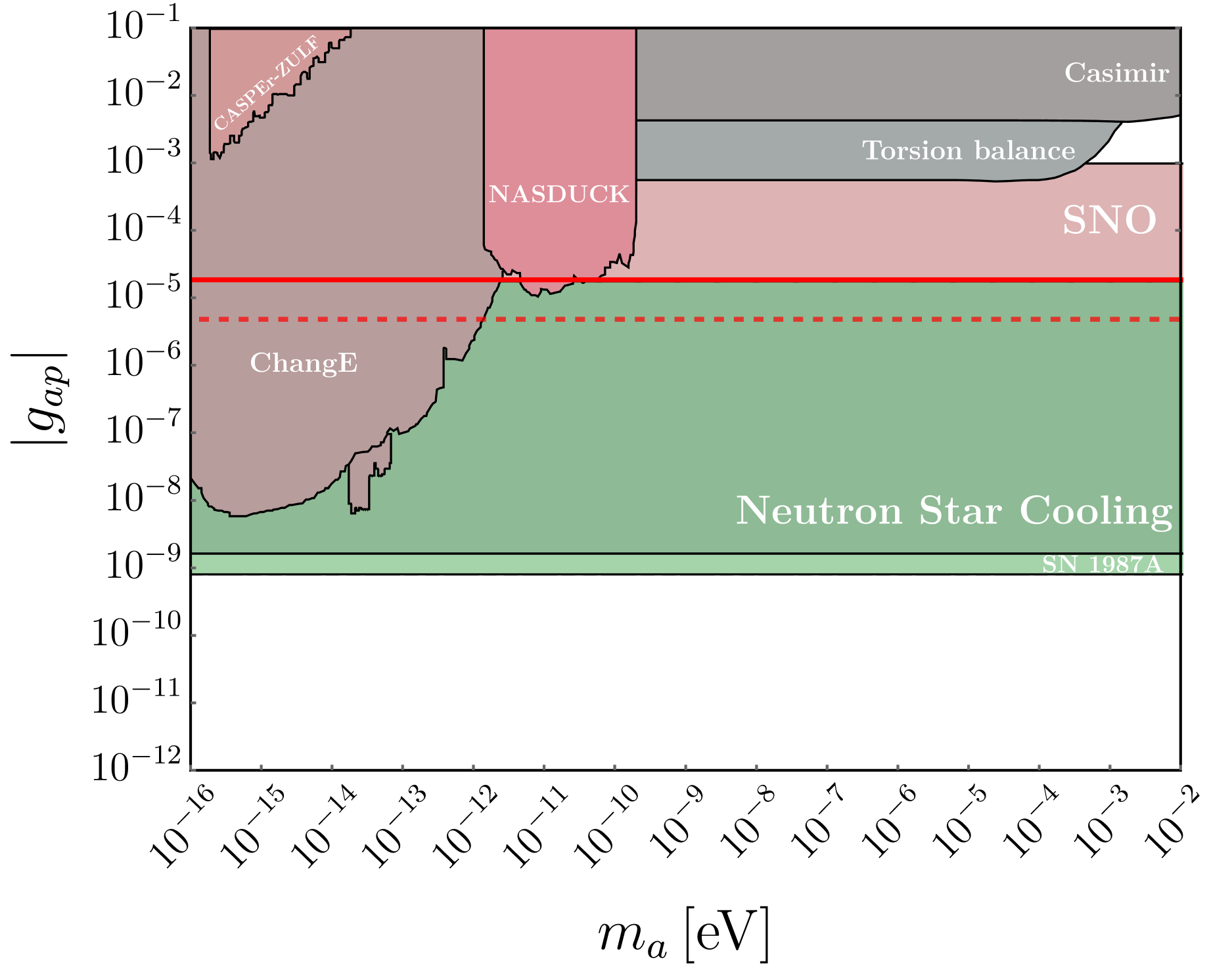} 
\caption{Exclusion limits for the ALP--neutron (left) and ALP--proton (right) couplings derived from various experiments (adapted from \cite{AxionLimits}). The red solid lines correspond to the bounds implied by our analysis of $K^\pm\to\pi^\pm a$ decays. The red dashed lines show a possible exclusion limit assuming an improvement of the theoretical uncertainty and the experimental bound by a factor of~3 each.}
\label{fig:Neutron_Bounds}
\end{figure}

The situation changes if we impose the assumption of flavor-universality at the high scale $\Lambda=4\pi f$. Compared with the numbers in Table~\ref{table:bounds}, we then find that the lower bound on $\Lambda_{\tilde c_d}^{\rm eff}$ is relaxed from 50\,TeV to 22\,TeV, whereas the bound on  $\Lambda_{\tilde c_u}^{\rm eff}$ remains almost unchanged. Using the second relation in \eqref{eq:cqqmuchi}, we then obtain to good approximation (with $N=n,p$)
\begin{align}
    g_{aN} < \frac{m_N}{2\Lambda_{\tilde c_d}^{\rm eff}} \left| g_0\pm g_A \right| ,
\end{align}
where the upper (lower) sign refers to the neutron (proton). Numerically, we find\footnote{To derive these bounds we have used RG evolution from $\Lambda=4\pi f$ (with $f=1$\,TeV) down to $\mu_\chi=1.6$\,GeV. Given the flatness of the line corresponding to $\Lambda_{\tilde c_d}^{\rm eff}$ in Figure~\ref{fig:Lambda_eff}, the choice of $\Lambda$ has only a very small impact on the results.} 
$|g_{an}|\lesssim 3.7\times 10^{-5}$ and $|g_{ap}|\lesssim 1.8\times 10^{-5}$.  We represent these bounds by solid red lines in Figure~\ref{fig:Neutron_Bounds} and compare them with other bounds from astrophysical measurements and non-accelerator experiments. For comparison, we show as a dashed lines the bounds one would obtain assuming that the theoretical uncertainties can be reduced by a factor of~3, and the experimental bound on the $K^\pm\to\pi^\pm a$ branching ratio can also be improved by a factor~3. Notably, the bounds derived from $K^\pm\to\pi^\pm a$ decays, first derived at LO in \cite{Bauer:2021mvw}, currently represent the only  particle-physics probes of the ALP--nucleons couplings in the mass range shown in Figure~\ref{fig:Neutron_Bounds}. In particular, one can see that these bounds can compete with several current bounds from non-accelerator and astrophysical probes. They are weaker than the bounds derived from neutron-star cooling and supernova observations (SN~1987A), which however rely on some not well tested assumptions about neutron stars, such as their core composition.

\section{Conclusions}
\label{sec:conclusions}

The rare decay processes $K^\pm\to\pi^\pm a$ offer an important probe of axions and axion-like particles (ALPs). This work provides a comprehensive analysis of these decays at next-to-leading order (NLO) in chiral perturbation theory. We have improved upon existing calculations \cite{Bauer:2021wjo,Bauer:2021mvw} by adding the complete set of one-loop diagrams as well as the contributions from low-energy constants appearing in the QCD and weak chiral Lagrangians at $\mathcal{O}(p^4)$, and by including isospin-breaking corrections to the leading-order amplitudes at first order in the quark-mass difference $(m_d-m_u)$. 

We have shown that including the ALP in the $\mathcal{O}(p^2)$ and $\mathcal{O}(p^4)$ chiral Lagrangians requires a non-trivial extension of the known operator bases. Interestingly, some new features emerge already at order $\mathcal{O}(p^2)$. First, in contrast to the SM case, the so-called ``weak mass term'' with coupling $g_8^\prime$ is non-redundant in the presence of an ALP. Second, a genuinely new octet operator with coupling $g_8^\theta$ must be included. At NLO, the situation becomes still more involved. In particular, in the presence of an ALP, three/nine new operators need to be added to the $\mathcal{O}(p^4)$ QCD/weak chiral Lagrangians. 

The $K^\pm\to\pi^\pm a$ decay amplitudes depend on both flavor-violating and flavor-conserving ALP couplings, defined at the low scale $\mu_\chi\approx 4\pi F_\pi$. NLO corrections proportional to the low-scale flavor-violating ALP couplings are governed by the $K^-\to\pi^-$ transition form factor $F_0^{K\to\pi}(q^2)$  evaluated at $q^2=m_a^2$. They depend on a single, reasonably well known QCD low-energy constant and range from $-3\%$ to $+8\%$ based on the value of $m_a$, with a small theoretical uncertainty. These corrections are therefore well under control. Conversely, for the flavor-conserving ALP couplings, NLO corrections are not only subject to the uncertainties in the (known) QCD low-energy constants, but also contain 17 independent new, unknown parameters: the two leading-order couplings $g_8^\prime$ and $g_8^\theta$, as well as 15 combinations of $\mathcal{O}(p^4)$ QCD and weak low-energy constants, which we set to zero at a scale $\mu_0\sim\mu_\chi$. The presence of these unknown low-energy constants implies a significant uncertainty, which we estimate by assessing the scale dependence of the amplitude under variations of $\mu_0$. With this in mind, we find that in most cases the corrections to the leading-order contributions for the couplings $g_8$, $g_8^\prime$, and $g_8^\theta$ range between a few percent up to $\pm 60\%$ and show only a weak dependence on the ALP mass. In the case of $i A^{G_8,\,(c^a_{dd}-c^a_{ss})}$, where the leading-order amplitude is strongly suppressed, the NLO contribution provides the dominant effect.

Using the NA62 upper bound on the branching ratio $\mathcal{B}(K^+\to \pi^+ X)$ and setting the unknown couplings $g_8^\prime$ and $g_8^\theta$ to zero, we have deduced upper bounds on the ALP couplings (or lower bounds on the effective scales $\Lambda_{c_i}^{\rm eff}=f/|c_i|$) for $m_a=0$ and 200\,MeV, both at the low scale $\mu_\chi$ and at the high scale $\Lambda=4\pi f$ where the ALP is generated. In the latter case, this is achieved by employing the solutions to the renormalization-group evolution equations of the ALP couplings \cite{Bauer:2020jbp,Chala:2020wvs}. In the well-motivated scenario of a flavor-universal ALP, we find that the NLO corrections weaken the bounds for $\tilde{c}_{GG}(\Lambda)$ and $\tilde{c}_{d}(\Lambda)$ by 30\,--\,40\%, while the bounds on $\tilde{c}_{BB}(\Lambda)$, $\tilde{c}_{WW}(\Lambda)$ and $\tilde{c}_{u}(\Lambda)$ remain almost unchanged with respect to those obtained at leading order. All bounds exhibit a weak logarithmic dependence on $\Lambda$ and become slightly stronger as $\Lambda$ is increased. It is important to note that some bounds depend on our choice of $g_8^\theta=0$ and $g_8^\prime=0$, which underlines the importance of determining the values of these yet unknown low-energy constants. As an example, we have shown how the bound on $\tilde{c}_{GG}(\Lambda)$ changes as a function of $f$ and $g_8^\theta$.

While the inclusion of NLO corrections to the $K^\pm\to\pi^\pm a$ decay amplitudes slightly weakens the bounds on ALP couplings compared to those deduced at leading order, our analysis confirms that these processes yield the strongest particle-physics constraints on ALP couplings in the mass range where $m_a\lesssim 350$\,MeV. Furthermore, at very low ALP masses our bounds can be used to derive constraints on the effective ALP couplings to nucleons, which are competitive with limits derived from astrophysical measurements and non-accelerator experiments. To the best of our knowledge, $K^\pm\to\pi^\pm a$ stands as the strongest particle-physics probe of the effective ALP interaction with nucleons in this mass range. 

Our work establishes a rigorous theoretical framework for further studies of the decays $K^\pm\to\pi^\pm a$ and other rare processes in the context of chiral perturbation theory. The methodology and results presented here could contribute to a more robust understanding of the weak decays involving ALPs and mesons, enhancing our ability to interpret experimental observations and probe the nature of these elusive particles. Looking forward, there are several directions that can be pursued. First of all, it would be interesting to study the neutral decay mode $K_L\to\pi^0 a$ and the semileptonic decay $\pi^-\to e^-\bar\nu_e\,a$ beyond the leading order \cite{Bauer:2021wjo,Bauer:2021mvw} in the chiral expansion. Additionally, a more quantitative investigation of the effects of the $\eta^\prime$ meson could be undertaken, which in our approach are included in the values of the low-energy chiral couplings. Finally, we stress that a better determination of the QCD low-energy constants as well as of the many completely unknown additional low-energy constants entering the expressions for the $K^\pm\to\pi^\pm a$ decay amplitudes would be needed to further refine our analysis.

\subsection*{Acknowledgments}

It is a pleasure to thank Gerhard Ecker, Gino Isidori, Stefan Scherer, and Marvin Schnubel for valuable discussion. The research of C.C., A.M.G.\ and M.N.\ was supported by the Cluster of Excellence \textit{Precision Physics, Fundamental Interactions, and Structure of Matter} (PRISMA$^+$, EXC 2118/1) within the German Excellence Strategy (Project-ID 390831469). C.C.\ would also like to thank Perimeter Institute for hospitality during the completion of this work.  This research was supported in part by Perimeter Institute for Theoretical Physics. Research at Perimeter Institute is supported by the Government of Canada through the Department of Innovation, Science and Economic Development and by the Province of Ontario through the Ministry of Research, Innovation and Science. The Feynman diagrams in this paper have been drawn with the Latex package \texttt{TikZ-Feynman} \cite{Ellis:2016jkw}. Results for the loop graphs have been cross-checked with \texttt{Package-X} \cite{Patel:2015tea,Patel:2016fam}.

\begin{appendix}

\section{Reduction of the KMW basis to the EKW basis}
\label{app:A}

\renewcommand{\theequation}{A.\arabic{equation}}
\setcounter{equation}{0}

In this appendix we work out in detail the reduction of the (redundant) octet basis 
\begin{align}
   \mathcal{L}_{\rm weak}^{(p^4)\,\text{KMW}} 
   = \sum_{i=1}^{48} E_i^8\,O_i^8
\end{align}
proposed by Kambor, Missimer and Wyler (KMW) \cite{Kambor:1989tz} and the octet basis 
\begin{align}
   \mathcal{L}_{\rm weak}^{(p^4)\,\text{EKW}} 
   = \frac{G_8 F^2}{2} \sum_{i=1}^{37} N_i\,W_i^8
\end{align}
introduced by Ecker, Kambor and Wyler (EKW) in \cite{Ecker:1992de}, focusing on the case of the SM extended by an ALP. In both cases one considers CP-invariant operators only. We start from the KMW basis, which consists of 48 operators $O_i^8$. In its original form, the basis contained an operator $O_{47}^8$ that is odd under CP and thus should have been omitted. The vanishing of the field-strength tensors $F_{\mu\nu}^L$ and $F_{\mu\nu}^R$ associated with the external ALP currents $l_\mu$ and $r_\mu$ implies that the 16 operators with $i=16,\dots,31$ are absent. Moreover, the fact that in the ALP model the object $\chi=2 B_0\,m_q$ is a diagonal matrix yields the algebraic identity
\begin{align}
   O_1^8 + O_3^8 + O_5^8 = 4 \braket{\lambda_6\,\chi\chi^\dagger} = 0 \quad
   \Rightarrow \quad O_5^8 = - O_1^8 - O_3^8 \,,
\end{align}
which we use to eliminate the operator $O_5^8$ from the basis. This leaves us with 30 operators. 

As explained in \cite{Kambor:1989tz}, the Cayley--Hamilton theorem can be used to eliminate one operator in the set $\{O_{10,\dots,14}^8\}$ and two operators in the set $\{O_{41,\dots,46}^8\}$. We choose to get rid of the operators
\begin{align}
\begin{aligned}
\label{eq:redundancies2}  
   O_{11}^8 &= - O_{10}^8 + \frac12 \left( O_{12}^8 + O_{13}^8 \right) + O_{14}^8 \,, \\
   O_{42}^8 &= O_{41}^8 - O_{44}^8 + \frac12\,O_{45}^8 \,, \\
   O_{43}^8 &= - 2 O_{41}^8 + \frac32\,O_{44}^8 + O_{46}^8 \,.
\end{aligned}
\end{align}
The equation of motion \eqref{eq:EOM} can be employed to express the operators with $i=32,33,34,38,40$ in terms of other operators, up to corrections of $\mathcal{O}(G_F^2)$, which we neglect. The relevant relations are
\begin{align}
\begin{aligned}
\label{eq:redundancies3}  
   O_{32}^8 &= - 2 O_4^8 \,, \\
   O_{33}^8 &= - 4 O_3^8 + \frac43\,O_4^8 - \frac83\,W_5^{\theta\,8} \,, \\
   O_{34}^8 &= 2 O_5^8 = - 2 O_1^8 - 2 O_3^8 \,, \\[2mm]
   O_{38}^8 &= 2 O_{15}^8 \,, \\
   O_{40}^8 &= 4 O_3^8 - \frac83\,O_4^8 + \frac{16}{3}\,W_5^{\theta\,8} \,.
\end{aligned}
\end{align}
A basis of operators $\{W_i^{\theta\,8}\}$ containing the $\theta$ field explicitly has been presented in Table~\ref{table:p4weak}. 

Further relations between operators can be found using integration by parts. Before presenting them, we collect some useful rules for the covariant derivative $D_\mu$. For objects $A$ and $B$ transforming as $A\to g_L A\,g_R^\dagger$ and $B\to g_R B g_L^\dagger$ under chiral transformations, and objects $C$ transforming as $C\to g_L C g_L^\dagger$, the covariant derivative (dropping the photon field for simplicity) is defined as 
\begin{align}
\begin{aligned}
\label{eq:Drules}  
   D_\mu A &= \partial_\mu A - i l_\mu A + i r_\mu A \,, \\
   D_\mu B &= \partial_\mu B - i r_\mu B + i l_\mu B \,, \\
   D_\mu C &= \partial_\mu C - i \left[ l_\mu,C \right] ,
\end{aligned}
\end{align}
where the last rule follows from the first two by setting $C=AB$. These definitions also hold for the objects containing spurion fields, e.g.\ $D_\mu S=\partial_\mu S-i\,[l_\mu,S]$. With these definitions, the covariant derivative obeys the product rule. It also follows that 
\begin{align}
\begin{aligned}
\label{eq:Drules_traces}  
   \partial_\mu \braket{C}
   &= \braket{\partial_\mu C} = \braket{D_\mu C} \,, \\
   \partial_\mu \braket{\lambda_6 C}
   &= \braket{\lambda_6\,\partial_\mu C} = \braket{\lambda_6 D_\mu C} 
    + i \braket{\lambda_6\left[ l_\mu,C \right]} .
\end{aligned}
\end{align}
Hence, when using integration by parts on the operators in the weak chiral Lagrangian, one can encounter operators involving the left-handed ALP current explicitly. Concretely, integrating by parts in $O_{35}^8$ and $O_{36}^8$, we obtain the relations
\begin{align}
\label{eq:O35O36} 
\begin{aligned}
   O_{35}^8 & = - \frac12\,O_{39}^8 - i \braket{\lambda_6 \left[ l_\nu,\{ L_\mu,W^{\mu\nu} \} \right]} , \\
   O_{36}^8 &= - \frac12\,O_{40}^8 
   - i \braket{\lambda_6 \left[ l_\mu, \{ L^\mu, W_\nu^{\,\nu} \} \right]} \\
   &= - 2 O_3^8 + \frac43\,O_4^8 - \frac83\,W_5^{\theta\,8} 
    + 2i \braket{\lambda_6 \left[ l_\mu, \{ L^\mu,P \} \right]} 
    - \frac43\,i \braket{\lambda_6 \left[ l_\mu,L^\mu \right]} \braket{P} , 
\end{aligned}
\end{align}
where $O_{40}^8$ has been reduced to other operators in \eqref{eq:redundancies3}, and we have used the equation of motion \eqref{eq:EOM} in the last step. Using that $[D_\mu,D_\nu]\,L_\rho=0$ in the ALP model, one can moreover show that
\begin{align}
\begin{aligned}
   O_{35}^8 
   &= - \frac14\,O_{40}^8 - 4 \braket{\lambda_6 (D_\mu L_\nu)(D^\mu L^\nu)} \\
   &\quad - \frac{i}{2} \braket{\lambda_6 \left[ l_\mu,\{ L^\mu,W_\nu^{\,\nu} \} \right]}
    - 2i \braket{\lambda_6 \left[ l_\mu, \{L_\nu,D^\mu L^\nu\} \right]} .
\end{aligned}
\end{align}
Using the identity 
\begin{align}
\label{eq:DL}
   D_\mu L_\nu = \frac14\,W_{\mu\nu} + \frac{i}{2} \left[ L_\mu,L_\nu \right] ,
\end{align}
which holds since the field-strength tensors associated with the ALP currents vanish, along with the equation of motion we find that
\begin{align}
\begin{aligned}
\label{eq:O35}
   O_{35}^8 
   &= - \frac14\,O_{39}^8 - \frac14\,O_{40}^8 + 2 O_{42}^8 - 2 O_{43}^8 
   + i \braket{\lambda_6 \left[ l_\mu,\{ L^\mu,P \} \right]}
    - \frac{2i}{3} \braket{\lambda_6 [l_\mu,L^\mu]} \braket{P} \\
   &\quad - \frac{i}{2} \braket{\lambda_6 \left[ l_\nu,\{L_\mu,W^{\mu\nu}\} \right]} 
    - \braket{\lambda_6 \left[ l_\mu,[L^2,L^\mu] \right]}
    + \mathcal{O}\bigg(\frac{a^2}{f^2}\bigg) \,.
\end{aligned}
\end{align}
In a similar way, using relation \eqref{eq:DL} it is possible to show that
\begin{align}
\label{eq:O37}
   O_{37}^8 
   = 2 O_{15}^8 + 4 O_{41}^8 - 4 O_{43}^8 
    - 4 \braket{\lambda_6 \left[ l_\mu,[L^2,L^\mu] \right]} , 
\end{align}
where $O_{43}^8$ has been reduced to other operators in \eqref{eq:redundancies2}. Next, the operators $O_i^8$ with $i=6,7,8$ can be rewritten in the form
\begin{align}
\label{eq:O678red}
\begin{aligned}
   O_6^8 &= \frac12\,O_{10}^8 - O_{11}^8 - \frac12\,O_{15}^8 + i \braket{\lambda_6 [ L_\mu, D^\mu S ]} , \\
   O_7^8 &= - \frac12\,O_{10}^8 - O_{11}^8 + \frac12\,O_{15}^8 + \braket{\lambda_6 \{ L_\mu, D^\mu P \}} , \\
   O_8^8 &= - \frac12\,O_{12}^8 + \braket{\lambda_6 L_\mu} \braket{D^\mu P} .
\end{aligned}
\end{align}
The terms involving $D_\mu S$ and $D_\mu P$ can be eliminated by once more using integration by parts and the product rule, which leads to 
\begin{align}
\begin{aligned}
\label{eq:O6O7O8}
   O_6^8 
   &= \frac12 \left( O_1^8 + O_3^8 + 3 O_{10}^8 - O_{12}^8 - O_{13}^8 - 2 O_{14}^8 - O_{15}^8 \right)
    + \braket{\lambda_6 \left[ l_\mu, [ L^\mu,S ]\right]} , \\
   O_7^8 
   &= O_3^8 - \frac13\,O_4^8 + \frac12 \left( O_{10}^8 - O_{12}^8 - O_{13}^8 - 2 O_{14}^8 + O_{15}^8 \right)
    - i \braket{\lambda_6 \left[ l_\mu, \{ L^\mu,P \}\right]} , \\
   O_8^8 
   &= \frac12 \left( O_4^8 - O_{12}^8 \right) - i \braket{\lambda_6 \left[ l_\mu,L^\mu \right]} \braket{P} .
\end{aligned}
\end{align}
Finally, working consistently to first order in $a/f$, it is not difficult to show that
\begin{align}
\label{eq:redundancies4}  
   O_9^8 = \frac12 \left( O_7^8 - O_6^8 \right) + \mathcal{O}\bigg(\frac{a^2}{f^2}\bigg) \,.
\end{align}

Combining the four relations in \eqref{eq:O35O36}, \eqref{eq:O35}, \eqref{eq:O37} with the last two equations in \eqref{eq:O6O7O8}, we can derive two relations not involving the ALP current $l_\mu$ explicitly. They can be solved to yield the non-trivial results
\begin{align}
\label{eq:redundancies5} 
\begin{aligned}
   O_{35}^8
   &= - 2 O_7^8 + \frac43\,O_8^8 + O_{10}^8 - \frac13\,O_{12}^8 - O_{13}^8 - 2 O_{14}^8 \\
   &\quad + \frac12\,O_{37}^8 + 6 O_{41}^8 - 7 O_{44}^8 + 2 O_{45}^8 - 2 O_{46}^8 - \frac83\,W_5^{\theta\,8} \,, \\
   O_{36}^8 
   &= - 2 O_7^8 + \frac43\,O_8^8 + O_{10}^8 - \frac13\,O_{12}^8 - O_{13}^8 - 2 O_{14}^8 + O_{15}^8 
   - \frac83\,W_5^{\theta\,8} \,.
\end{aligned}
\end{align}
They can be used to eliminate these two operators from the basis. Note that the first relation is consistent with equation (A.2) in the appendix of \cite{Ecker:1990kz}.

Altogether, the relations \eqref{eq:redundancies2}, \eqref{eq:redundancies3}, \eqref{eq:redundancies4} and \eqref{eq:redundancies5} eliminate 11 additional operators, leaving a final set containing 19 basis operators. We give their definitions in Table~\ref{table:p4weak_KMW}. Operators containing four factors of $L_\mu$ do not contribute to the $K^-\to\pi^- a$ decay amplitude. Moreover, the three relations in \eqref{eq:O6O7O8} and the two equations in \eqref{eq:O37} and \eqref{eq:O35O36} allow us to express the operators $O_i^8$ with $i=6,7,8,37,39$ (the operators below the horizontal line in the right portion of the table) in terms of other operators $O_j^8$ plus operators involving the left-handed ALP current $l_\mu$, showing that these operators are redundant in the SM. In the absence of new physics, the basis of weak octet operators thus contains only the first 14 operators in the table.

\begin{table}[t]
\centering
\renewcommand{\arraystretch}{1.2} 
\begin{tabular}{|c|c||c|c|}
\hline 
$i$ & $O_i^8$ & $i$ & $O_i^8$ \\
\hline
\hline
1 & $\braket{\lambda_6 S^2}$
 & 41 & $\braket{\lambda_6 L^2 L^2}$ \\
2 & $\braket{\lambda_6 S} \braket{S}$
 & 44 & $\braket{\lambda_6 L^2} \braket{L^2}$ \\
3 & $\braket{\lambda_6 P^2}$
 & 45 & $\braket{\lambda_6 \{ L_\mu,L_\nu \}} \braket{L^\mu L^\nu}$ \\
4 & $\braket{\lambda_6 P} \braket{P}$
 & 46 & $\braket{\lambda_6 L_\mu} \braket{L^\mu L^2}$ \\
10 & $\braket{\lambda_6 \{ S,L^2 \}}$
 & 48 & $i\epsilon_{\mu\nu\rho\sigma} \braket{\lambda_6 L^\mu} \braket{L^\nu L^\rho L^\sigma}$ \\
\cline{3-4} 
12 & $\braket{\lambda_6 L_\mu} \braket{\{ L^\mu,S \}}$
 & 6 & $i\braket{\lambda_6 [ L_\mu,(D^\mu\chi) \Sigma^\dagger+\Sigma D^\mu\chi^\dagger ]}$ \\
13 & $\braket{\lambda_6 S} \braket{L^2}$
 & 7 & $\braket{\lambda_6 \{ L_\mu,i\big((D^\mu\chi) \Sigma^\dagger-\Sigma D^\mu\chi^\dagger\big) \}}$ \\
14 & $\braket{\lambda_6 L^2} \braket{S}$
 & 8 & $\braket{\lambda_6 L_\mu} \braket{i\big((D^\mu\chi) \Sigma^\dagger-\Sigma D^\mu\chi^\dagger\big)}$ \\
15 & $i\braket{\lambda_6 [ P,L^2 ]}$
 & 37 & $i\braket{\lambda_6 [ L_\mu L_\nu,W^{\mu\nu} ]}$ \\
 & & 39 & $\braket{\lambda_6 W_{\mu\nu} W^{\mu\nu}}$ \\
\hline
\end{tabular}
\caption{Operators $O_i^8$ forming a basis of weak-interaction octet operators at $\mathcal{O}(p^4)$ for the SM extended by an ALP, where we use the same numbering as in \cite{Kambor:1989tz}. The last five operators only exist in the presence of an ALP and can be eliminated in the SM.}
\label{table:p4weak_KMW}
\end{table}

We are now in a position to express the operators $W_i^8$ of the EKW basis shown in the left portion of Table~\ref{table:p4weak} in terms of linear combinations of the operators $O_i^8$ in the KMW basis. The relevant relations are
\begin{align}
   W_1^8 &= O_{41}^8 \,, \quad
    & W_9^8 &= O_{15}^8 \,, \notag\\[1mm]
   W_2^8 &= - 2 O_{41}^8 + \frac32\,O_{44}^8 + O_{46}^8 \,, \quad 
    & W_{10}^8 &= O_1^8 \,, \notag\\
   W_3^8 &= \frac12\,O_{45}^8 \,, \quad
    & W_{11}^8 &= O_2^8 \,, \notag\\[1mm]
   W_4^8 &= O_{46}^8 \,, \quad
    & W_{12}^8 &= - O_3^8 \,, \notag\\[3mm]
   W_5^8 &= O_{10}^8 \,, \quad
    & W_{13}^8 &= - O_4^8 \,, \label{eq:EckerToKambor1} \\[1mm]
   W_6^8 &= \frac12\,O_{12}^8 \,, \quad
    & W_{24}^8 &= - \frac12\,O_4^8 + O_8^8 + \frac12\,O_{12}^8 \,, \notag\\[1mm]
   W_7^8 &= O_{13}^8 \,, \quad
    & W_{28}^8 &= O_{48}^8 \,, \notag\\[2.5mm]
   W_8^8 &= O_{14}^8 \,, \quad
    & \notag   
\end{align}
and 
\begin{align}
\begin{aligned}
   W_{19}^8 &= \frac12\,O_{15}^8 - \frac14\,O_{37}^8 + 3 O_{41}^8 - \frac32\,O_{44}^8 - O_{46}^8 \,, \\
   W_{20} &= O_7^8 - \frac23\,O_8^8 - \frac12\,O_{10}^8 + \frac16\,O_{12}^8 + \frac12\,O_{13}^8 + O_{14}^8 \\
   &\quad - \frac14\,O_{37}^8 - \frac14\,O_{39}^8 - 3 O_{41}^8 + \frac72\,O_{44}^8 - O_{45}^8 + O_{46}^8 \,, \\
   W_{21}^8 &= - \frac12\,O_1^8 - \frac12\,O_3^8 + O_6^8 - \frac32\,O_{10}^8 + \frac12\,O_{12}^8 
    + \frac12\,O_{13}^8 + O_{14}^8 + \frac12\,O_{15}^8 \,, \\
   W_{23}^8 &= - O_3^8 + \frac13\,O_4^8 + O_7^8 - \frac12\,O_{10}^8 + \frac12\,O_{12}^8 
    + \frac12\,O_{13}^8 + O_{14}^8 - \frac12\,O_{15}^8 \,.
\end{aligned}
\label{eq:EckerToKambor2}
\end{align}
As explained in Section~\ref{sec:weak_Lagrangian_p4}, the remaining 18 operators $W_i^8$ either vanish ($i=14,\dots,18,25,26,27,29,\dots,37$) or are redundant ($i=22$) in our model.

Inverting the above relations, we obtain
\begin{align}
\begin{aligned}
   O_1^8 &= W_{10}^8 \,, \quad
    & O_{14}^8 &= W_8^8 \,, \\[1mm]
   O_2^8 &= W_{11}^8 \,, \quad
    & O_{15}^8 &= W_9^8 \,, \\[1mm]
   O_3^8 &= - W_{12}^8 \,, \quad
    & O_{37}^8 &= 4 W_1^8 - 4 W_2^8 + 2 W^{8}_{9} - 4 W^{8}_{19} \,, \\[1mm]
   O_4^8 &= - W_{13}^8 \,, \quad
    & O_{41}^8 &= W_1^8 \,, \\
   O_8^8 &= - W_6^8 - \frac12\,W_{13}^8 + W_{24}^8 \,, \quad
    & O_{44}^8 &= \frac43\,W_1^8 + \frac23\,W_2^8 - \frac23\,W_4^8 \,, \\[1mm]
   O_{10}^8 &= W_5^8 \,, \quad
    & O_{45}^8 &= 2 W_3^8 \,, \\[1mm]
   O_{12}^8 &= 2 W_6^8 \,, \quad
    & O_{46}^8 &= W_4^8 \,, \\[1mm]
   O_{13}^8 &= W_7^8 \,, \quad
    & O_{48}^8 &= W_{28}^8 \,, 
\end{aligned}
\end{align} 
and
\begin{align}
\begin{aligned}
   O_6^8 &= \frac32\,W_5^8 - W_6^8 - \frac12\,W_7^8 - W_8^8 -\frac12\,W_9^8 + \frac12\,W_{10}^8 - \frac12\,W_{12}^8 + W_{21}^8 \,, \\
   O_7^8 &= \frac12\,W_5^8 - W_6^8 - \frac12\,W_7^8 - W_8^8 + \frac12\,W_9^8 - W_{12}^8 + \frac13\,W_{13}^8 + W_{23}^8 \,, \\
   O_{39}^8 &= \frac83\,W_1^8 + \frac{40}{3}\,W_2^8 - 8 W_3^8 - \frac{16}{3}\,W_4^8 - 4 W_{12}^8 + \frac83\,W_{13}^8 \\ 
   &\quad + 4 W_{19}^8 - 4 W_{20}^8 + 4 W_{23}^8 - \frac83\,W_{24}^8 \,. 
\end{aligned} 
\end{align} 
These expressions allow one to read off the relations between the low-energy constants in the two bases, as well as those between the corresponding anomalous dimensions.

\section{Explicit results for the amplitude}
\label{app:B}

\renewcommand{\theequation}{B.\arabic{equation}}
\setcounter{equation}{0}

In this section we collect our analytic results for the NLO contributions to the $K^- \to \pi^- a$ decday amplitude, using the parameterization in \eqref{eq:amplitudeLOandNLO}. To write our results in a compact form, it is convenient to define the following functions:
\begin{align}
 s_1 (x,y)  & =  \sqrt{\lambda \left(1,x,\frac{4-y}{3}\right)}\,\ln  \frac{3\,\sqrt{\lambda
   \left(1,x,\frac{4-y}{3}\right)}-3 \, x-y + 7}{2 
   \sqrt{3 \, (4-y)}} \,, \nonumber \\
 s_2 (x,y)  &= s_2 (y) =\sqrt{\lambda \left(1,\frac{4-y}{3},y\right)} \,\ln \frac{3 \sqrt{\lambda
   \left(1,\frac{4-y}{3},y\right)}+2 \,y+1}{2  \sqrt{3 \, y (4-y)}}  \,,  \notag \\
  s_3 (x,y)  & = s_3 (y) = \ln \frac{4-y}{3}    \,, \nonumber \\[2mm]
 s_4 (x,y)  &  = s_4 (y)=\ln y   \,,  \\
 s_5 (x,y)  & = s_5 (y)=\sqrt{\lambda \left(1,\frac{4-y}{3},y\right)} \,\ln \frac{3\sqrt{\lambda
   \left(1,\frac{4-y}{3},y\right)}- 4 \, y+ 7}{2 
   \sqrt{3\,(4-y)}}   \,,  \nonumber \\
 s_6 (x,y)  & = s_6(y) = \sqrt{1-4 \, y}  \,\ln \frac{\sqrt{1-4  \, y}+1}{2 \sqrt{y}}   \nonumber   \,, \\
  s_7 (x,y)  & = \sqrt{\lambda (1,x,y)}\, \ln \frac{\sqrt{\lambda (1,x,y)}-x+y+1}{2 \sqrt{y}}  \ \,,  \nonumber   
\end{align}
where only $s_1$ and $s_7$ depend on the argument $x$. Here $\lambda(a,b,c)$ is the K\"allén function, defined as
\begin{align}
   \lambda(a,b,c) = a^2 + b^2 + c^2 - 2\,a\,b - 2\,a\,c - 2\,b \,c \,.    
\end{align}
In the limit $x\to 0$ we find the relations
\begin{align}
\begin{aligned}
   s_1(0,y) &= - \frac16 \left(1-y\right) s_3(y) \,, \\
   s_7(0,y) &= - \frac12 \left(1-y\right) s_4(y) \,,
\end{aligned}
\end{align} 
while for $y\to 0$ we obtain
\begin{align}
\begin{aligned}
   s_2(y) &\overset{y\to 0}{=} - \frac16\,\big[ \ln 12 + s_4 (y) \big] \,, \\
   s_3(0) &= - 6\,s_5(0) \,, \\
   s_6(y) &\overset{y\to 0}{=} - \frac12\,s_4 (y) \,.
\end{aligned}
\end{align} 
Using these relations, one can show that our results are finite for $x\to 0$ and $y\to 0$.

\subsection[$\mathcal{A}^{\rm FC}$ at NLO]{\boldmath $\mathcal{A}^{\rm FC}$ at NLO} 

In the following, we collect the contributions $A^{G, {c}_{\rm ALP}}$ entering the $\mathcal{A}^{\rm FC}$ term in \eqref{eq:amplitudeLOandNLO} at NLO. As explained in the main text, we compute these contributions only for the octet couplings  ($G_{8}$,  $G_{8}^\theta$ and  $G_{8}^\prime$). Below, we provide explicit expressions for each of these contributions. The  low-energy constants $\hat{L}_{i,r}$, $\hat{L}_{i,r}^\theta$, $\hat{N}_{i,r}^\prime$, and $\hat{N}_{i,r}^{\theta \prime}$ appearing in this appendix are understood to be evaluated at the scale $\mu$.  

\subsubsection[$G_{8}$ contribution]{\boldmath $G_{8}$ contribution} 

For the $G_{8}$ contribution proportional to $\tilde{c}_{GG}$, we find 
{\small \begin{align}
 i A_{\text{NLO}}^{G_{8}, \tilde{c}_{GG}}    &=  \frac{m_{K^-}^2}{8 \pi^2 F^2_{\pi}}\bigg\{     \ln \frac{\mu}{m_K} g^{G_{8}, \tilde{c}_{GG}}(x,y)  +  f_0^{G_{8}, \tilde{c}_{GG}} (x,y) + \sum_{i=1}^7 s_i  (x,y) f_i^{G_{8}, \tilde{c}_{GG}}(x,y)  \nonumber \\
 & \quad- \frac{4 (1-y)(1-x)}{(4-y -3  x )^2} \bigg[\hat{L}_{5,r} \left(4 +9 y - 4  y^2  - x-5 x y -3  x^2\right) \nonumber  \\
 & \quad + 4  \hat{L}_{4,r} (2+y) (4-y-3  x)    -\hat{L}_{7,r}
  \frac{6 y (4- y^2)- 4 x (2+ 11 y- 4 y^2 )+6 x^2 (2+y)}{1-x} \nonumber \\
 & \quad  - 2 \hat{L}_{8,r} \frac{1-y}{1-x} \left( y (2+y)- 2 x
   (1+ 2 y )+ 3 x^2\right) \bigg]  \nonumber  \\ 
&\quad +  \frac{2(1-y)}{(4-y-3 x)} \bigg[  x (2+ y -3 x) (\hat{L}^{\theta}_{1,r} - \hat{L}^{\theta}_{2,r}  )  \notag \\
& \quad +4 \left(  \hat{N}^{\prime}_{5,r}+2 \hat{N}^{\prime}_{8,r}+\hat{N}^{\prime}_{9,r} - \hat{N}^{\theta\prime}_{1,r}  + \hat{N}^{\theta\prime}_{2,r}  -2\hat{N}^{\theta\prime}_{3,r}-2\hat{N}^{\theta\prime}_{4,r} \right) \nonumber \\
& \quad+ y \left( 6  \hat{N}^{\prime}_{5,r}+6 
  \hat{N}^{\prime}_{6,r}+4  \hat{N}^{\prime}_{8,r}-8  \hat{N}^{\prime}_{10,r}-12   \hat{N}^{\prime}_{12,r} 
 -12  \hat{N}^{\prime}_{13,r} -3  \hat{N}^{\theta\prime}_{1,r}+3N^{\theta\prime}_{2,r}
 \right. \nonumber \\ 
 &  \hspace{10.2 cm} \left.
 -6 \hat{N}^{\theta\prime}_{3,r}-2  \hat{N}^{\theta\prime}_{4,r} \right) \nonumber \\
& \quad -  y^2 \left(  2  \hat{N}^{\prime}_{5,r}+3 \hat{N}^{\prime}_{6,r}-4 \hat{N}^{\prime}_{10,r}-4 \,\hat{N}^{\prime}_{12,r}-6  \hat{N}^{\prime}_{13,r}  - \hat{N}^{\theta\prime}_{1,r} +N^{\theta\prime}_{2,r}-2 N^{\theta\prime}_{3,r}  -\hat{N}^{\theta\prime}_{4,r}\right)   \nonumber \\
&\quad - x  \left(  2   \hat{N}^{\prime}_{5,r} + 2  
  \hat{N}^{\prime}_{6,r}+8  \hat{N}^{\prime}_{8,r}+4  \hat{N}^{\prime}_{9,r}-4   \hat{N}^{\prime}_{12,r} -4  \hat{N}^{\prime}_{13,r} +  \hat{N}^{\theta\prime}_{1,r}  +7 N^{\theta\prime}_{2,r}-14 \hat{N}^{\theta\prime}_{3,r}  \right. \nonumber \\
 & \hspace{10.3 cm}   \left.-6  \hat{N}^{\theta\prime}_{4,r} - 4  \hat{N}^{\theta\prime}_{5,r} \right) \nonumber  \\
& \quad- x  y \left(
   4 \hat{N}^{\prime}_{5,r}+\hat{N}^{\prime}_{6,r}+4 \hat{N}^{\prime}_{8,r}-4  N^{\prime}_{10,r}-4   \hat{N}^{\prime}_{12,r}-2  
   \hat{N}^{\prime}_{13,r}  - 
 4  \hat{N}^{\theta\prime}_{1,r}  +2   \hat{N}^{\theta\prime}_{2,r}   \right . \nonumber \\
   &  \hspace{9cm} \left. 
 -4   \hat{N}^{\theta\prime}_{3,r}
 -3   \hat{N}^{\theta\prime}_{4,r}
 + \hat{N}^{\theta\prime}_{5,r} \right) \nonumber \\
& \hspace{5.0 cm} - x^2 \left(  2 \,\hat{N}^{\prime}_{5,r}   - 3 \hat{N}^{\theta\prime}_{1,r} - 3    \hat{N}^{\theta\prime}_{2,r}+ 6  \hat{N}^{\theta\prime}_{3,r}+ 3   \hat{N}^{\theta\prime}_{5,r} \right) \nonumber  \bigg]  \bigg\} \,, \\
\end{align} }
\noindent
where the functions $g^{G_8,\tilde{c}_{GG}}(x,y)$ and $f_{0,1,\dots ,7}^{G_8,\tilde{c}_{GG}}(x,y)$ are given by
\begin{align}
g^{G_8,\tilde{c}_{GG}}(x,y) &=  - \frac{4 \,(1-y)}{ 3 (4-y-3  x)^2} \left[  2 (24 + 36 y - 10 y^2 + y^3)  \right. \nonumber  \\
 & \quad \left. - x(108 + 139 y - 16 y^2)  + 3x^2 (29 + 23 y)- 27 x^3  \right] \,, \nonumber  \\ 
f_0^{G_8,\tilde{c}_{GG}}(x,y) &= - \frac{2(1-y)}{3  x  y(4-y-3 x)}  \left[ 5 y  (1-y)^2 + x \left(2-7 y +16 y^2 -5 y^3 \right) \right. 
\nonumber \\
& \hspace{5.5 cm}\left. -x^2 \left(2+5 y+ 8 y^2\right) + 9 x^3 y \right]  \,,  \nonumber \\ 
f_1^{G_8,\tilde{c}_{GG}} (x,y)     &= -\frac{(1-y) (1-x)}{3 x\,(4-y-3 x)} \left( 2  (1+y) -3  x + \frac{(1-y)^2}{x} \right) \,,   \nonumber \\
f_2^{G_8,\tilde{c}_{GG}} (x,y)     &=-\frac{4\,y(1-y)(2-x-y)}{3\,( 4 - y-3x  ) } \,,  \nonumber\\
f_3^{G_8,\tilde{c}_{GG}} (x,y)     &= - \frac{1}{18  x^2 y ^2 (4-y-3  x)^2} \left[ -y^2 (4-y) (1-y)^4 \right. \nonumber  \\
& \quad+x y^2 (1-y)^2 \left(83-54 y + 14 y^2 - y^3 \right) \nonumber \\[2mm]
& \quad-x^2 \left(4-93 y+255 y^2 -432 y ^3 -169 y^4 +333 y^5 -130 y^6  +16 y^7\right) \nonumber \\[2mm]
&\quad +x^3 \left(7-190 y +403 y^2 -747
   y^3  -107 y^4 +266 y^5 -64 y^6 \right) \nonumber \\[2mm]
& \quad\left. -3 x^4 \left(1-30 \,y  +25 y^2 -44 y^3 -40 y^4+ 16y^5 \right) + 27 x^5 y^2 (1-y)  \right]\,,   \nonumber \\
f_4^{G_8,\tilde{c}_{GG}} (x,y)  &=  \frac{1}{18 x^2 y ^2 (4-y-3  x)^2} \left[  -27 y^2 (4-y)(1-y)^4  \right. \nonumber \\
& \quad+ 9 x y ^2 (1-y)^2 \left(25 - 42 y  + 26\,y^2 - 3 y^3\right) \nonumber \\[2mm]
&\quad -x^2 \left(36-261\,y+423\,y^2-1534 y ^3  + 1283 y^4  -153
\,y^5  -26 y^6 +16 y^7\right)\nonumber \\[2mm]
& \quad  +x^3 \left(63-414 y +135 y^2-1625 y^3  +1323 y^4 +150 y^5 -64 y^6\right) \nonumber \\[2mm]
&\quad \left. -3 x^4 \left(9-54 y  -87 y^2 -128 y^3 +172 y^4 + 16 y^5 \right) -135 x^5 y^2 (1-y)  \right]\,,    \nonumber \\
f_5^{G_8,\tilde{c}_{GG}}(x,y) & =  -\frac{1-y}{3 y^2(4-y-3\,x)} \left( 1-7 y^2+4 y^3 - x (1-7 y + 4 y^2) \right) \,,  \nonumber \\
f_6^{G_8,\tilde{c}_{GG}}(x,y) & = -\frac{1-y}{y(4-y-3 x)}  \left(\frac{1}{y}-4+y+ x \left(3-\frac{1}{y}\right) \right) \,,  \nonumber \\
f_7^{G_8,\tilde{c}_{GG}}(x,y) & = -\frac{(1-x)(1-y)}{x\,(4-y-3  x)} \left( \frac{3 (1-y)^2}{x}+2 (1+y)-5 x\right) \,.  
\end{align} 
Similarly, for the contribution proportional to $c_{uu}^{a}$, we find 
{\small
\begin{align}
 i A_{\rm NLO}^{G_{8}, c_{uu}^{a}}    &=  \frac{m_{K^-}^2}{8 \pi^2 F^2_{\pi}}\bigg\{ \ln \frac{\mu}{m_K} g^{G_8, c^a_{uu}}(x,y)  +  f_0^{G_{8}, c_{uu}^{a}} (x,y) + \sum_{i=1}^7 s_i  (x,y) f_i^{G_{8}, c_{uu}^{a}}(x,y) \nonumber  \\
 &\quad + \frac{y(1-y)}{2 (4-y-3  x)^2}  \bigg[  4  \hat{L}_{4,r}  (2+y)  (4-y-3  x)  \nonumber   \\[2mm]   
 &\quad-  \hat{L}_{5,r}  \left(4 -7 y + 3 y^2 -x (17-5 y )+ 12 x^2   \right) - 24 \hat{L}_{7,r}  \left(4 - y^2 - 2 x (4-y) + 3 x^2 \right)  \nonumber \\[2mm]   
 &   \hspace{7.3cm} -8  \hat{L}_{8,r}  \left(2 - y -y^2 - 3 x (1-y) \right)  \bigg]  \nonumber  \\
 & \quad +   \frac{y(1-y)}{4-y-3 x} \bigg[ -3 \hat {N}^{\prime}_{5, r}  -2 \hat{N}^{\prime}_{6, r} -2 \hat{N}^{\prime}_{8, r} +3
   \hat {N}^{\prime}_{9, r}  +8 \hat{N}^{\prime}_{10, r}   +12 \hat {N}^{\prime}_{12, r} +12 \hat {N}^{\prime}_{13, r}   \nonumber  \\
 & \hspace{2.5cm} +y
   \left(\hat {N}^{\prime}_{5, r}  +2 \hat {N}^{\prime}_{6, r}  -\hat{N}^{\prime}_{8, r} -\hat{N}^{\prime}_{9, r}  -4 \hat{N}^{\prime}_{10, r}   -4\hat {N}^{\prime}_{12, r} -6 \hat{N}^{\prime}_{13, r}\right)  \nonumber \\
     & \hspace{4.7 cm}  +  x  \left(
    \hat{N}^{\prime}_{5, r}
    -3 \hat {N}^{\prime}_{9, r} 
    -6  \hat {N}^{\prime}_{10, r} -6  \hat {N}^{\prime}_{12, r}  -6  \hat {N}^{\prime}_{13, r}   \right) \bigg] \bigg\}\,,   
 \end{align} }  
with   
\begin{align}
g^{G_8, c^a_{uu}}(x,y) & =  \frac{y\,(1-y)}{6 (4 -3 x - y)^2}  \left[  4 y (25 - 4 y) + 3 x (16 -35 y) - 27 x^2  \right] \,, \nonumber   \\
f_0^{G_8, c^a_{uu}}(x,y) & =  \frac{(1-y) \left( (1-y)^2 (8+ 3 y) -2 \, x \left(10 -27 y+ 5 \, y^2 \right) + x^2 (12-45 \, y)  \right)}{12 \, x (4-3 \, x -y)} \,, \nonumber \\ 
f_1^{G_8, c^a_{uu}}(x,y) & = - \frac{y\,(1-y)(1-x)}{ 6 \, x^2(x-y) (4 -3 \, x - y)}     \left((1-y)^2 +2 \, x (1+y) -3 \, x^2 \right) \,, \nonumber  \\ 
f_2^{G_8, c^a_{uu}}(x,y) & = \frac{y(1-y)(4-3 \, x-2 \, y )}{ 3( 4 -3 \, x - y) }  \,,   \nonumber  \\ 
 f_3^{G_8, c^a_{uu}} (x,y)& = 
 -\frac{1}{72 x^2 \, y  (4-y-3  x)^2} \left[   8 y -34 \, y^2+56 y^3-44 \, y^4 +16 y^5-2 \, y ^6 \right. \nonumber\\
&\quad + x \left(8 -200 y +496
y^2 -454 \, y^3 +182 \, y ^4 -34 \, y^5  +2 \, y ^6 \right) \nonumber\\[2mm]
& \quad- x^2
\left(26 -589 y+1256 y^2+147 y^3-635 y^4+259 y^5-32 \, y ^6\right) \nonumber \\[2mm]
&\quad+x^3 \left(27 -633 y+1464 \, y^2+393 y^3-639 y^4 +144 \, y^5\right) \nonumber \\[2mm]
& \quad \left. -  x^4 \left(9-225 y+360 y^2+ 324 \, y^3-144 \, y^4\right)  \right] \,, \nonumber \\
f_4^{G_8, c^a_{uu}} (x,y)  &= \frac{1}{72 x^2 \, y  (4-y-3  x)^2} \left[  72 \, y -270 y^2+351 y^3-144 \, y^4-54 \, y^5+54 \, y^6 \right.  \nonumber\\
&\quad -9 y^7   + x \left(72-360 y +873 y^2 -1125
   y^3  +711 y^4 -171 y^5\right) \nonumber\\[2mm] 
& \quad  -x^2 \left( \nonumber 234-837 y+1259 y^2+ 449 y^3-864 \, y^4+223 y^5-32 \, y ^6\right) \nonumber \\[2mm]
& \quad  +x^3 \left(243-783
   y+777 y^2+1902 y^3-1203 y^4 + 144 \, y^5\right)\nonumber \\[2mm]
& \quad  \left.  -   x^4 \left(81-243 y+153 y^2 + 801 y^3 - 144  y^4\right)  \right] , \nonumber \\[2mm]  
f_5^{G_8, c^a_{uu}}(x,y) & = \frac{(1-y)\left( 2  +3 y+5 y^2 -  6 y^3  -  x \left(5+8 y-12  y ^2\right) +  x^2 (3- 6 y)   \right) }{ 12  y\,(4 -3  x - y)(x-y)} \,,  \nonumber   \\[2mm] 
f_6^{G_8, c^a_{uu}}(x,y) & =\frac{(1-y) \left(2 +3 y -y^2 -x (5+ 2  y)+ 3  x^2  \right)}{4  y\, (4 -3  x - y)(x-y)}  \,, \nonumber   \\[2mm]
f_7^{G_8, c^a_{uu}}(x,y) & = -  \frac{y (1-y)  \left(2 -3 y +y^3  + x \left(1 +10 y -3 y^2\right)  -5 x^2 (2+y) + 7 x^3 \right) }{4 \, x^2 (x-y) (4 -3  x - y)} \,.  \nonumber  \\ 
\end{align}
The contribution proportional to $(c_{dd}^{a}+c_{ss}^{a})$ is given by 
{\small
\begin{align}
 i A_{\rm NLO}^{G_{8}, c_{dd}^{a} +  c_{ss}^{a}}    &= \frac{m_{K^-}^2}{8 \pi^2 F^2_{\pi}}\bigg\{ \ln \frac{\mu}{m_K} g^{G_8, c^a_{dd}+c^a_{ss}}(x,y)  +  f_0^{G_{8}, c^a_{dd}+c^a_{ss}} (x,y)  \nonumber \\
 &\quad+ \sum_{i=1}^7 s_i  (x,y) f_i^{G_{8}, c^a_{dd}+c^a_{ss}}(x,y) \nonumber  \\
 &\quad + \frac{ (1-y)}{4 (4-3 \, x -y)^2}  \bigg[ 4\, \hat{L}_{4,r} (2+y)  \left(16-3 \, x (12-y) + 18 x^2 - y^2\right) \nonumber \nonumber\\[2mm]
 & \quad +\hat{L}_{5,r} \left( 16 + y \, (32 - 9 y -3 y^2) -x (44 + 61 y -21 y^2)   + 6 x^2 (5 + 3\, y)  \right) \nonumber \\[2mm]
  &\quad -48 \hat{L}_{7,r} \left(3 \, x^2 - x
   \left(2 +7 y - 3 y^2  \right) + y \left( 4 - y^2 \right)\right)\nonumber \\
 & \quad+  8 \hat{L}_{8,r} (1-y) \left(3 \, x (2 + 3 y) - 9 x^2-2 \, y  (2+y)\right) \bigg]   \nonumber   \\ 
  &\quad - \frac{1}{2 \, (4-y-3  x)} \bigg[ 4 \left(\hat{N}^{\prime}_{5, r}-2 \hat{N}^{\prime}_{6, r}  +2 \hat{N}^{\prime}_{8, r} + \hat{N}^{\prime}_{9, r} \right)   \nonumber \\
   & \quad +y \,  \left( 5
   \hat{N}^{\prime}_{5, r} +22 \hat{N}^{\prime}_{6, r} -2 \hat{N}^{\prime}_{8, r} -7 \hat{N}^{\prime}_{9, r} -16 \hat{N}^{\prime}_{10, r}  -24 
   \hat{N}^{\prime}_{12, r} -24   \hat{N}^{\prime}_{13, r}  \right) \nonumber \\
   &\quad  -y^2 \left(12 \hat{N}^{\prime}_{5, r} + 20 \hat{N}^{\prime}_{6, r} + 5 \hat{N}^{\prime}_{8, r} - 4 \hat{N}^{\prime}_{9, r}  - 24 \hat{N}^{\prime}_{10, r} -32 \hat{N}^{\prime}_{12, r} -36 \hat{N}^{\prime}_{13, r} \right) \nonumber \\
   &\quad  +y^3 \left(3
   \hat{N}^{\prime}_{5, r}+6 \hat{N}^{\prime}_{6, r}-\hat{N}^{\prime}_{8, r}-\hat{N}^{\prime}_{9, r}-8 (\hat{N}^{\prime}_{10, r}  +  \hat{N}^{\prime}_{12, r})-12 \, \hat{N}^{\prime}_{13, r} \right)  \nonumber  \\
  & \quad - 2 \, x \left(4 \hat{N}^{\prime}_{5, r} 
  +6 \hat{N}^{\prime}_{8, r}
  -\hat{N}^{\prime}_{9, r}
  -6 \hat{N}^{\prime}_{12, r}
  -6 \hat{N}^{\prime}_{13, r} \right)    \nonumber \\
&\quad  + x  y \left(5 \hat{N}^{\prime}_{5, r}+6 \hat{N}^{\prime}_{8, r}-\hat{N}^{\prime}_{9, r}+6 \hat{N}^{\prime}_{10, r} 
    -6 \hat{N}^{\prime}_{12, r}-12 \hat{N}^{\prime}_{13, r} \right) \nonumber \\
& \quad  + x  y^2 \left(3 \hat{N}^{\prime}_{5, r}+6 \hat{N}^{\prime}_{8, r}-\hat{N}^{\prime}_{9, r}-6 \hat{N}^{\prime}_{10, r}-6 \hat{N}^{\prime}_{12, r}\right) - 6 \, x^2   \hat{N}^{\prime}_{9, r} (1-y)\bigg] \bigg\} \,,
\end{align}} 
with
\begin{align}
g^{G_8, c^a_{dd}+ c^a_{ss}}(x,y) & = \frac{1-y}{12  (4-y-3  x)^2} \left[  192 + 4 y (72 + 5 y - 2 y^2)  \right.  \nonumber  \\
 & \left. \hspace{3.5 cm} - 3 x (144 +284 y - 9 y^2 -3 x (44 + 59 y) + 54 x^2)  \right] \,,  \nonumber  \\ 
 f_0^{G_8, c^a_{dd}+ c^a_{ss}}(x,y) & = \frac{1-y}{24  x   y   (4-y-3  x)} \left[  y (1-y)^2  (12+ 7
    y) + 6  x \left( 4 - 2  y ^3 + 9 y^2- 11 y\right) \right.  \nonumber \\ 
 & \hspace{6.3cm} \left. - 3  x^2 \left(8 -8   y + 15   y^2\right) + 54   x^3 y  \right]   \,,   \nonumber  \\
f_1^{G_8, c^a_{dd}+ c^a_{ss}}(x,y) & = - \frac{1-y}{24 x^2 (x-y) (4-y-3  x)} \left[y^2 (1-y)^2 + 2  x \left(2  y ^2+y-1\right)   \right. \nonumber \\
& \hspace{4.5 cm}\left.  -x^2 \left(4  y^2+14  y+1\right) +12 x^3 (1+y)-9 x^4 
   \right] \,,  \nonumber  \\
f_2^{G_8, c^a_{dd}+ c^a_{ss}}(x,y) & =  \frac{y(1-y)  (8 - 4  y - 3  x )}{6  ( 4-y -3  x )} \,, \nonumber   \\
f_3^{G_8, c^a_{dd}+ c^a_{ss}}(x,y) & =  \frac{ 1}{144 x^2  y ^2 (4-y-3  x)^2} \left[ - y^3 (4-y) (1-y)^4  \right.  \nonumber \\
&\quad + x y (1-y)^2  \left(8 -14  y +82  y ^2 -41 y^3  + 7 y^4 \right) \nonumber \\[2mm] 
& \quad - x^2 \big(24-532
   y +1196 y^2 -34  y^3 -2118 y^4  +1747 y^5\nonumber \\[2mm]  
   & \hspace{1.7cm} -563 y^6 + 64  y^7\big)\nonumber \\[2mm]
& \quad + 3  x^3 \left( 14- 371 y + 850 y^2- 655 y^3 - 281
   y^4 + 343 y^5- 80 y^6\right)\nonumber \\[2mm]
& \quad \left. - 9 x^4 \left(16 y^5-53 y^4+12  y ^3+46 y^2-59 y+2\right) +   81 x^5 y^2  (1-y)    \right]\,,  \nonumber  \\
 f_4^{G_8, c^a_{dd}+ c^a_{ss}}(x,y) & =  \frac{ 1}{144 x^2  y ^2 (4-y-3  x)^2} \big[   18  y^2 (1-y)^4  \left( 8+2  y - y^2 \right)    \nonumber \\
 & \quad- 9  x  y (1-y)^2  \left(8+38 y -57
   y^2+36 y^3-y^4\right) \nonumber \\[2mm] 
&\quad+ x^2 \big(216 - 1332 y + 2250 y^2 - 3463 y ^3 + 1442 y^4  + 954 y^5 \nonumber \\[2mm]  
   & \hspace{1.7cm}- 347 y^6 + 64 y^7\big)\nonumber \\[2mm]
&\quad  - 3  x^3 \left(126 - 747 y + 726 y^2 - 1578 y^3+ 1174 y^4+ 307 y^5- 80 y^6\right)\nonumber \\[2mm]
& \quad + 9 x^4 \big( 18 - 99 y - 18 y^2 - 155 y^3 + 238 y^4 + 16 y^5\big)  \nonumber \\
 & \quad  + 405 x^5 y^2 (1-y)   \big] \,,  \nonumber \\[2mm]
 f_5^{G_8, c^a_{dd}+ c^a_{ss}}(x,y) & =   \frac{ 1-y}{24  y^2 (x-y) (4-y-3  x)} \left[ - y \left(8+3
   y-37 y^2+ 18 y^3 \right)  \right.  \nonumber \\
 &  \hspace{2.7cm} \left. +x \left(6+ 11 y-76 y^2 + 36 y^3\right) - 3  x^2 \left(2 -13 y + 6 y^2\right) \right] \,,  \nonumber  \\
 f_6^{G_8, c^a_{dd}+ c^a_{ss}}(x,y) & = -  \frac{ (1-y)}{8 y^2 (x-y) (4-y-3  x)}  \left[y \left(8-21 y+5 y^2\right)\right. \nonumber\\
 & \left. \hspace{5.2cm} - x \left(6-13 y - 10 y^2\right)+  x^2 (6-15 y)\right]\,,   \nonumber  \\
 f_7^{G_8, c^a_{dd}+ c^a_{ss}}(x,y) & = \frac{ 1-y}{8  x^2 (x-y) (3  x+y-4)} \left[  2  y 
   \left(2-3 y+y^3\right)+ x^3 (16+ 19 y)  \right. \nonumber \\
& \quad \left. -x \left(6  - 3 y  - 14  y^2+ 7 y^3\right)  +x^2 \left(5-30 y + y^2\right)  - 15 x^4  \right] \,.  
\end{align}
For the contribution proportional to $(c_{dd}^{a}-c_{ss}^{a})$, we find
{\small
\begin{align}
 i A_{\rm NLO}^{G_{8}, c_{dd}^{a} -  c_{ss}^{a}}    &= \frac{m_{K^-}^2}{8 \pi^2 F^2_{\pi}}\bigg\{  \ln \frac{\mu}{m_K} g^{G_8, c^a_{dd} - c^a_{ss}}(x,y)  +  f_0^{G_{8}, c^a_{dd} - c^a_{ss}} (x,y) \nonumber  \\
 &\quad + \sum_{i=1}^7 s_i  (x,y) f_i^{G_{8}, c^a_{dd} - c^a_{ss}}(x,y) \nonumber \\
&\quad - \frac{ x}{4 (4-3  x -y)^2}  \bigg[ 4 \hat{L}_{4,r}  (2+y) \left(4 +7 y -2  y ^2 -3  x (y+5)+ 9 x^2 \right)   \nonumber \\[2mm] 
& \quad - \hat{L}_{5,r}     \left(4  - y (31 +19 y -10 y^2) + 3 x (1 + 24 y -y^2 - 3 x - 9 x y)   \right)  \nonumber  \\[2mm] 
 & \hspace{4.3 cm}  + 24  (2 \hat{L}_{7,r} + \hat{L}_{8,r})      (1-y)^2     (2+y-3x) \bigg]   \nonumber \\ 
   &\quad - \frac{1}{2  (4-y-3  x)} \bigg[ 4 \left(\hat{N}^{\prime}_{5, r}+2 \hat{N}^{\prime}_{6, r}-\hat{N}^{\prime}_{9, r}-4 \hat{N}^{\prime}_{10, r} -4
   \hat{N}^{\prime}_{11, r}  +\hat{N}^{\prime}_{21, r}+\hat{N}^{\prime}_{23, r} \right)  \nonumber \\ 
     & \quad-y \left(9 \hat{N}^{\prime}_{5, r}+18
   \hat{N}^{\prime}_{6, r}-9 \hat{N}^{\prime}_{9, r}-36 \hat{N}^{\prime}_{10, r}-28 \hat{N}^{\prime}_{11, r}  +9 \hat{N}^{\prime}_{21, r}+9
   \hat{N}^{\prime}_{23, r} \right)\nonumber \\
   & \quad   +6  y^2 \left(\hat{N}^{\prime}_{5, r}+2 \hat{N}^{\prime}_{6, r}-\hat{N}^{\prime}_{9, r}-4
   \hat{N}^{\prime}_{10, r}-\hat{N}^{\prime}_{11, r} +\hat{N}^{\prime}_{21, r}+\hat{N}^{\prime}_{23, r}\right)\nonumber \\
    &\quad  -y^3
\left(\hat{N}^{\prime}_{5, r}+2 \hat{N}^{\prime}_{6, r}-\hat{N}^{\prime}_{9, r}-4 \hat{N}^{\prime}_{10, r}+8 \hat{N}^{\prime}_{11, r}  +\hat{N}^{\prime}_{21, r}+\hat{N}^{\prime}_{23, r} \right) +2  y^4 \hat{N}^{\prime}_{11, r} \nonumber \\ 
   & \quad - x \left(6 \hat{N}^{\prime}_{5, r} +2 \hat{N}^{\prime}_{8, r}
+2\hat{N}^{\prime}_{9, r}
-12 \hat{N}^{\prime}_{10, r}
-12 \hat{N}^{\prime}_{11, r}+12 \hat{N}^{\prime}_{12, r} +12 \hat{N}^{\prime}_{13, r} \right. 
\nonumber \\
&  \hspace{7.1 cm} \left.  
-4\hat{N}^{\prime}_{20, r}
 -\hat{N}^{\prime}_{21, r}
+7\hat{N}^{\prime}_{23,r}\right)\nonumber \\
   & \quad + x y \left(\hat{N}^{\prime}_{5, r}-5 \hat{N}^{\prime}_{8, r}-7
   \hat{N}^{\prime}_{9, r}-18 \hat{N}^{\prime}_{10, r}-18 \hat{N}^{\prime}_{11, r} \right. \nonumber \\
   &\hspace{3.2 cm}  \left. +18 \hat{N}^{\prime}_{12, r}+24 \hat{N}^{\prime}_{13, r}+3
   \hat{N}^{\prime}_{20, r}+\hat{N}^{\prime}_{21, r}+3 \hat{N}^{\prime}_{23, r}\right)\nonumber \\
   & \quad- x y^2 \left(\hat{N}^{\prime}_{5, r}+2  
   \hat{N}^{\prime}_{8, r}-3  \hat{N}^{\prime}_{9, r}-6  \hat{N}^{\prime}_{10, r}+6 \hat{N}^{\prime}_{12, r} +12 \hat{N}^{\prime}_{13, r}+\hat{N}^{\prime}_{20, r} \right. \nonumber \\
   & \hspace{5.1 cm} \left. +2  \hat{N}^{\prime}_{21, r}+2  \hat{N}^{\prime}_{23, r} \right)   \nonumber \\
& \quad + 6  x  y^3 \hat{N}^{\prime}_{11, r}  + x^2 \left(6 \hat{N}^{\prime}_{5, r}+6 \hat{N}^{\prime}_{8, r}+6 \hat{N}^{\prime}_{9, r}-7 \hat{N}^{\prime}_{20, r}-3
   \hat{N}^{\prime}_{21, r}+3 \hat{N}^{\prime}_{23, r} \right)\nonumber \\
   & \quad   + x^2  y   \left(3 \hat{N}^{\prime}_{8, r}-2 \hat{N}^{\prime}_{20, r}  +3  \hat{N}^{\prime}_{21, r}+ 3   \hat{N}^{\prime}_{23, r} \right)   +  3  x^3 \hat{N}^{\prime}_{20, r}  \bigg]  \bigg\} \,,
 \end{align}}
where  
\begin{align}
g^{G_8, c^a_{dd}- c^a_{ss}}(x,y) & = \frac{1}{36  (4 - 3  x -y)^2} \left[-2 (31+ 26 y) \left(4-5
   y + y^2\right)^2 \right.\nonumber  \\
 & \quad  + 12  x \left(68 - 138 y - 146 y^2 + 143 y^3 - 26 y^4 \right)   \nonumber  \\[2mm] 
   & \quad\left.+9 x^2 \left(74 + 143 y + 126 y^2 - 52 y^3\right) -27  x^3 (29+ 33 y)+243  x^4  \nonumber \right] \,, \\ 
f_0^{G_8, c^a_{dd}- c^a_{ss}}(x,y) & =  -\frac{1}{72  x  y  (4-y-3  x)} \left[ 4  y   (4-y) (1-y)^3 \right. \nonumber  \\
&  \quad + x \left(72-271
   y+460 y^2-347 y^3+86 y^4\right) \nonumber  \\[2mm]
   & \quad-3   x^2 \left(24- 109 y+ 112  y ^2-51 y^3\right)    \left.  -9  x^3 y  (9 + 8 y)  + 81  x^4  y    \right]\,,  \nonumber  \\ 
f_1^{G_8, c^a_{dd}- c^a_{ss}}(x,y) & = -\frac{1}{144  x^2    (x-y) (4-y-3  x)} \left[ y (4-y) (1-y)^3  \right. \nonumber \\
& \quad - x \left(16 -54  y +53 y^2 -20 y^3 +  5 y^4\right)  \nonumber \\[2mm]
   &\quad- x^2 \left(17 +43 y -38 y^2+14  y^3\right) \nonumber\\[2mm]
  &\quad \left.   + 3  x^3 \left(23 +4  y+  6 y^2\right)     - 9  x^4 (7+ 3 y) +  27 x^5  \right] \,,  \nonumber  \\
f_2^{G_8, c^a_{dd} - c^a_{ss}}(x,y) & =  -\frac{x y   (1-y) }{2  (4-y-3  x)}  \,,  \nonumber \\
f_3^{G_8, c^a_{dd} - c^a_{ss}}(x,y) & = -\frac{ 1}{864  x^2  y ^2 (4 - y - 3  x)^2} \left[  (4-y)^2 (1-y)^4
   y^2 \right. \nonumber \\
   &  - 2  x y (1-y)^2  \left(24+110 y-89 y^2+19 y^3-y^4\right)   \nonumber \\[2mm]
& \quad- 3  x^2 \left(48-1120 y+ 4715 y^2  - 6236 y^3 + 3792
   y^4   \right.  \nonumber \\[2mm]
   & \hspace{6.8cm} \left. - 1136 y^5 + 161 y^6 -  8 y^7 \right) \nonumber \\[2mm]
   &  \quad + 18  x^3
   \left(14 -375 y+1438 y^2-1788 y^3+923 y^4 -236 y^5+ 24  y^6\right) \nonumber \\[2mm]
   & \quad   - 27  x^4 \left(4 -118 y+ 258 y^2 - 254  y^3 + 141 y^4 - 40 y^5\right)  \nonumber \\[2mm] 
   &  \hspace{6.0 cm} \left. + 162  x^5 y^2 (3- 2  y )  + 243   x^6 y^2  \right] \,,  \nonumber  \\
 f_4^{G_8, c^a_{dd} - c^a_{ss}}(x,y) & =   - \frac{1}{96  x^2  y ^2 (4-y-3  x)^2} \left[   3 y^2  (4-y)^2 (1-y)^4  \right.  \nonumber\\
 & \quad+ 6  x  y  (1-y)^2 
   \left(8-18 y+52  y ^2-40 y^3+7 y^4\right) \nonumber \\[2mm]
   &\quad + 3  x^2 \big( 48 - 352  y  + 801 y^2  - 906 y^3  + 836 y^4 - 722  y ^5 \nonumber \\[2mm]  
   & \hspace{1.7cm}+ 247 y^6- 24  y^7\big)  \nonumber \\[2mm]  
   & \quad- 2   x^3 \big(126-783
   y +1572  y ^2-1261 y^3+1944 y^4\nonumber \\[2mm]  
   & \hspace{1.7cm} -1290  y^5+232  y ^6\big)  \nonumber \\[2mm]  
   & \quad  + 3   x^4 \left(36 - 198 y + 458 y^2  + 228  y ^3 + 561 y^4 - 248
   y^5\right) \nonumber \\[2mm]
   &\quad \left.   - 18   x^5 y^2 (52  y+23) + 135   x^6 y^2  \nonumber \right] \,,  \\ 
 f_5^{G_8, c^a_{dd} - c^a_{ss}}(x,y) & =  \frac{1-y}{24  y^2  (x-y) (4-y-3  x)} \left[ y \left(4+3
   y -9 y^2 +2  y ^3\right)  \right. \nonumber \\
 &\quad \left. - x  \left(6+7 y -48 y^2+20 y^3\right) +  3  x^2 \left(2-13 y+6 y^2\right) \right]\,,   \nonumber  \\
 f_6^{G_8, c^a_{dd} - c^a_{ss}}(x,y) & =  \frac{1-y}{8   y^2  (x-y)(4-y-3  x)} \left[ y \left(4-21
   y +21 y^2-4 \, y^3\right)  \right. \nonumber \\
 & \quad\left. - x  \left(6-17 y+6 y^2+8 y^3\right) + 3  x^2 \left(2-5 y+4  y^2\right) \right]\,,   \nonumber  \\
 f_{7}^{G_8, c^a_{dd} - c^a_{ss}}(x,y) & =  \frac{1}{16  x^2 (x-y) (4-y-3  x)} \left[ y  (4-y) (1-y)^3  \right. \nonumber  \\
   & \quad+ x   y  \left(8  +7 y -26 y^2+11
   y^3\right) - x^2 \left(3+25 y -48 y^2+8 y^3\right) \nonumber  \\[2mm]
   & \hspace{2.3 cm}\left. +  x^3 \left(7 - 26 y  - 20 y^2\right) + x^4 (11+ 31 y) - 15 x^5    \right] \,. 
\end{align}
\vspace{0.5 cm}
Finally, the contribution proportional to $(c_{dd}^{v}-c_{ss}^{v})$ is given by
{\small
\begin{align}
 i A_{\rm NLO}^{G_{8}, c_{dd}^{v} -  c_{ss}^{v}}    &= \frac{m_{K^-}^2}{8 \pi^2 F^2_{\pi}}\bigg\{ \ln \frac{\mu}{m_K} g^{G_8, c^v_{dd} - c^v_{ss}}(x,y)  +  f_0^{G_{8}, c^v_{dd} - c^v_{ss}} (x,y) \nonumber  \\
&\quad+ \sum_{i=1}^7 s_i  (x,y) f_i^{G_{8}, c^v_{dd} - c^v_{ss}}(x,y)  -  (1-x+y)   \bigg[ \hat{L}_{4,r} (2+y) +\hat{L}_{5,r}  \frac{1 + 3 y}{4} \bigg]  \nonumber   \\ 
&\quad + \frac{1}{2 (4-y-3  x)} \left[ 4 \left(2 \hat{N}^{\prime}_{5, r}+2 \hat{N}^{\prime}_{8, r}-4 \hat{N}^{\prime}_{10, r}-4
   \hat{N}^{\prime}_{11, r}+\hat{N}^{\prime}_{21, r}+\hat{N}^{\prime}_{23, r} \right) \nonumber \right. \\
& \quad  +y
   \left(6 \hat{N}^{\prime}_{5, r}+10 \hat{N}^{\prime}_{8, r}+4 \hat{N}^{\prime}_{10, r}-4 \hat{N}^{\prime}_{11, r}-16 \hat{N}^{\prime}_{12, r}-9 \hat{N}^{\prime}_{21, r}-9 \hat{N}^{\prime}_{23, r} \right)  \nonumber   \\
 & \quad  - y^2 \left(2 \hat{N}^{\prime}_{5, r}-\hat{N}^{\prime}_{8, r}-2 \hat{N}^{\prime}_{11, r}-4 \hat{N}^{\prime}_{12, r}-6 \hat{N}^{\prime}_{21, r}-6 \hat{N}^{\prime}_{23, r} \right)\nonumber \\
&\quad -y^3 \left(\hat{N}^{\prime}_{8, r}+\hat{N}^{\prime}_{21, r}+\hat{N}^{\prime}_{23, r} \right) \nonumber \\
& \quad-  x \left(14 \hat{N}^{\prime}_{5, r}+ 14 \hat{N}^{\prime}_{8, r}-12
\hat{N}^{\prime}_{10, r}-12 \hat{N}^{\prime}_{11, r} -4 \hat{N}^{\prime}_{20, r}-\hat{N}^{\prime}_{21, r}+7 \hat{N}^{\prime}_{23, r}\right) \nonumber \\
  &  \quad- x  y  \left(4 \hat{N}^{\prime}_{5, r}  + 11 \hat{N}^{\prime}_{8, r} - 6 \hat{N}^{\prime}_{11, r} - 12 \hat{N}^{\prime}_{12, r}  -3 \hat{N}^{\prime}_{20, r} -\hat{N}^{\prime}_{21, r} -3 \hat{N}^{\prime}_{23, r} \right) \nonumber  \\
 & \quad -  x  y^2  \left(2 \hat{N}^{\prime}_{8, r}+ \hat{N}^{\prime}_{20, r}+ 2 \hat{N}^{\prime}_{21, r}+ 2 \hat{N}^{\prime}_{23, r} \right) \nonumber \\
& \quad + x^2 \left( 6 \hat{N}^{\prime}_{5, r}+6 \hat{N}^{\prime}_{8, r}-7 \hat{N}^{\prime}_{20, r} -3 \hat{N}^{\prime}_{21, r} +3
   \hat{N}^{\prime}_{23, r}\right)  \nonumber \\ 
& \quad \left. + x^2  y     \left( 3 \hat{N}^{\prime}_{8, r}-2 \hat{N}^{\prime}_{20, r}+3 \hat{N}^{\prime}_{21, r} +3\hat{N}^{\prime}_{23, r} \right)   + 3  x^3 \hat{N}^{\prime}_{20, r}  \right] \bigg\}  \,,  
\end{align}}  
with
\begin{align}
g^{G_8, c^v_{dd}- c^v_{ss}}(x,y) & =  \frac{13}{18} -\frac{35}{9} y -\frac{5}{2} y^2  + \frac{23}{12} x    + \frac{13}{4}  x y -\frac{3}{4} x^2  \nonumber \,, \\ 
 f_0^{G_8, c^v_{dd}- c^v_{ss}}(x,y) & =  -\frac{1}{72} \bigg( \frac{(1-y)^2 (17+ 13 y)}{x} +  1+ 54  y - 7 y^2 \notag\\&\hspace{2cm} -x (45+ 33 y -27 x) \bigg)    \nonumber \,,   \\
f_1^{G_8, c^v_{dd}- c^v_{ss}}(x,y) & = - \frac{  (7-y) (1-y)^2+  x \left(11+ 18 y- 5 y^2 \right) - 3  x^2 (y+9) + 9 x^3}{144  x^2}   \nonumber  \,, \\
f_{2,5,6}^{G_8, c^v_{dd} - c^v_{ss}}(x,y) & = 0    \,, \nonumber  \\
f_3^{G_8, c^v_{dd} - c^v_{ss}}(x,y) & =  \frac{1}{864  x^2} \left(  (7-y)
   (1-y)^3 -  4 x \left( 34 - 49 y + 17 y^2 - 2  y ^3 \right)  \right. \nonumber \\
& \left. \hspace{2.3 cm}  -  6 x^2 \left( 59 - 3 y^2 - 8 y\right) + x^3 (72-54  y) +  27 x^4  \right) \nonumber \,,  \\
 f_4^{G_8, c^v_{dd} - c^v_{ss}}(x,y) & = -\frac{1-x+y}{32  x^2}  \left( 3 (1-y)^3- x (1-y^2) - x^2 (7+33  y) + 5 x^3  \right)  \nonumber \,,  \\
 f_7^{G_8, c^v_{dd} - c^v_{ss}}(x,y) & =   -\frac{3}{16  x^2 } (1-y)^2 (1+ y)  + \frac{1}{16  x} \left(1-10 y+y^2\right)  + \frac{7}{16}  (1+y) -\frac{5}{16} x \,.  
\end{align}

\subsubsection[$G_{8}^\theta$ contribution]{\boldmath $G_{8}^\theta$ contribution} 

By construction, the $G_{8}^\theta$ contribution is proportional to the single ALP coupling $\tilde{c}_{GG}$. It is given by
{\small
\begin{align}
 i A_{\rm NLO}^{G_{8}^\theta, \tilde{c}_{GG}}    &=  \frac{m_{K^-}^2}{8 \pi^2 F^2_{\pi}}\bigg\{    (1-y) (4 + 8  y - 3  x )   \ln \frac{\mu}{m_K} \nonumber \\
 & \quad+ \frac{1}{6x}(1-y)(4x(1+y) - 9 x^2 + 5(1-y)^2) \nonumber\\
 &\quad+ \frac{s_1(x,y)}{12 }  \frac{1-y}{x} \left(\frac{(y-1)^2}{x}+ 2  (1+y) -3  x\right) \nonumber  \\
 & \quad-  \frac{s_3(y)}{72} \left(\frac{(1-y)^4}{x^2}- (19-5 y) \frac{(1-y)^2}{x}-9 x (1-y)-3  \left(15+ 10 y-y^2\right) \right)  \nonumber \\
 & \quad + \frac{s_4(y)}{8} \left( \frac{3 (1-y)^4}{x^2}   - \frac{(1+y) (1-y)^2}{x} - 7 - 26 y  +25 y^2+ 5 x (1-y) \right)  \nonumber \\
  &\quad+  \frac{s_7(x,y)}{4} \frac{1-y}{x}  \left( 3\frac{(1-y)^2}{x} + 2  (1+y)-5 x \right)  \nonumber\\
  &\quad+  (1-y) \big(2(4+2  y )   \hat{L}_{4,r} +   (1+ 3 y) \hat{L}_{5,r} \big) \bigg\}  \,.   
\end{align} } 

\subsubsection[$G_{8}^\prime$ contribution]{\boldmath $G_{8}^\prime$ contribution} 

Finally, the contribution proportional to $G_{8}^\prime$ is non-zero only for two combinations of ALP couplings, $(c_{dd}^{a} -  c_{ss}^{a})$ and $(c_{dd}^{v} -  c_{ss}^{v})$. The corresponding NLO contributions to $\mathcal{A}^{\rm FC}$ are given by
{\small
\begin{align}
 i A_{\rm NLO}^{G_{8}^\prime, c_{dd}^{a} -  c_{ss}^{a}}    &= \frac{m_{K^-}^2}{8 \pi^2 F^2_{\pi}} \frac{(1-y)^2}{2} \bigg\{  - \left(4+\frac{16}{3}  y-3  x\right) \ln \frac{\mu}{m_K} \notag  \\ 
 &\quad - \frac{5 (1-y)^2}{6 x} - \frac23 (1+ y)+ \frac32 x  - \frac{s_1 (x,y)}{12}   \left(\frac{(1-y)^2}{x^2} + 2 \frac{ 1+y}{x}-3\right)   \notag \\
 & \quad+ \frac{s_3(y)}{72}\! \bigg(  \frac{ (1-y) ( 1 - 2  y + y^2) }{x^2} - \frac{(1-y)(19 - 5 y)}{x}  \notag\\&\hspace{2cm}- 9  \frac{\left(5+14  y-3  y^2\right)}{1-y}-9  x \bigg) \notag   \\
  & \quad +\frac{s_4(y)}{8}  \left(-3 \frac{(1-y)^3}{x^2}+\frac{1-y^2}{x} +  \frac{7+26 y- 17 y^2}{1-y} -5 x\right) \notag \\
  & \quad-\frac{s_7(x,y)}{4} \left(-5+2 \frac{ 1+y}{x}+ 3 \frac{(1-y)^2}{x^2} \right)  \notag\\&\quad-   2 (2 + y) \hat{L}_{4,r} +  x \hat{L}_{5,r} - 4 y \hat{L}_{8,r}   \bigg\}\,, 
 \end{align}}
 {\small
\begin{align}
 i A_{\rm NLO}^{G_{8}^\prime, c_{dd}^{v} -  c_{ss}^{v}}    &= \frac{m_{K^-}^2}{8 \pi^2 F^2_{\pi}} \bigg\{  \left(2 + \frac{16}{3} y- \frac32 x \right)  \ln \frac{\mu}{m_K}\notag  \\
 & \quad- \frac34 x + \frac{5}{12 x}(1-y)^2 + \frac{1+y}{3} + \frac{s_1 (x,y)}{24} \left( \frac{(1-y)^2}{x^2} + 2  \frac{1 + y}{x} - 3 \right) \notag \\
 &\quad + \frac{s_3 (y)}{144} \left( - \frac{(1-y)^3}{x^2} + \frac{19 - 24  y +5  y^2}{x} + 3  (15 - 7 y) + 9 x \right) \notag \\
 &\quad + \frac{s_4(y)}{16} \left(  3 \frac{ (1-y)^3}{x^2} - \frac{1-y^2}{x} - 7 - 33 y  + 5  x \right) \notag \\
 & \quad+ \frac{s_7(x,y)}{8} \left(  3 \frac{(1-y)^2}{x^2} +2 \frac{ 1+ y}{x} -5 \right) \notag\\&\quad +  (2+y) \hat{L}_{4,r} + \frac{4 y -x}{2} \hat{L}_{5,r} -2 y \hat{L}_{8,r}  \bigg\}  \,. \notag\\
 \end{align}}

\subsection[$\mathcal{A}^{\rm FV}$ at NLO]{\boldmath $\mathcal{A}^{\rm FV}$ at NLO} 

Finally, for the contribution proportional to flavor-violating ALP couplings between strange and down quarks we obtain
 {\small
\begin{align}
 i A_{\rm NLO}^{\rm FV}    &= \frac{m_{K^-}^2}{8 \pi^2 F^2_{\pi}} \bigg\{   \frac32 x \ln \frac{\mu}{m_K}  - \frac{5}{12} \frac{(1-y)^2}{x} - \frac{1+y}{3}+ \frac34 x \nonumber \\
 & \quad - \frac{s_1(x,y)}{24 }  \left(\frac{(1-y)^2}{x^2} +2 \frac{ (1+y)}{x}-3 \right)   \notag \\
 &\quad + \frac{s_3 (y)}{144} \left( \frac{(1-y)^3}{x^2} -  \frac{ 19-5  y}{x} (1-y)  - 3  \frac{15+ 10 y-y^2}{1-y}  - 9  x   \right) \notag \\
 & \quad - \frac{s_4(y)}{16 }   \left( 3\frac{ (1-y)^3}{x^2} - \frac{ 1-y^2}{x} - \frac{7-6 y+ 7 y^2}{1-y}+ 5  x  \right)  \notag\\
 &\quad -\frac{s_7 (x,y)}{8}  \left(3 \frac{ (1-y)^2}{x^2}+2 \frac{1+y}{x}-5 \right)  + \frac{1}{2} x  \hat{L}_{5,r} \bigg\}\,.    
 \end{align}}
\end{appendix}

\section{Remarks on mixing near degeneracy}
\label{app:C}

Consider two real neutral fields $\phi_1$, $\phi_2$. For simplicity we consider only mass mixing. The relevant Lagrangian reads
\begin{align}
\mathcal{L} = \frac12 (\partial^\mu \Phi)^T (\partial_\mu \Phi) - \frac12 \Phi^T M^2 \Phi\,,
\end{align}
where $\Phi = (\phi_1,\phi_2)^T$, and 
\begin{align}
M =  \begin{pmatrix}
 m_{1,0}^2 & \delta \\
 \delta & m_{2,0}^2 
\end{pmatrix}.
\end{align}
We want to find an orthogonal matrix 
\begin{align}
R=  \begin{pmatrix}
 \cos \theta & -\sin \theta \\
 \sin \theta &  \cos \theta 
\end{pmatrix},
\end{align}
such that $R^{T} M^2 R = \mathrm{diag}(m_1^2,m_2^2)$. The physical states will then be related to the Lagrangian ones via $\Phi = R\,\Phi^{\mathrm{phys}}$. The condition for diagonalization is given by
\begin{align}
\tan 2 \theta = \frac{2 \delta}{ m_{1,0}^2-  m_{2,0}^2}\,,
\label{eq:tan_2theta}
\end{align}
and the eigenvalues read 
\begin{align}
\begin{aligned}
m_1^2 & = m_{1,0}^2 \cos^2 \theta +  m_{2,0}^2 \sin^2 \theta + 2 \,  \delta \sin \theta \cos \theta \,,  \\
m_2^2 & = m_{2,0}^2 \cos^2 \theta +  m_{1,0}^2 \sin^2 \theta - 2 \, \delta \sin \theta \cos \theta \,.    \\
\end{aligned}
\end{align}
Since the tangent has period $\pi$, we can choose for $\theta$ any sector with opening angle $\pi/2$, such as $\theta \in [ -\pi/4,\pi/4]$. In this domain, we find
\begin{align}
\begin{aligned}
\sin \theta & = \mathrm{sign}\left( \frac{\delta}{m_{1,0}^2-  m_{2,0}^2}\right) \frac{1}{\sqrt{2}} \sqrt{1 - \frac{|m_{1,0}^2-  m_{2,0}^2|}{\sqrt{(m_{1,0}^2-  m_{2,0}^2)^2+4 \delta^2}}} \,,\\
\cos \theta & = \frac{1}{\sqrt{2}}\sqrt{1 +\frac{|m_{1,0}^2-  m_{2,0}^2|}{\sqrt{(m_{1,0}^2-  m_{2,0}^2)^2+4 \delta^2}}}\,.
\end{aligned}
\end{align}
In the limit where $m_{1,0}$ goes to $m_{2,0}$, one finds $\cos \theta, |\sin \theta| \to 1/\sqrt{2}$, where the sign of the sine depends on whether $m_{1,0}$ approaches $m_{2,0}$ from above or below. This sign change is the reason why some of the amplitudes in Section \ref{sec:amplitude_LO} are not continuous in $m_{a,0} = \tilde{m}_{\pi^0}$.

For $m_{1,0} \neq m_{2,0}$ and $ \left| \frac{\delta}{m_{1,0}^2-m_{2,0}^2} \right| \ll 1 $, we have the expansion 
\begin{align}
\cos \theta &\approx 1, &\sin \theta &\approx  \frac{\delta}{m_{1,0}^2-m_{2,0}^2}\,,
\end{align}
which leads to
\begin{align}
\begin{aligned}
\phi_1 & = \phi_1^{\mathrm{phys}} + \frac{\delta}{m_{1,0}^2-m_{2,0}^2}  \, \phi_2^{\mathrm{phys}} \,, \\
\phi_2 & = \phi_2^{\mathrm{phys}} - \frac{\delta}{m_{1,0}^2-m_{2,0}^2}\,  \phi_1^{\mathrm{phys}} \,,
\label{eq:field_definitions}
\end{aligned}
\end{align}
with $m_1 \approx m_{1,0} $ and $m_2 \approx m_{2,0}$.

\renewcommand{\theequation}{C.\arabic{equation}}
\setcounter{equation}{0} 

\bibliographystyle{JHEP}
\bibliography{references}

\end{document}